%% file: Final_mnras_R2_paper.tex
\documentclass[a4paper,fleqn,usenatbib]{mnras}

\usepackage[T1]{fontenc}
\usepackage{ae,aecompl}

\usepackage{graphicx}	
\usepackage{amsmath}	
\usepackage{amssymb}	
\usepackage{lmodern,amsmath,amssymb}

\usepackage[switch]{lineno}

\usepackage{enumitem}

\title[Ion density distribution in comet C/2016 R2]{A 
physico-chemical model to study the ion density distribution in the 
inner coma of  comet C/2016 R2 (Pan-STARRS)}

\author[S. Raghuram et al.]{
Susarla Raghuram,$^{1}$\thanks{E-mail: \href{raghuramsusarla@gmail.com} 
{raghuramsusarla@gmail.com}}
Anil Bhardwaj,$^{1}$
Damien Hutsemékers,$^{2}$
Cyrielle Opitom,$^{3}$  \newauthor 
Jean Manfroid,$^{2}$ and
Emmanuel Jehin$^{2}$ 
\\
$^{1}$Physical Research Laboratory, Ahmedabad, 380009, India.\\
$^{2}$STAR Institute - University of Liège, Allée du 6 Août 
19c, B-4000 Liège, Belgium.\\
$^{3}$ESO (European Southern Observatory) - Alonso de 
Cordova 3107, Vitacura, Santiago Chile.
}

\date{Accepted 2020 December 4. Received 2020 December 4; in original form 
2020 August 23}

\pubyear{2020}

\begin{document}
\label{firstpage}
\pagerange{\pageref{firstpage}--\pageref{lastpage}}
\maketitle

\begin{abstract}
The recent observations  show that   comet C/2016 R2 (Pan-Starrs) has a 
unique and peculiar composition when compared with several other comets 
observed at 2.8 au heliocentric distance. Assuming solar resonance fluorescence 
is the only excitation source, the  observed  ionic emission intensity ratios 
are  used to constrain the corresponding neutral abundances in this comet.   We 
developed a physico-chemical model to study the  ion density distribution in 
the inner coma of this comet by accounting for photon and electron 
impact ionization of neutrals, charge exchange and proton transfer reactions 
between ions and neutrals, and electron-ion thermal recombination reactions. 
Our calculations show that CO$_2^+$ and CO$^+$ are the  
major ions in the inner coma, and close to the surface of nucleus CH$_3$OH$^+$, 
CH$_3$OH$_2^+$ and O$_2^+$ are also important ions. By considering various 
excitation sources, we also studied the emission 
mechanisms of different excited states of CO$^+$, CO$_2^+$, N$_2^+$, and 
H$_2$O$^+$.  We  found that the photon and electron impact ionization and 
excitation of corresponding neutrals significantly contribute to the  observed 
ionic emissions for radial distances smaller than 300 km and at larger 
distances,  solar resonance fluorescence is the major excitation source. Our 
modelled ion emission intensity ratios are consistent with the ground-based 
observations. Based on the modelled emission processes, we suggest that   the 
observed ion emission intensity ratios  can be used to derive the neutral 
composition in the cometary coma only when the ion densities are significantly 
controlled by photon and photoelectron impact ionization of neutrals 
rather than by the ion-neutral chemistry.     
\end{abstract}



\section{Introduction}
Cometary nuclei are formed due to the agglomeration of different icy grains  
and dust particles in the outer reaches of {the} Solar 
nebula. As the comet approaches the Sun, sublimation of these ices  causes a 
giant gaseous transient atmosphere around the nucleus, which is called a 
cometary coma.  At a given heliocentric distance, sublimation rate of the 
nucleus is the key deciding factor that determines the dynamical activity of 
the cometary coma.  Most of the observations have shown that during 
the peak activity of a comet, water is the most dominant component  of 
{a} cometary coma, whereas at larger heliocentric distances CO and 
CO$_2$ are the  {dominant} species. The interaction of solar radiation 
with cometary species drives a chain of chemical reactions in the inner coma 
and also leads to various spectroscopic emissions.  
Remote observation of different spectroscopic emissions from neutrals and ions 
is a potential tool to study the global composition  {of} comets and the  
dynamical activity of cometary plasma \citep{Feldman04,Bockelee04}.  For an 
active comet, like 1P/Halley, the solar wind starts to interact with cometary 
neutrals and ionic species more than a  million kilometres away from the 
nucleus which leads to complex structures in the cometary plasma and magnetic 
fields \citep{Ip04}. {Depending} on the cometary activity and 
solar wind conditions, various dynamical plasma boundaries, such as bow shock, 
the cometopause, the collisionopause, and the diamagnetic cavity,  
manifest in the cometary coma. A detailed description of these boundaries is 
provided in a tutorial by \cite{Cravens89}. 

The study of cometary plasma composition  {has been} subjected to a great 
interest after the ion mass spectrometer onboard Giotto spacecraft detected 
many peaks in the mass range 12 and 120 amu \citep{Balsiger86, Krankowsky86, 
Mitchell87, Altwegg93}. By developing photochemical models, numerous studies 
focused  on comet 1P/Halley  explained the observed ion distribution 
in a 
water-dominated coma. \citep{Allen87, 
Wegmann87, Schmidt88, Cravens89, Ip90, Gan90, Bhardwaj90, Bhardwaj96, 
Bhardwaj99a, Haider93, Haberli95, Haider97, Haider05,Rubin09, Cordiner14}. 
By making two years of observations, the recent
Rosetta space mission on comet 67P/Churyumov-Gerasimenko has revolutionized 
our understanding  {of} the activity of the cometary coma.  During the 
Rosetta observation period, continuous measurements around the nucleus were 
helpful to study the evolution of ion and neutral distribution and also the 
driving photochemical processes in the coma. Several modelling works on this 
comet have shown that ion composition in the coma  {varies} based on the 
sublimation rate  of the nucleus \citep{Vigren13, Fuselier15, Fuselier16, 
Galand16, Vigren17, Heritier17, Heritier18, Beth19}. All these studies show 
that solar photons are the primary energy source that determine the ion 
composition in the inner coma. Solar extreme ultraviolet photons having 
 {an} energy more than 12 eV  {ionize} H$_2$O and 
 {produce} H$_2$O$^+$, and the collisions among these species quickly 
 {lead} to the formation of H$_3$O$^+$. The sublimated parent species 
such as CH$_3$OH, NH$_3$, HCN, HCOOH, CH$_3$CHO, have high proton affinities 
compared to that of H$_2$O,  {causing} the loss of H$_3$O$^+$ in the 
inner coma. \cite{Haider05} developed a comprehensive chemical network to study 
the ion distribution in comet 1P/Halley.  Their calculations show that  
NH$_4^+$ is the most dominant ion in the inner coma followed by H$_3$O$^+$ and 
CH$_3$OH$_2^+$ ions. Similarly, the model calculations of \cite{Heritier17} on 
comet 67P/Churyumov-Gerasimenko  {showed} that  NH$_4^+$, 
CH$_3$OH$_2^+$, H$_3$O$^+$, H$_3$S$^+$, HCNH$^+$ are the important ions 
in the inner coma. They also  {showed} that the densities of these ions  
vary with the relative mixing ratios of corresponding proton affinity 
species coming from the nucleus.  {Even if} the mixing ratios of parent 
species, which  {have} high proton affinity, are very low (<2\%), 
 {they} can 
play a significant role in modifying the ionospheric composition of the inner 
coma.  Hence, the ion distribution in the cometary coma  essentially depends  
on the neutral composition and  photochemical reactions.

 {The main volatile constituent of the inner coma, H$_2$O, cannot be 
detected in the visible due to the lack of 	electronic transition, which is 
also the case for CO$_2$ and N$_2$.} Several ground based observatories studied 
the composition of cometary plasma by making spectroscopic observation of 
different ions. The observed emission intensities of these ions 
were used to constrain their respective parent species abundances in the coma. 
H$_2$O$^+$ has been observed in various comets from ground-based observatories 
\citep{Delsemme79, Disanti90, Haberli97,  Wegmann99}. Spectroscopic 
emissions from CO$^+$ have been used to study the dynamics of the plasma in 
cometary ion tails \citep{Ip04, Arpigny64, Krishnaswamy79, Swamy86, Combi80, 
Larson80, Cochran91, Cochran00}. So far a few detections of N$_2^+$ emissions 
on comets from ground based observations are reported \citep{Wyckoff89, Lutz93, 
Korsun08, Korsun14}. But the presence of N$_2$ was more conclusive in comet 
 {67P/Churyumov-Gerasimenko} from Rosetta in situ measurements 
\citep{Rubin15}. As discussed before, besides the composition of parent 
species, the collisional  chemistry  can significantly change the ion 
composition in the inner coma. Hence, the knowledge of formation and loss 
processes of different ions is essential to constrain their 
 {corresponding} neutral abundances based on the observed ionic emission 
intensities.


Several  recent observations  {of} comet C/2016 R2 (Pan-Starrs) - 
hereafter C/2016 R2 - have shown that the cometary coma  is  predominantly 
composed of CO and remarkably depleted in water, when it was at 2.75 au away 
from the Sun \citep{Biver18,Cochran18, Wierzchos18, Opitom19, McKay19, 
Kumar20}. Several multi-wavelength observations  {of} this comet have 
shown that  {it} has substantially low H$_2$O production rate, which is 
contrary to many cometary observations made at this heliocentric distance 
\citep{McKay19,Biver18}.  {\cite{Cochran18} were the first to report the 
strong emission features of CO$^+$ comet tail bands (A$^2\Pi$ $\rightarrow$ 
X$^2\Sigma$) and N$_2^+$ first negative bands (B$^2\Sigma^+$ $\rightarrow$ 
X$^2\Sigma$)  in comet C/2016 
R2, followed  by \cite{Biver18} and \cite{Opitom19}. Besides these emissions,  
\cite{Opitom19} also observed  CO$_2^+$ 
Fox-Duffendack-Barker band system ($\widetilde{A}^2\Pi$ $\rightarrow$ 
$\widetilde{X}^2\Pi$) in their high resolution optical spectra  {of} this 
comet.} \cite{Kumar20} reported a tentative detection of 
H$_2$O$^+$($\widetilde{A}^2A_1$ $\rightarrow$ $\widetilde{X}^2B_1$) (0-8-0) 
emission in their low resolution observed spectra. But they also claim that the 
detection is not conclusive due to the strong blend of CO$^+$ doublet emission. 
However, it should be noted that these H$_2$O$^+$ emissions are not seen in  
 {other} high resolution optical observations. Assuming that 
 {solar resonance florescence} is the only excitation source, the 
measured ionic emission intensity 
ratios are used to constrain their respective neutral composition  {in} 
the coma. As discussed before, besides the neutral composition, the  ion 
density in the  coma is strongly controlled by ion-neutral chemical reactions.  
In this case, the conversion of observed ion emission intensity ratios into 
corresponding neutral density  {ratios} is not straightforward.


In the light of  recent observations  {of} comet C/2016 R2 
\citep{Opitom19}, we aim to explore the photochemistry of CO-dominated and 
water-depleted comet and also the emission mechanisms of different ions. 
We developed a physico-chemical model for comet C/2016 R2 to study the ion 
density distribution in the inner coma by incorporating  different excitation 
sources and various chemical reactions. Using our model, we also 
studied  various photochemical processes of different ionic 
emissions in this comet, which were observed by various ground based 
observatories. We describe the model inputs such as the neutral composition of 
the coma, the atomic and molecular parameters, the chemical 
network of ion-neutral chemistry, and the photochemical reactions of excited 
states of ions in Section~\ref{sec:input}. In this section, we also explain the 
calculation of radial density profiles of different ions and the emission 
intensity profiles of the excited states of CO$_2^+$, CO$^+$, N$_2^+$, and 
H$_2$O$^+$ as a function of the nucleocentric projected distance. The modelled 
production and loss mechanisms of different ions and their density 
distribution, volume emissions rates of excited states, and intensity ratios  
of different ionic emissions are presented in 
Section~\ref{sec:results}. The inferences of the model calculations are 
discussed in Section~\ref{sec:discuss}. We 
summarize the current work and draw the conclusions in 
Section~\ref{sec:sum_con}. 

\vspace{-15pt}
\section{Observations}
\label{sec:data}
Observations of comet C/ 2016 R2 were carried out on 2018 February 11--16 with 
the Ultraviolet-Visual Echelle Spectrograph (UVES) mounted on the 8.2 m UT2 
telescope of the European Southern Observatory Very Large Telescope. The slit 
width of 0.44\arcsec\ gives a resolving power R $\simeq$ 80000. Full account of 
the observations and data reductions is given in \cite{Opitom19}. 
\cite{Opitom19} used the measured fluxes for  N$_2^+$ 
(B$^2\Sigma^+$$\rightarrow$X$^2\Sigma$) (0-0), CO$_2^+$ 
($\widetilde{A}^2\Pi$$\rightarrow$$\widetilde{X}^2\Pi$) (0-0), and 
CO$^+$ (A$^2\Pi$$\rightarrow$X$^2\Sigma$) (2-0) band emissions, at 
respective wavelengths 391, 351 and 425 nm, to derive the ion density ratios in 
the coma. They also considered H$_2$O$^+$ 
($\widetilde{A}^2A_1$$\rightarrow$$\widetilde{X}^2B_1$) (0-8-0) band emission  
to constrain the upper limit of the H$_2$O$^+$ abundance in the coma. Assuming 
solar resonance fluorescence is the only excitation mechanism, the observed 
emission flux ratios are used to derive ionic abundance ratios. They derived 
ionic ratios of N$_{2}^{+}$/CO$^{+}$, CO$_{2}^{+}$/CO$^{+}$, and 
H$_{2}$O$^{+}$/CO$^{+}$ in the cometary coma  {of} 0.06$\pm$0.01,  
1.1$\pm$0.3, and  $<$0.4, respectively. These ratios were computed using 
intensities averaged over the full slit  {length}. For the UVES blue arm 
spectra where the ionic emissions of interest are located, the slit extends 
over $\sim$1.5 $\times$ 10$^4$~km at the comet distance. Surface brightnesses 
were also measured on the two-dimensional spectra, cutting the slit in chunks 
as done in \cite{Raghuram20} for the [OI] spectral lines (which 
are located in the red arm spectra). For the blue settings, the spectra were 
rebinned along the spatial dimension so that the final pixel size projected 
onto the comet corresponds to 0.125\arcsec\ (about 220 km). Seven spatial 
chunks were defined, the central one ranging from -3 to +3 pixels, and the 
other ones corresponding to the pixel ranges [-24,-18], [-17,-11], [-10,-4], 
[4,10], [11,17], and [18,24]. The intensity ratios are given in 
Table~\ref{tab:ratios} as a function of the projected nucleocentric distance 
(radius). The radius is given as the central value in each subslit plus or 
minus the range divided by two. The measurements done on each side of the comet 
in a given range of nucleocentric distances were averaged. The errors are 
dominated by the uncertainties from the fitting procedure \citep{Opitom19}. 
There is no significant variation of the I(N$_{2}^{+})$/I(CO$^{+}$) intensity 
ratio with the nucleocentric distance, while the I(CO$_{2}^{+}$)/I(CO$^{+}$) 
ratio apparently decreases by a factor 2 at $\sim$5000~km from the nucleus.

\begin{table} 
	\caption{The measured ionic emission intensity ratios on comet C/2016 R2 
		as a function of the projected distance
	when it was at 2.75 au heliocentric distance}
    \vspace{-0.2cm}
	\label{tab:ratios}
	\centering
	\begin{tabular}{lcc}
		\hline\hline
		Radius             & I(N$_{2}^{+})$/I(CO$^{+}$)  & 
		I(CO$_{2}^{+}$)/I(CO$^{+}$) \\
		($10^3$ km)        & &  \\
		\hline
		0.3875 $\pm$ 0.3875  & 1.14 $\pm$ 0.2 & 0.17 $\pm$ 0.04 \\
		1.550 $\pm$ 0.775    & 1.18 $\pm$ 0.2 & 0.17 $\pm$ 0.04 \\
		3.100 $\pm$ 0.775    & 1.19 $\pm$ 0.2 & 0.14 $\pm$ 0.04 \\
		4.650 $\pm$ 0.775    & 1.24 $\pm$ 0.2 & 0.08 $\pm$ 0.04 \\
		\hline
	\end{tabular}
\end{table}

\vspace{-15pt}
\section{Model inputs and calculations} \label{sec:input}
The detailed description of model calculations is provided in our earlier work 
\citep{Bhardwaj12, Raghuram13, Raghuram14, Decock15, Raghuram16, 
Raghuram20}. The model inputs such as the heliocentric and geocentric 
distances, the nucleus sublimation rate, and the gaseous composition 
of  C/2016 R2  {coma} are the same as described in our recent work 
\citep{Raghuram20}. Here we briefly describe the neutral distribution of coma, 
the atomic and molecular parameters, and the chemical network used for 
calculating the ion density profiles and emission intensity  {profiles}  
of various ions.

\vspace{-15pt}
\subsection{Neutral distribution}
We consider eight neutral species viz., H$_2$O, CO, CO$_2$, 
N$_2$, CH$_3$OH, CH$_4$, NH$_3$, and O$_2$ as the primary composition of the 
cometary coma. Abundances of these species are taken from the  
various ground-based observations which were made when the comet was at 
 {a} heliocentric distance of 2.8 au. At this heliocentric distance, 
\cite{Biver18} observed that CO is the most dominant species in the coma with a 
gas production rate of 1.1 $\times$ 10$^{29}$ s$^{-1}$. We have 
taken the relative abundances of other neutral species viz., H$_2$O, CO$_2$, 
N$_2$, CH$_3$OH, and CH$_4$  as 0.3\%, 18\%, 7\%,
1.1\%, and 0.6\%, respectively, with respect to CO production 
rate, from the observations of \cite{Biver18} and \cite{McKay19}.

 {In a water-dominated comet, NH$_3$ can play an important role in 
modifying the ion-density distribution in the inner coma due to high proton 
affinity.  In comet 1P/Halley, with a 1.5\% mixing ratio in the coma (relative 
to water production rate), \cite{Haider05} showed that  NH$_3$ quickly reacts 
with water ion and produces NH$_4^+$ as the dominant ion  in the inner coma. 
Similarly, the modelling works of \cite{Vigren13} and \cite{Heritier17} in 
comet 67P/Churyumov-Gerasimenko also showed that  NH$_3$ and CH$_3$OH react 
with H$_2$O$^+$ and produce protonated ions. To explore the role of NH$_3$ and 
CH$_3$OH in the ion-neutral chemistry of CO-dominated coma, we incorporated the 
abundance of these species  based on the observations of  \cite{McKay19} and 
\cite{Biver18}. We considered 0.01\% of NH$_3$ relative to CO production 
rate in the model  based on the derived upper limit by \cite{McKay19}.}

 {The detection of O$_2$ in comets 67P/Churyumov-Gerasimenko and 
1P/Halley suggest that this species might be a common and abundant primary 
species \citep{Bieler15,Rubin15}. To explore the role of this neutral species 
in determining the O$_2^+$ ion density in the C/2016 R2 coma, we  assumed
1\% of O$_2$ abundance with respect to CO production rate \citep{Raghuram20}. 
However, we show that the inclusion of this species in the model with a large 
abundance  does not influence the modelled ion-density profiles.}

 {Other species such as HCN, C$_2$H$_6$ and H$_2$CO are also detected  in 
this comet. However, their production rates are smaller by more than three 
orders 
of magnitude compared to that of  CO  \citep{McKay19}. Due to their low 
relative abundances in comet C/2016 R2, they do not play any significant role 
in the ion-neutral chemistry. Hence, we did not account for these species in 
our model.} 

Density profiles for the primary neutral species are determined 
using the Haser's distribution formula and the model calculations are done 
under spherical symmetric assumption \citep{Haser57}. The neutral gas 
expansion velocity profile is taken from the hydro-dynamical calculations of
\cite{Ip83} for the CO-dominated  coma. We assumed that the electron 
temperature profile 
is the same as the thermal temperature profile  derived by \cite{Ip83} and 
the 
impact of this assumption on the modelled emission intensity ratios will be 
discussed later. Our baseline model input parameters are tabulated in 
Table~\ref{tab:baseline}.

\begin{table}
	\renewcommand{\thefootnote}{\fnsymbol{footnote}}
	\caption{Summary of the baseline input parameters used in the model}
	\vspace{-0.2cm}
	\begin{tabular}{llllll}
	\hline\hline
	 CO production rate	& Q$_{CO}$ = 1.1 $\times$ 10$^{29}$ s$^{-1}$ \\
	 Neutral composition\footnotemark[2] & H$_2$O(0.3\%), CO$_2$(18\%)\\ 
	                                     &N$_2$(7\%), 
	 CH$_3$OH(1.1\%),  \\
	 &  CH$_4$(0.6\%), 
	    NH$_3$(0.01\%), \\
	 &  and O$_2$(1\%)  \\
	 Heliocentric distance & 2.8 au \\
	 Geocentric distance & 2.44 au \\
	 Neutral gas expansion velocity & \cite{Ip83} \\
    \hline	
    \end{tabular}
   \footnotemark[2]{The values in the brackets are the abundances of the 
   species relative to the CO production rate}
   \label{tab:baseline}
\end{table}

\subsection{Atomic and molecular parameters}
\subsubsection{Cross sections}
The photon absorption and ionization cross sections of various neutrals
are taken from the compilation of \cite{Huebner92}, which are accessible from a 
web link \href{https://phidrates.space.swri.edu}
{(https://phidrates.space.swri.edu)}. Electron impact ionization and 
excitation cross sections of neutral species are compiled from different works 
\citep{Itikawah2o,Itikawaco2,Itikawao2,Shirai01,Tabata06,LiuCO,
Srivastava96,Straub97,Rao92}.

The branching ratios for the photoionization of  H$_2$O, N$_2$, CO$_2$, and CO, 
producing the respective excited H$_2$O$^+$($\widetilde{A}^2$A$_1$), 
N$_2^+$(B$^2\Sigma^+_u$), CO$_2^+$($\widetilde{A}^2\Pi_{u,3/2}$) and 
CO$^+$(A$^2\Pi$) are taken from \cite{Avakyan98}. Based on the measured photon 
branching ratio, we assumed that about 50\% of H$_2$O$^+$ is produced in the 
$\widetilde{A}^2$A$_1$ excited state for electron impact ionization.
Later we discuss the impact of this assumption on the modelled intensity 
profile.  The electron impact cross sections for CO$_2^+$, CO$^+$, and N$_2^+$ 
producing in the  $\widetilde{A}^2\Pi_{u,3/2}$, A$^2\Pi$, and B$^2\Sigma^+_u$ 
excited states, respectively, are taken from \cite{Shirai01} and 
\cite{Tabata06}. We obtained Frank-Condon factors and branching ratios for the 
observed electronic transition of CO$^+$ (2-0), CO$_2^+$(0-0), N$_2^+$(0-0) and 
H$_2$O$^+$(8-0) from different theoretical works \citep{McCallum71, Kim99, 
Judge73, Arqueros82, Jain66, Lutz93, Lofthus77, Lutz87}. We multiplied these 
branching ratios and Frank-Condon factors with corresponding volume emission 
rates, which are calculated for photon and electron impact excitation of 
neutral, to calculate the corresponding band emission intensities of the ions.

The solar resonance fluorescence efficiencies (g-factors) of  H$_2$O$^+$, 
CO$^+$, N$_2^+$, and CO$_2^+$ for corresponding ionic band emissions  
are taken as 4.2 $\times$ 10$^{-3}$ (g$_{H_2O^+}$), 3.55 $\times$ 10$^{-3}$ 
(g$_{CO^+}$),  7 $\times$ 10$^{-2}$ (g$_{N_2^+}$), and  4.96 $\times$ 10$^{-4}$ 
(g$_{CO_2^+}$) photons s$^{-1}$ mol$^{-1}$ from \cite{Lutz93}, 
\cite{Magnani86}, \cite{Lutz93}, and \cite{Kim99}, 
  respectively.  {These excitation 
rate factors are scaled as a function of inverse square of heliocentric 
distance of the comet.}

\vspace{-10pt}
\subsubsection{Chemical network}
We considered various ionization processes of neutrals by photons and 
photoelectrons which produce different ions in the coma. Various 
chemical reactions such as charge exchange, proton transfer, and thermal 
recombination are compiled from the literature. Most of these reactions are 
taken from the UMIST Rate2012 data base 
\citep[][\href{http://udfa.ajmarkwick.net}
{http://udfa.ajmarkwick.net}]{McElroy13} and their accuracy is within 
25\%. The chemical network  used to model ion density distribution in 
C/2016 R2 is presented in Tables~\ref{tab:chem-net} and 
\ref{tab:recomb_rect} of Appendix~\ref{sec:appa}. We calculated various photon 
and electron impact initiated photochemical reaction rates that 
produce the excited states of CO$_2^+$, CO$^+$, H$_2$O$^+$, and N$_2^+$ ions, 
which  are tabulated in  Table~\ref{tab:photfrq} of Appendix~\ref{sec:appb}.

\subsection{Calculations}
\subsubsection{Ion density profiles}
The degradation of the solar radiation and calculation of suprathermal electron 
flux in the cometary coma was described in our earlier work 
\citep{Bhardwaj90, Bhardwaj96, Bhardwaj12, Raghuram20, Raghuram20b, 
Raghuram13, Bhardwaj99a}. Using the modelled solar photon and the  
suprathermal electron flux profiles and corresponding cross sections, we 
determined the volume production rates of different ions for photoionization 
and electron impact ionization of neutrals. Besides the photon and electron 
impact ionization, various chemical reactions are also accounted for to 
determine radial ion density profiles.  Based on the observed neutral 
composition, we modelled density distribution of  thirteen ions viz., 
H$_2$O$^+$, H$_3$O$^+$, CO$_2^+$, CO$^+$, O$_2^+$, N$_2^+$, NH$_3^+$, 
CH$_3$OH$^+$, CH$_4^+$, CH$_3$OH$_2^+$, NH$_4^+$, C$^+$, and O$^+$ in the 
cometary coma. We solved the  following time-dependent spherical continuity 
equation for all the ions simultaneously to determine the steady state ion 
density profiles :

\begin{equation}
\frac{\partial n_i}{\partial t} + \frac{1}{r^2} \frac{\partial (n_i v_i 
	r^2)}{\partial r} = P_i - n_i L_i
\label{eq:conti}
\end{equation}
where $n_i$, $P_i$, and L$_i$ are the ion density, total production rate, 
and  loss frequency of the $ith$ ion at radial distance $r$. $v_i$ is the mean 
ion outflow velocity which is assumed to be same as neutral velocity.

\vspace{-12pt}
\subsubsection{Emission intensities of CO$_2^+$, CO$^+$, N$_2^+$, and 
H$_2$O$^+$}
We accounted for photoionization, electron impact ionization of neutrals, and 
solar resonance fluorescence excitation mechanisms to calculate the volume 
emission rates ($VER$) of excited states of ions.   To account for the 
resonance fluorescence excitation mechanism, the modelled ion density profiles 
are multiplied with corresponding g-factors. The total volume emission rates of 
CO$_2^+$($\widetilde{A}^2\Pi_{u,3/2}$), N$_2^+$(B$^2\Sigma^+_u$),
H$_2$O$^+$($\widetilde{A}^2$A$_1$), and CO$^+$(A$^2\Pi$)  are given by

\begin{eqnarray}
VER_{(CO_2^{+*})} & = & I_{f1} [CO_2] + g_{CO_2^+} [CO_2^+]\\
VER_{(N_2^{+*})} & = & I_{f2}[N_2] + g_{N_2^+} [N_2^+]\\
VER_{(H_2O^{+*})} & = & I_{f3}[H_2O] + g_{H_2O^+} [H_2O^+]\\
VER_{(CO^{+*})} & = & I_{f4}[CO_2] + I_{f5} [CO] + g_{CO^+} [CO^+]
\end{eqnarray}
where species in the brackets are corresponding neutral and ion densities.
Species in the parenthesis are the excited states of  ions.  
I$_{f1}$ to I$_{f5}$ are the total ionization and excitation frequencies of the 
  excited states of the ions via photon and electron impact
ionization of the respective neutrals. 
The volume emission rates are integrated along the line of sight and surface 
brightness profiles are determined as a function of the nucleocentric projected
distance. The ion emission intensity ratios are determined from the modelled 
surface brightness profiles and  compared with the observations.

\section{Results}
\label{sec:results} 

\subsection{Production and loss  mechanisms of CO$_2^+$}
\label{sec:pl_co2p}
\begin{figure}
	\includegraphics[trim=0 35 0 0, clip, 
	width=\columnwidth]{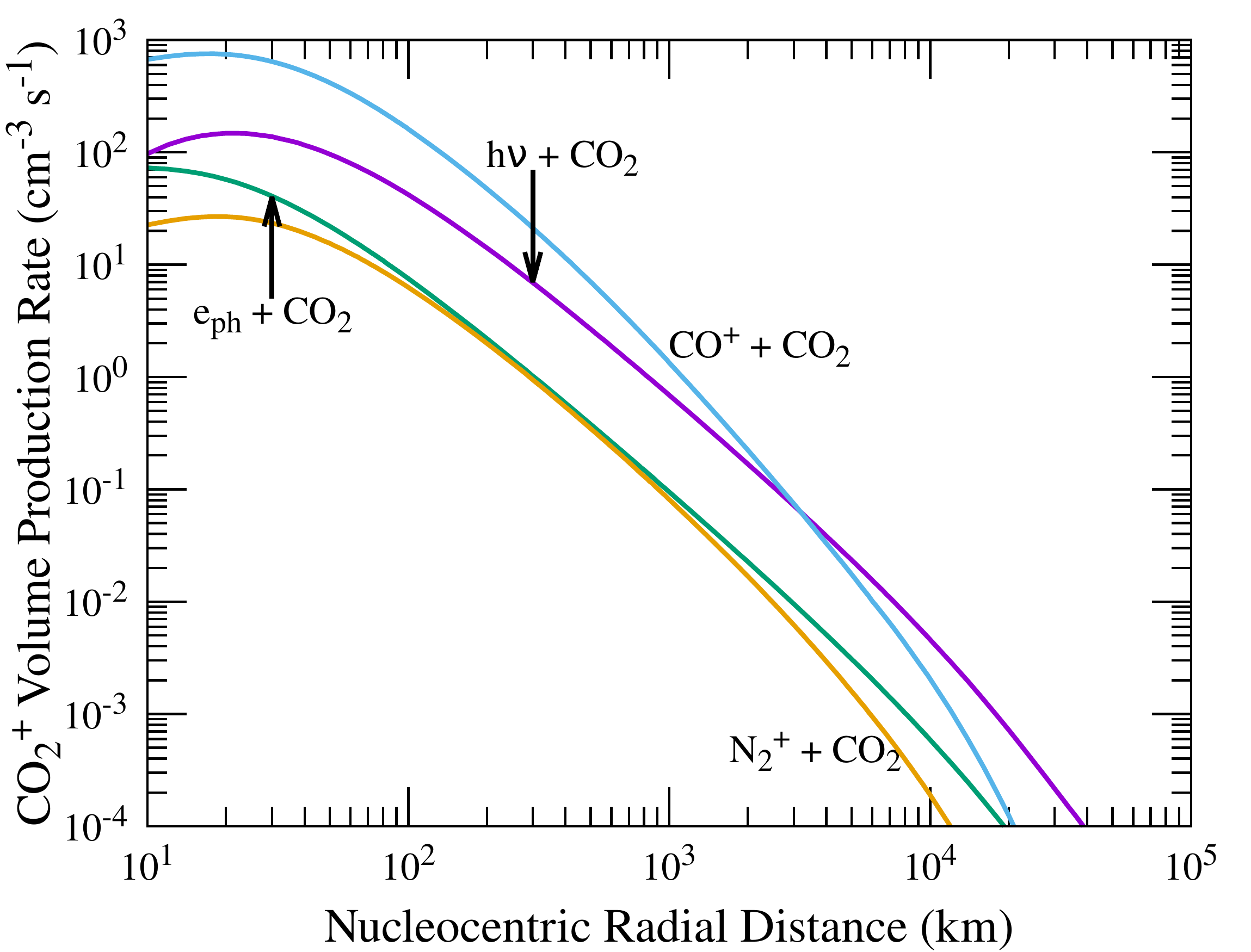}
	\includegraphics[width=\columnwidth]{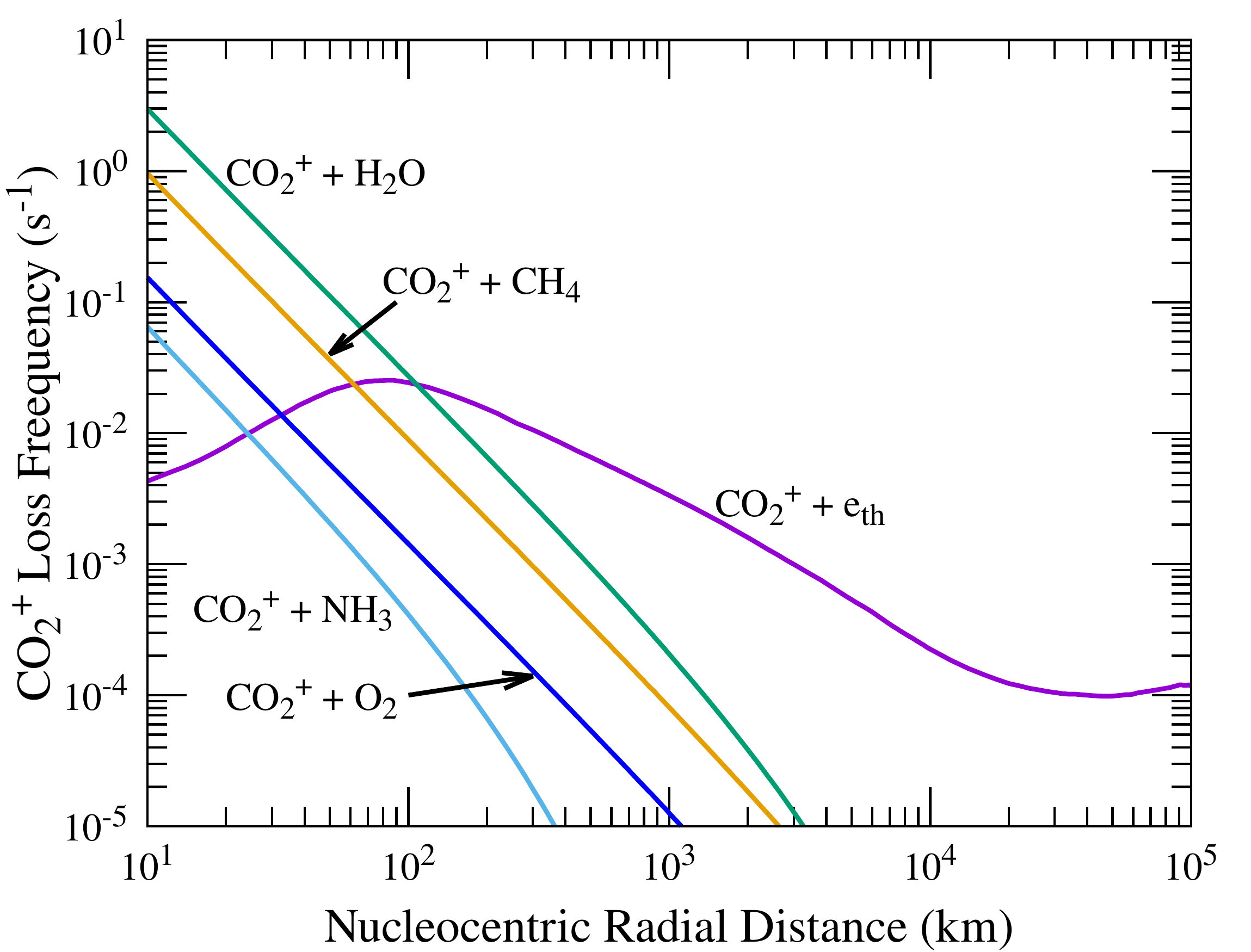}
	\vspace{-0.5cm}
	\caption{Modelled production rate  (top panel) and loss frequency profiles 
		(bottom panel) of CO$_2^+$ in the coma of comet  C/2016  R2. 
		Calculations are done at 2.8 au heliocentric distance using a CO 
		production rate of 1.1 $\times$ 10$^{29}$ s$^{-1}$. The relative 
		abundances of other species viz., H$_2$O, CO$_2$, N$_2$, CH$_3$OH, 
		CH$_4$, O$_2$, and NH$_3$ are taken as 0.3\%, 18\%, 7\%, 1.1\%, 0.6\%, 
		1\%, and 0.01\%, respectively, with respect to CO production rate.
		h$\nu$, e$_{ph}$, and e$_{th}$ represent solar photon, photoelectron, 
		and thermal electron, respectively.
		\label{fig:pr_co2p}}
\end{figure}

The modelled volume production rate and loss frequency profiles of CO$_2^+$ 
for 
different 
photochemical reactions are presented in the respective top and bottom panels 
of 
Figure~\ref{fig:pr_co2p}. Photoionization of CO$_2$ and charge exchange between 
CO$^+$ and CO$_2$ are the important production sources of CO$_2^+$ in 
this comet. Electron impact 
ionization of CO$_2$ and charge exchange of N$_2^+$ with CO$_2$ 
contribute little to the total formation of CO$_2^+$. The modelled loss 
processes 
presented in the lower panel of Figure \ref{fig:pr_co2p} show that 
 for radial distances 
smaller than 100 km, the collisions with H$_2$O  significantly remove CO$_2^+$  
in the coma. Above this distance, thermal recombination 
is  the 
significant loss source for CO$_2^+$.

\subsection{Production and loss  mechanisms of CO$^+$}
\label{sec:pl_cop}
\begin{figure}
	\includegraphics[trim=0 35  0 0, clip, 
	width=\columnwidth]{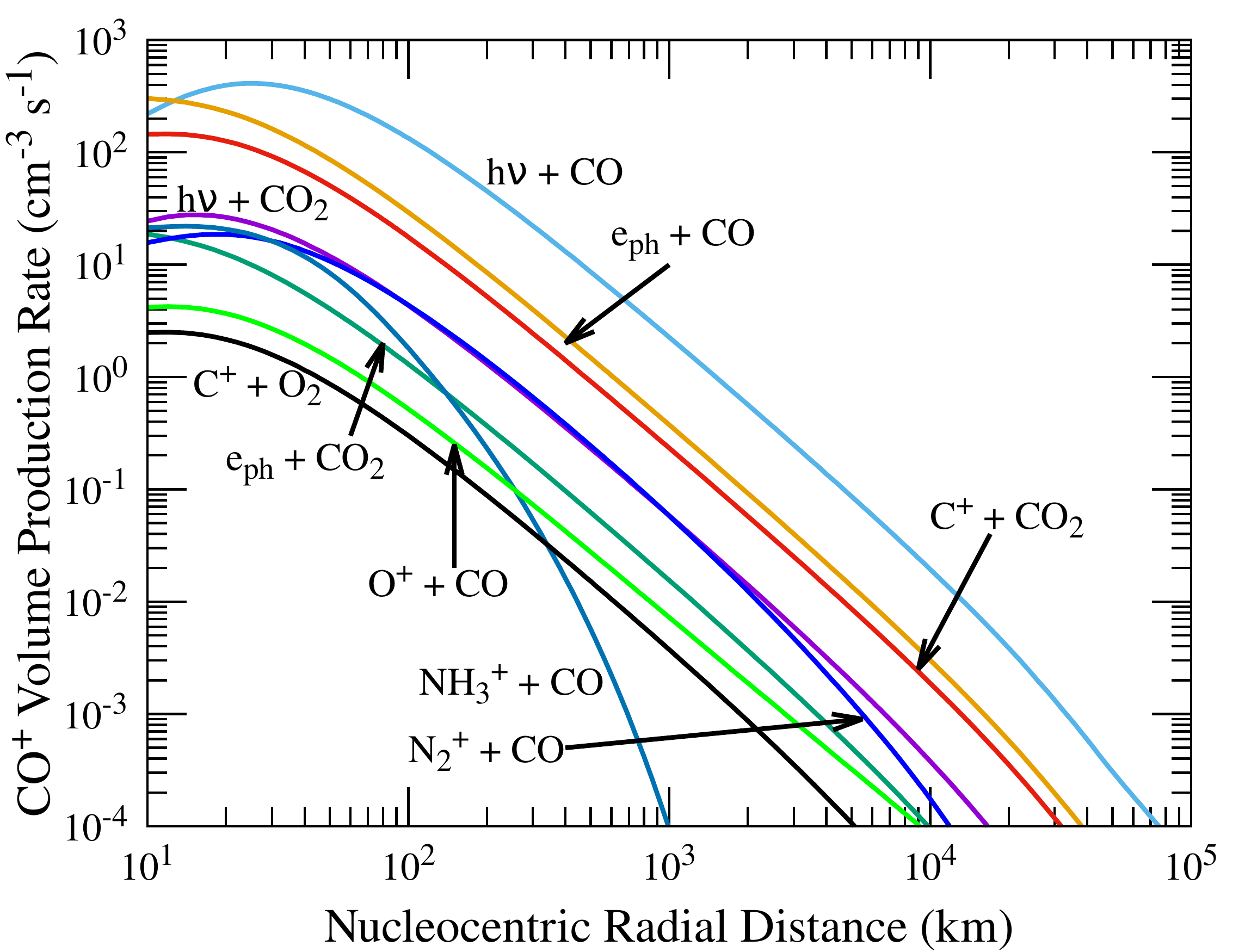}
	\includegraphics[width=\columnwidth]{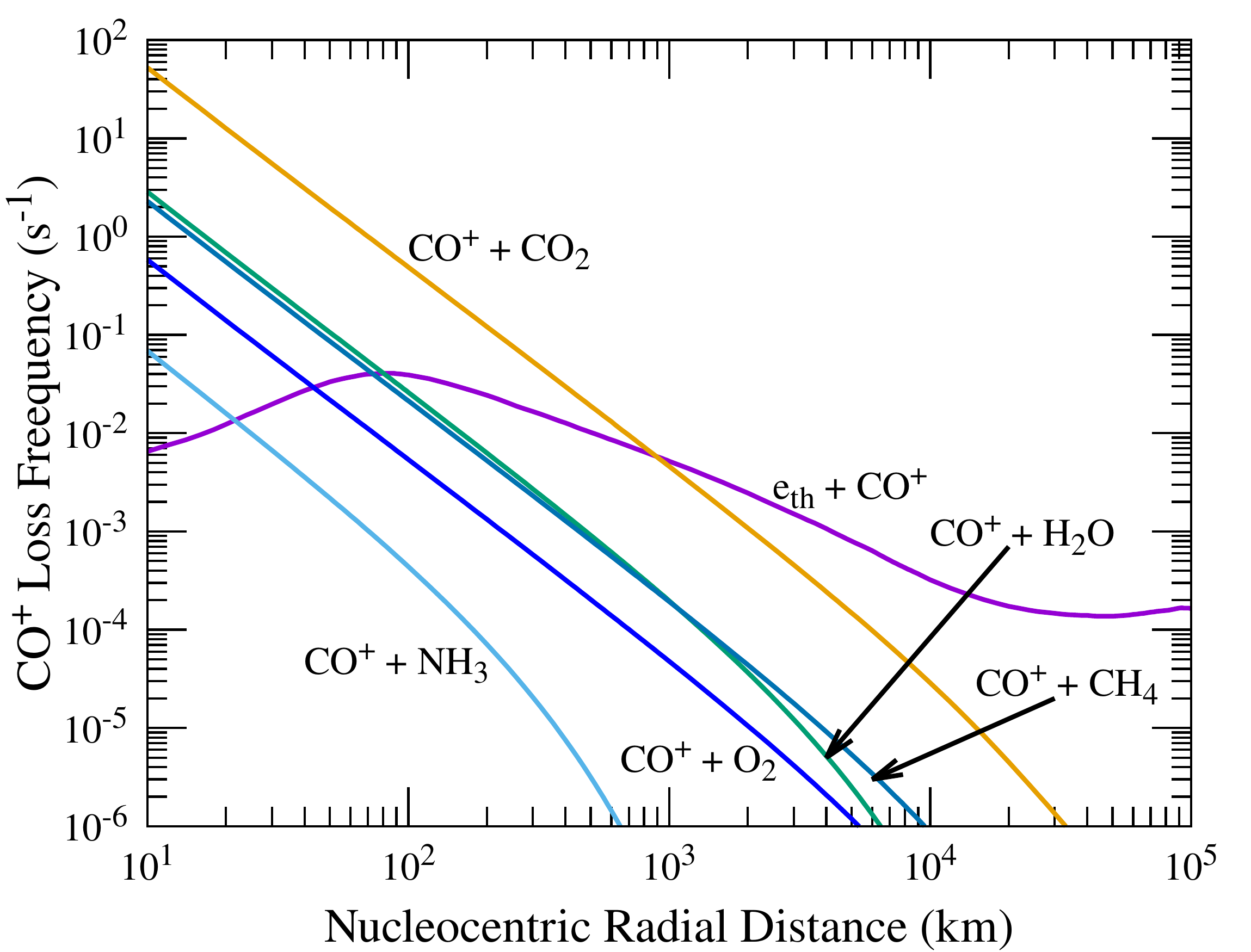}
	\vspace{-0.5cm}
	\caption{Modelled production rate (top panel) and loss frequency 
		(bottom panel) profiles of CO$^+$ in the coma of comet C/2016 R2. 
		Input conditions are the same as explained in  
		Figure~\ref{fig:pr_co2p}. h$\nu$, e$_{ph}$, and e$_{th}$ 
		represent solar photon, photoelectron, and thermal electron, 
		respectively. 
		\label{fig:pr_cop}}
\end{figure}

Figure~\ref{fig:pr_cop} shows the modelled  production rate and loss frequency 
profiles 
of CO$^+$ for different photochemical processes. As shown in the top panel of 
this 
Figure, photoionization 
of CO is the major formation channel of CO$^+$ in the cometary coma. Close to 
the surface of the nucleus, electron 
impact ionization of CO and the collisions between C$^+$ and CO$_2$ are also 
important production sources of CO$^+$.  The formation rates of CO$^+$ via 
dissociative 
ionization of CO$_2$ by photons and photoelectrons, and other charge 
exchange 
reactions are smaller by more than an order of magnitude  compared to that 
of photoionization of CO. The calculated loss frequency profiles in the bottom 
panel of Figure~\ref{fig:pr_cop} show that the collisions with CO$_2$ is major 
loss source of CO$^+$ for  radial distances smaller than 10$^3$ 
km. Above this distance, thermal recombination is the significant loss source 
of CO$^+$. Other charge exchange reactions play a minor role in the removal 
of   
CO$^+$.

\subsection{Production and loss  mechanisms of H$_2$O$^+$}
\begin{figure}
	\includegraphics[trim=0 35  0 0, clip,
	width=\columnwidth]{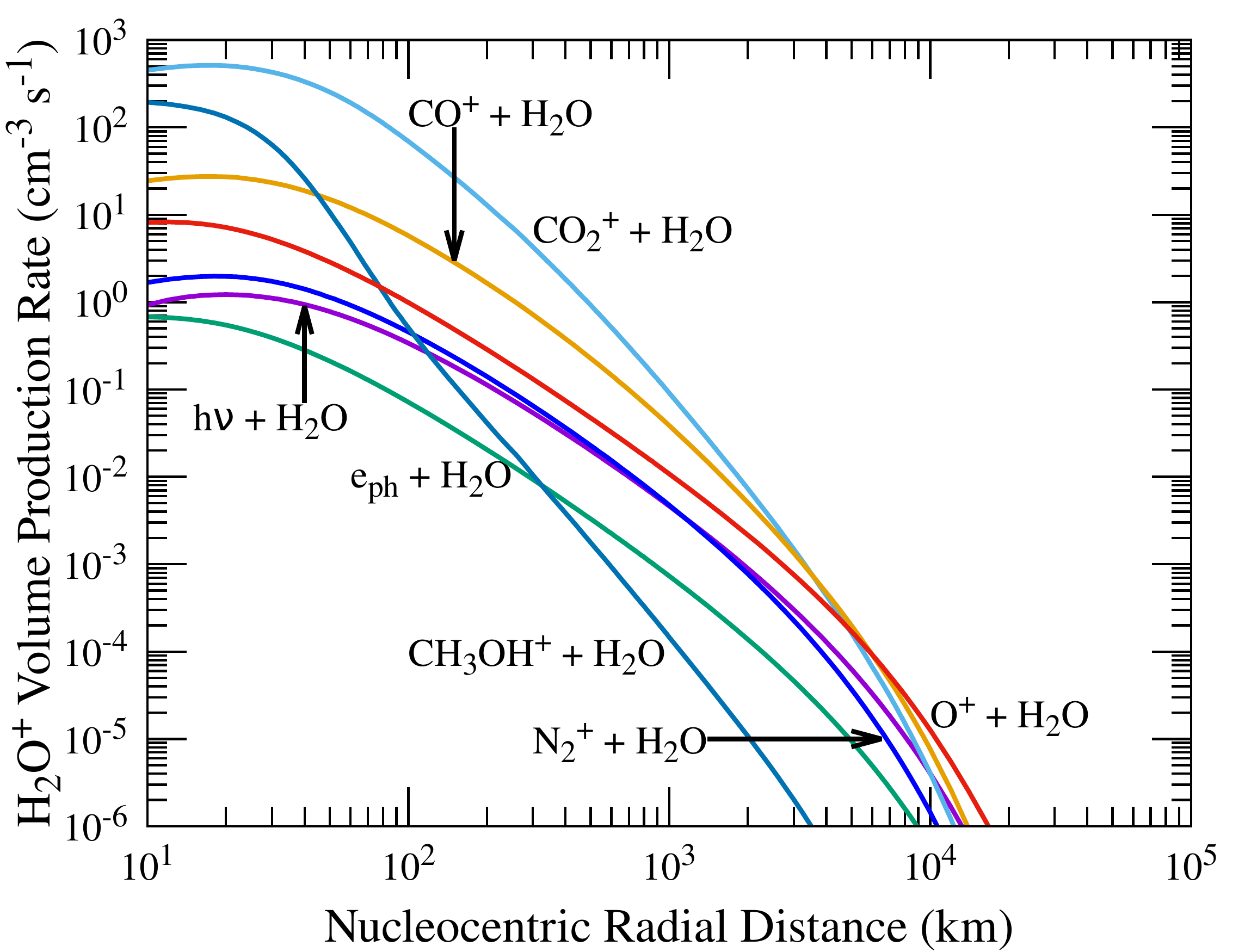}
	\includegraphics[width=\columnwidth]{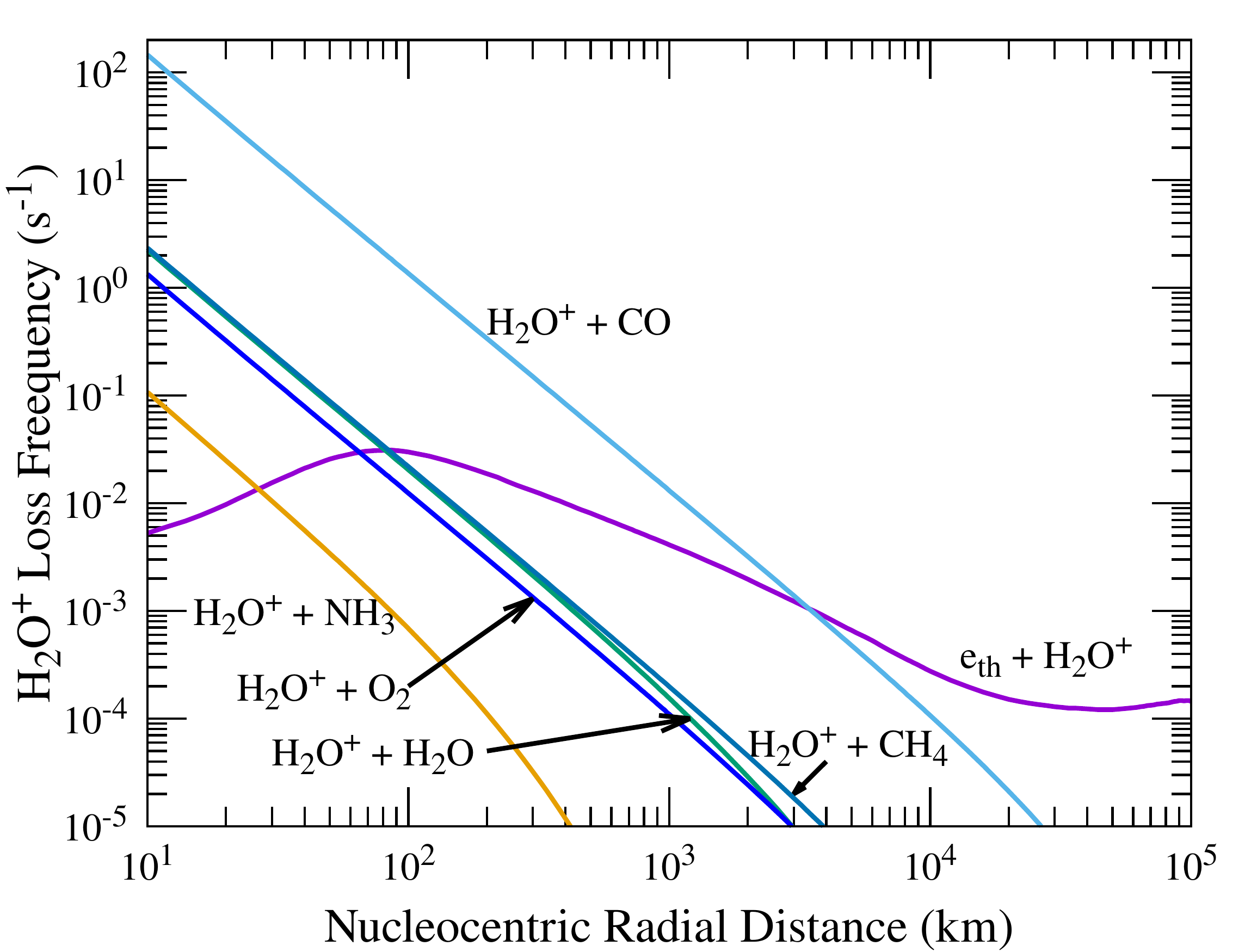}
	\vspace{-0.5cm}
	\caption{Modelled production rate (top panel) and loss frequency (bottom 
		panel) profiles of H$_2$O$^+$ via various mechanisms in the coma of 
		comet  C/2016  R2. Input conditions are the same as 
		explained in Figure~\ref{fig:pr_co2p}. h$\nu$, e$_{ph}$, and e$_{th}$ 
		represent solar photon, photoelectron, and 
		thermal electron, respectively.  \label{fig:pr_h2op}}
\end{figure}

Generally in comets, which are H$_2$O dominated, the main production source of 
H$_2$O$^+$ is  photoionization and electron impact ionization of H$_2$O. Our 
model calculations in the top  panel of Figure~\ref{fig:pr_h2op} show that in 
comet 
C/2016 R2, charge exchange between CO$_2^+$ and  H$_2$O is the major production 
of 
source of H$_2$O$^+$ in the inner 
coma. Close to the surface of the  nucleus, charge exchange between 
CH$_3$OH$^+$ and H$_2$O is also another important formation source of 
H$_2$O$^+$.  It 
can be noticed in this 
figure that for radial distances smaller than 1000 km, the formation rate of 
H$_2$O$^+$ via photoionization of H$_2$O is smaller by more than two orders of 
magnitude compared to the production rate of H$_2$O$^+$ due to charge exchange 
between CO$_2^+$ and H$_2$O. At larger radial distances (about 10$^4$ km), 
charge exchange between O$^+$ and H$_2$O significantly produces H$_2$O$^+$ and 
several 
other sources are also involved in the production of this ion.  As shown in the 
bottom panel of Figure~\ref{fig:pr_h2op}, the collisions 
between H$_2$O$^+$ and CO, which lead to the formation of HCO$^+$, is the major 
loss source of H$_2$O$^+$ for  
radial distances below 3000 km and  other collisional loss frequencies are 
smaller by more than an order of magnitude compared to the former. Above this 
radial distance thermal recombination is a significant loss source of 
H$_2$O$^+$.

%

\subsection{Production and loss  mechanisms of N$_2^+$}
\begin{figure}
	\includegraphics[trim=0 35  0 0, clip,
	width=\columnwidth]{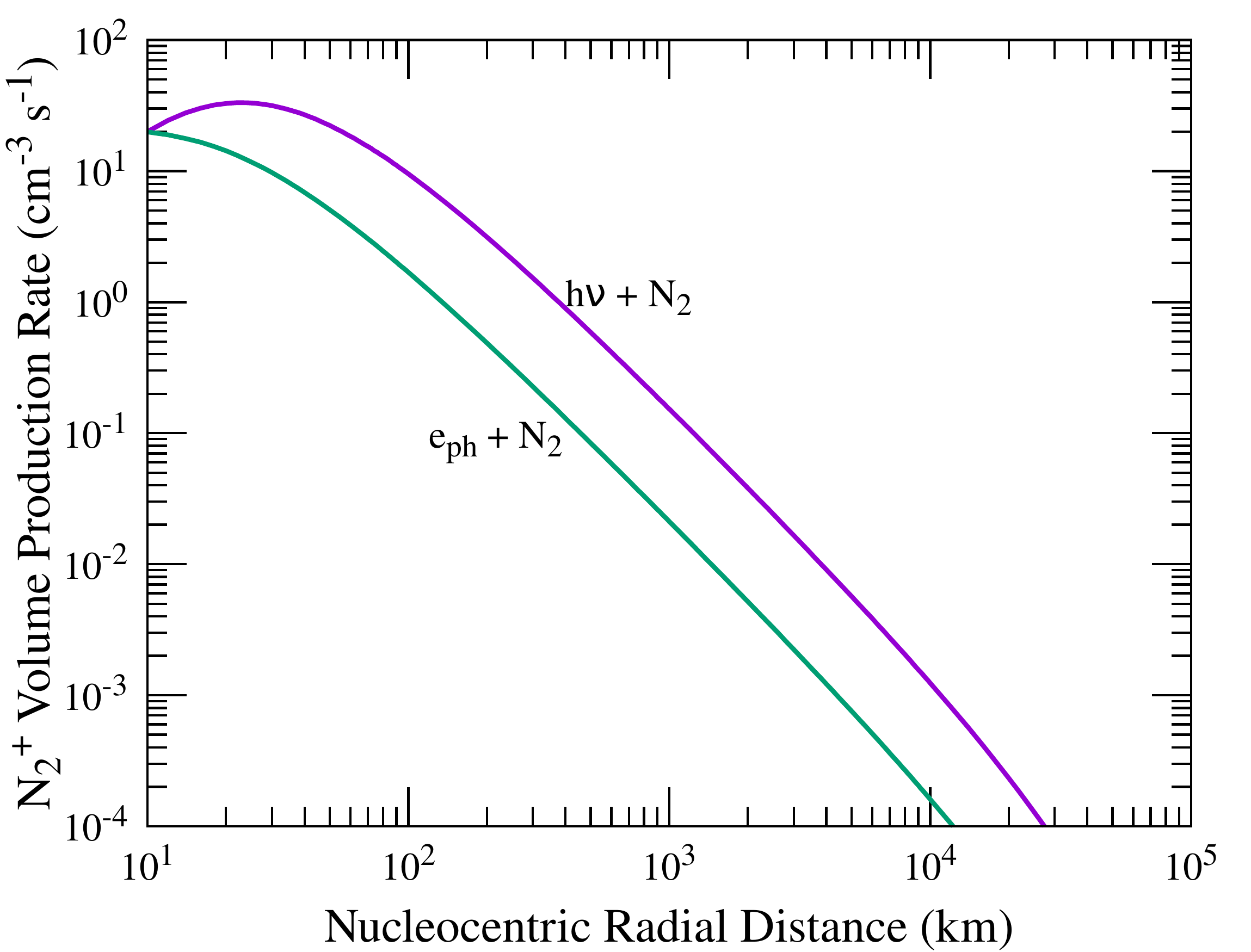}
	\includegraphics[width=\columnwidth]{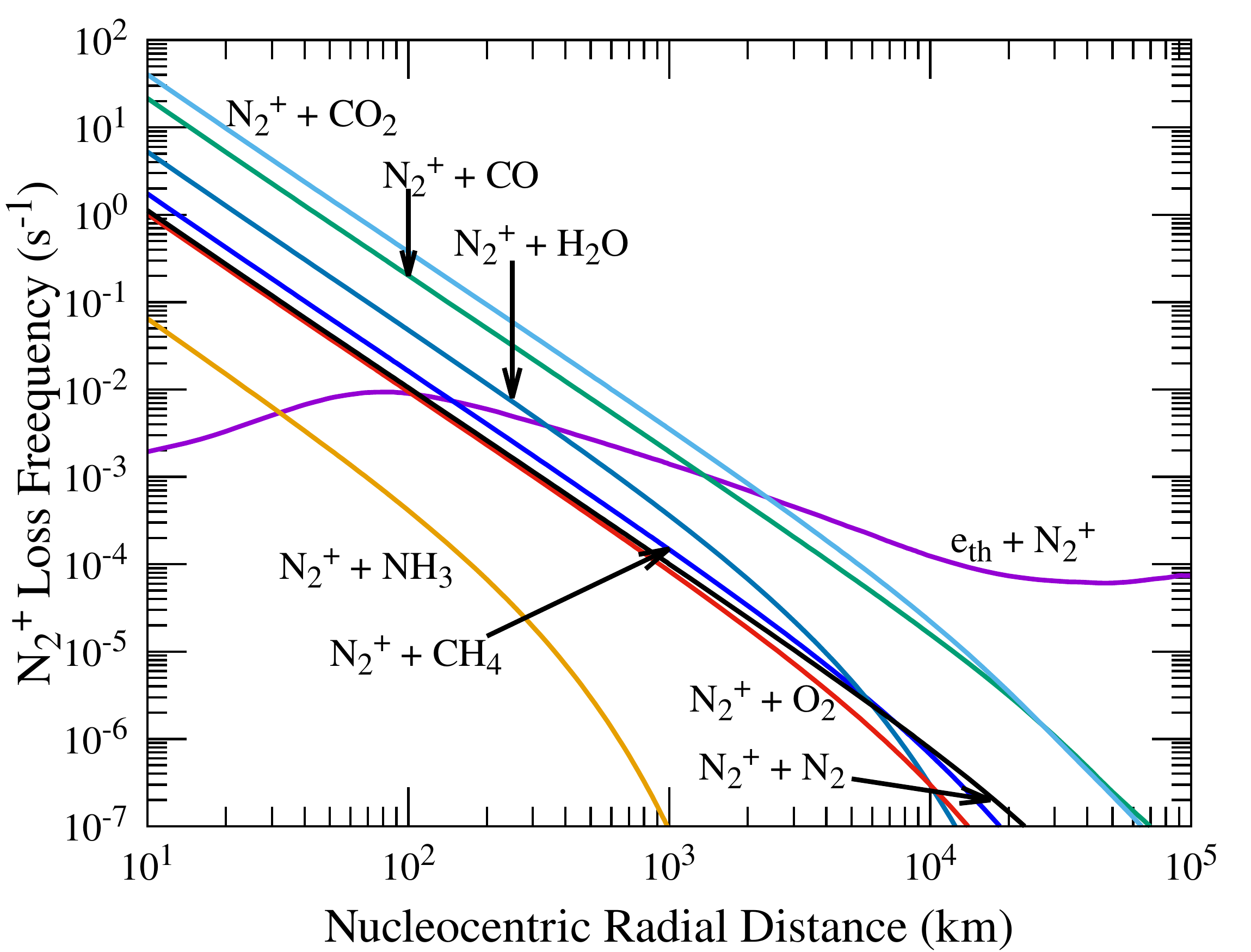}
	\vspace{-0.5cm}
	\caption{Modelled production rate (top panel) and loss frequency (bottom 
		panel) profile of N$_2^+$ in the coma of comet C/2016 R2. Input 
		conditions are same as explained in Figure~\ref{fig:pr_co2p}. h$\nu$, 
		e$_{ph}$, and e$_{th}$ represents solar photon, photoelectron, and 
		thermal electron, respectively.  \label{fig:pr_n2p}}
\end{figure}

The modelled production rate profiles in the top panel of 
Figure~\ref{fig:pr_n2p} show that N$_2^+$ is produced in the coma of C/2016 R2 
via  photon and 
electron impact ionization of N$_2$. Modelled loss profiles in the bottom panel 
of this 
Figure show that  
several collisional  mechanisms are involved in the removal of N$_2^+$ in the 
inner coma.  The 
charge exchange of N$_2^+$ with CO and CO$_2$ are the significant loss 
mechanisms for radial distances below 3000 km and above this radial distance 
thermal recombination is the major loss source for this ion. 

%

\subsection{Production and loss  mechanisms of CH$_3$OH$^+$}
\begin{figure}
	\includegraphics[trim=0 35  0 0, clip,
	width=\columnwidth]{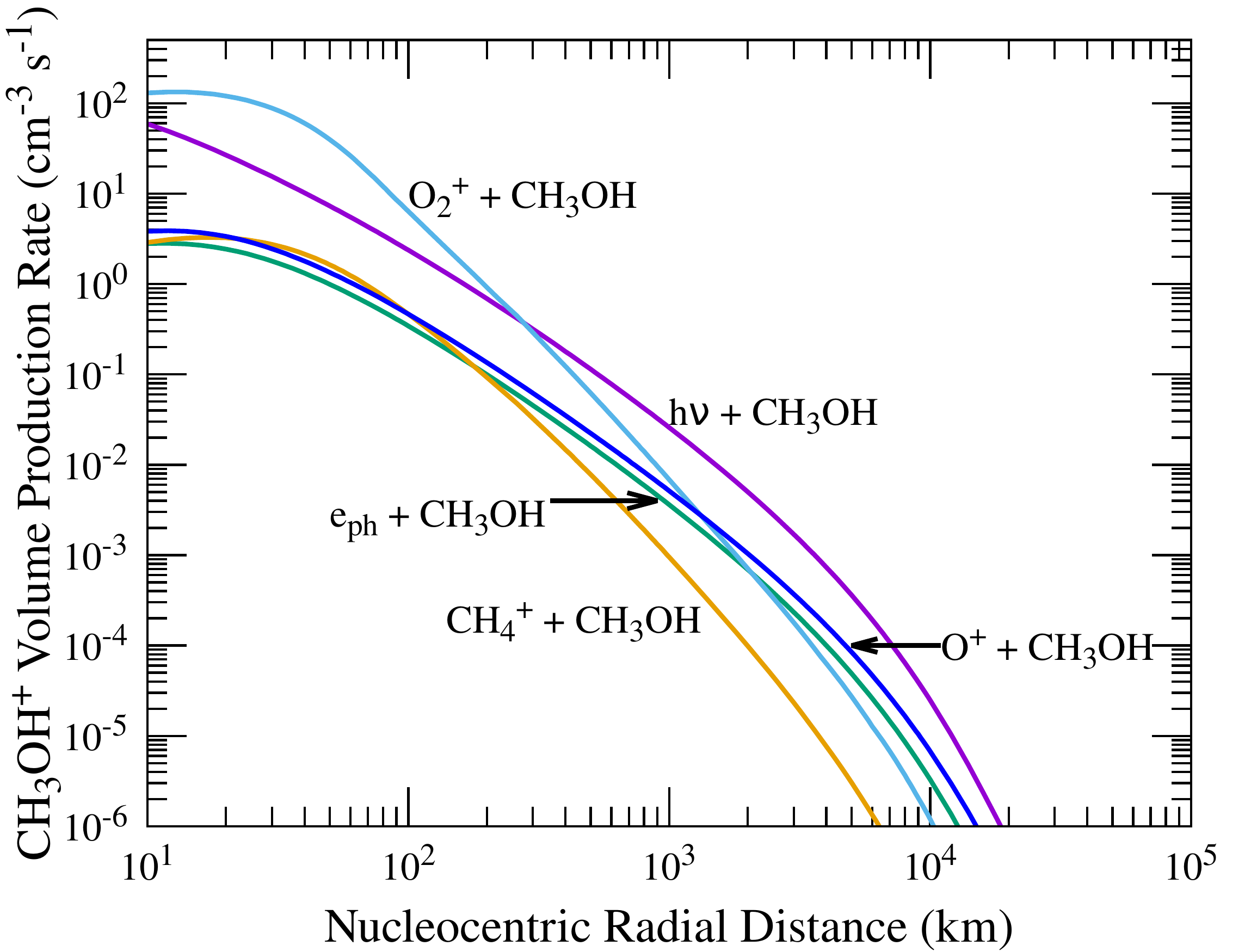}
	\includegraphics[width=\columnwidth]{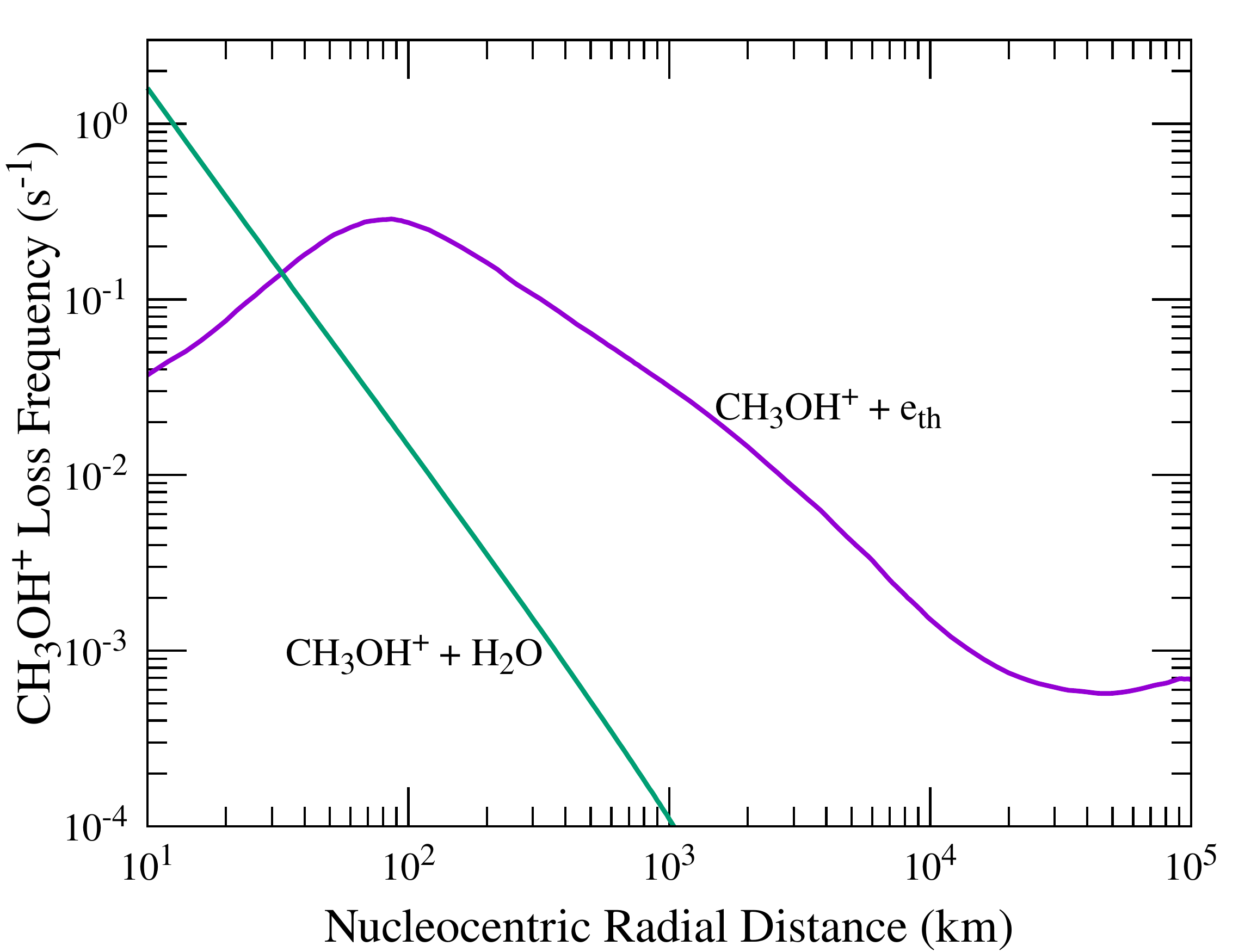}
	\vspace{-0.5cm}
	\caption{Modelled production rate (top panel) and loss frequency profiles
		(bottom panel) of CH$_3$OH$^+$ in the coma of comet C/2016 R2. Input 
		conditions are 	same as explained in Figure~\ref{fig:pr_co2p}. h$\nu$, 
		e$_{ph}$, and e$_{th}$ represent solar photon, photoelectron, and 
		thermal electron, respectively.  
		\label{fig:pr_ch3ohp}}
\end{figure}

The modelled production rate and loss frequency profiles, which 
are presented in the respective top and bottom panels of 
Figure~\ref{fig:pr_ch3ohp}, show that photoionization of CH$_3$OH is the 
most important formation source of CH$_3$OH$^+$ for radial distances above 300 
km. 
But  below this radial distance, the charge exchange  between O$_2^+$ and 
CH$_3$OH is an important source mechanism. As shown 
in the bottom panel, proton 
transfer between CH$_3$OH$^+$ and H$_2$O removes this ion for radial 
distances smaller than 50 km, above which thermal recombination takes over as
 the major loss source.  

\subsection{Production and loss  mechanisms of NH$_3^+$}
\begin{figure} 
	\includegraphics[trim=0 35  0 0, clip,
	width=\columnwidth]{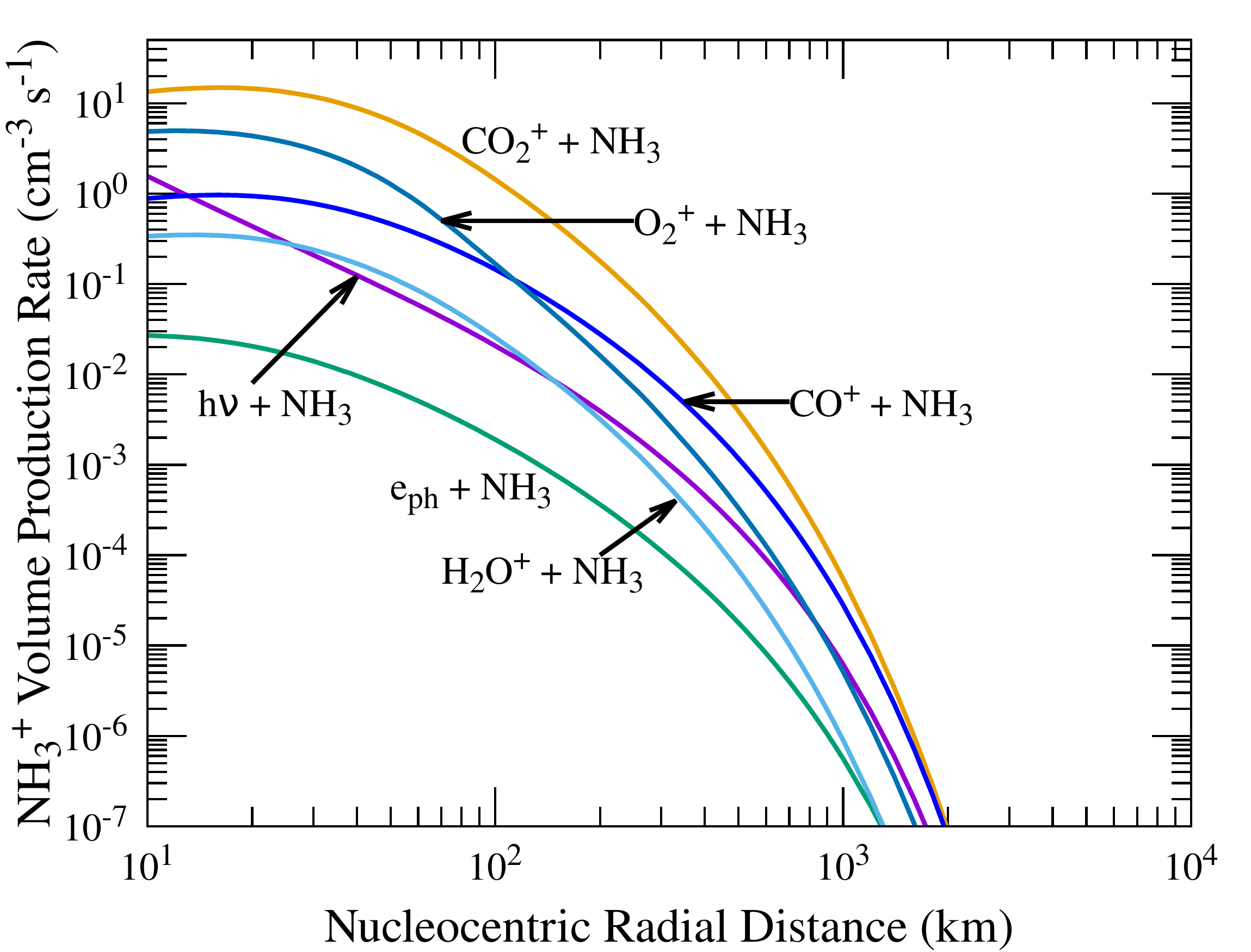}
	\includegraphics[width=\columnwidth]{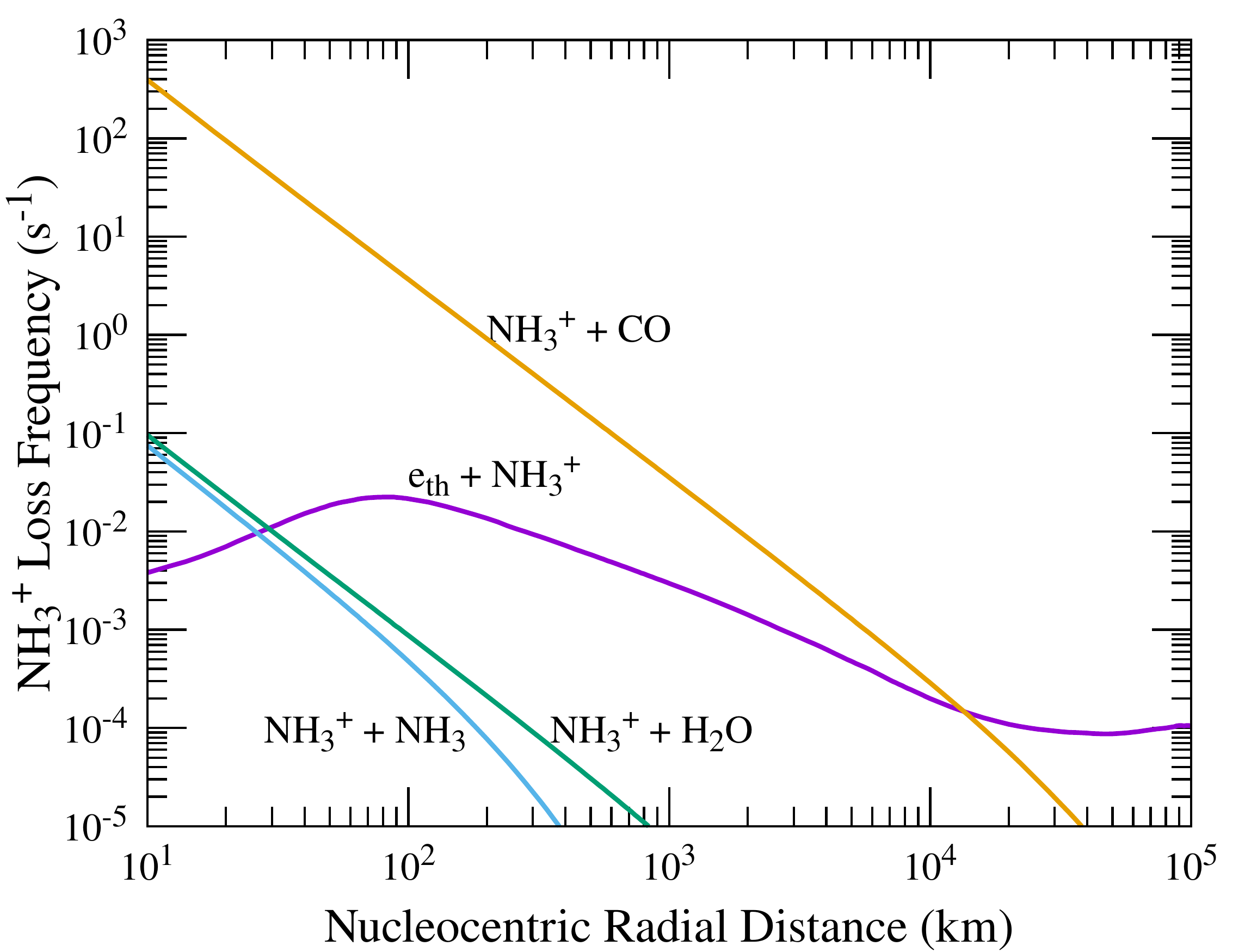}
	\vspace{-0.5cm}
	\caption{Modelled production rate (top panel) and loss frequency (bottom 
		panel) profiles of NH$_3^+$ in the coma of comet R2. Input conditions 
		are same 
		as explained in Figure~\ref{fig:pr_co2p}. h$\nu$, e$_{ph}$, and 
		e$_{th}$ represent solar photon, photoelectron, and 
		thermal electron, respectively.  \label{fig:pr_nh3p}}
\end{figure}

Modelled production and loss rate profiles of NH$_3^+$ for different  
mechanisms are presented in top and bottom panels of Figure~\ref{fig:pr_nh3p}, 
respectively. Collisions of NH$_3$ with CO$_2^+$, CO$^+$, and O$_2^+$ are the 
dominant 
production sources of NH$_3^+$ for 
radial distances below 10$^3$ km.  The formation rate of  NH$_3^+$ due to 
photoionization of 
NH$_3$ is  smaller by more than a factor of 5 compared  to that from the 
 charge exchange between CO$_2^+$ and NH$_3$. For radial distances below 
 10$^4$ km, 
collisions with CO is the most significant loss process of NH$_3^+$ and at 
larger radial distances  thermal recombination removes this ion in the inner 
coma of 
comet C/2016 
R2. 

%

\subsection{Production and loss  mechanisms of CH$_4^+$}
\begin{figure} 
	\includegraphics[trim=0 35  0 0, clip,
	width=\columnwidth]{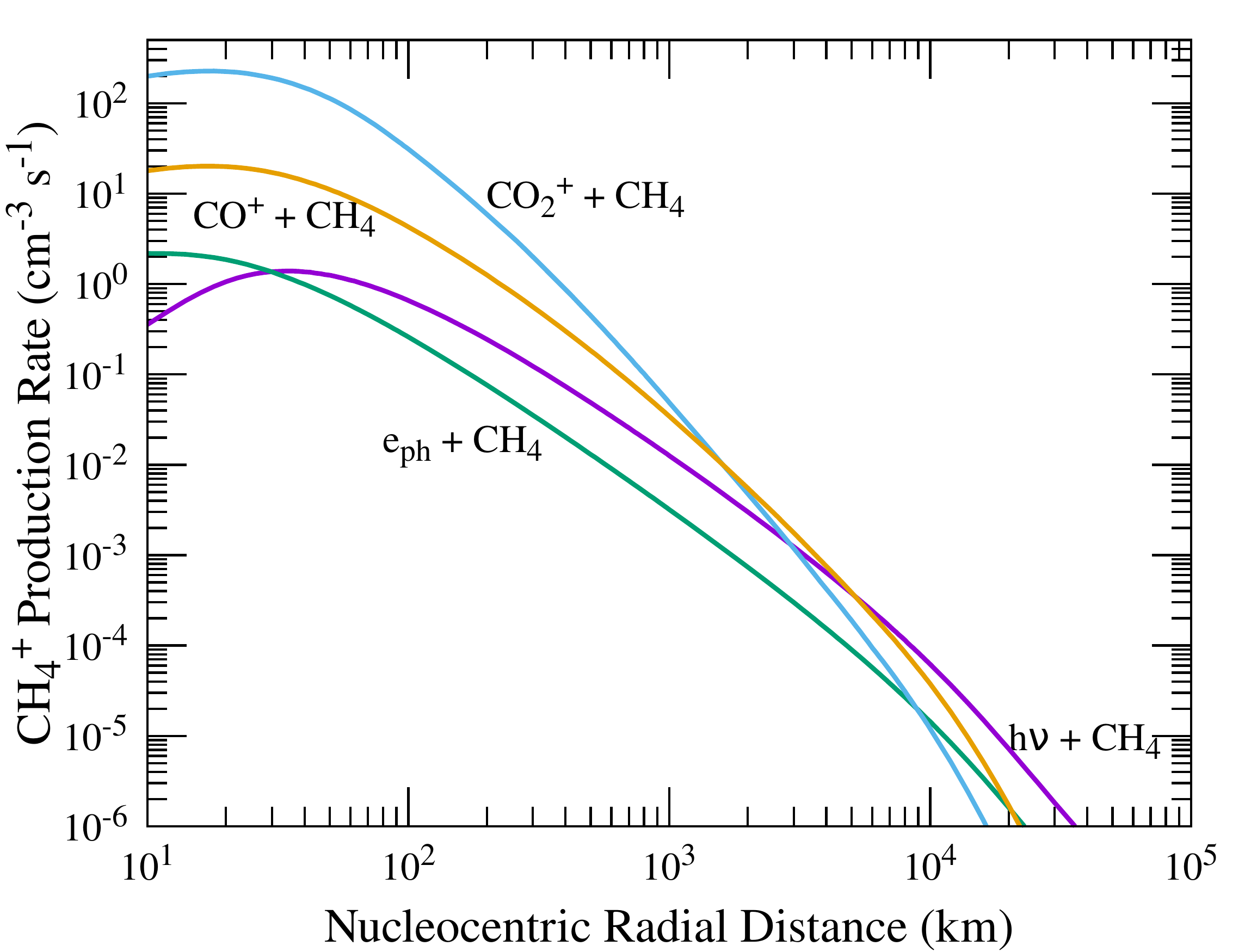}
	\includegraphics[width=\columnwidth]{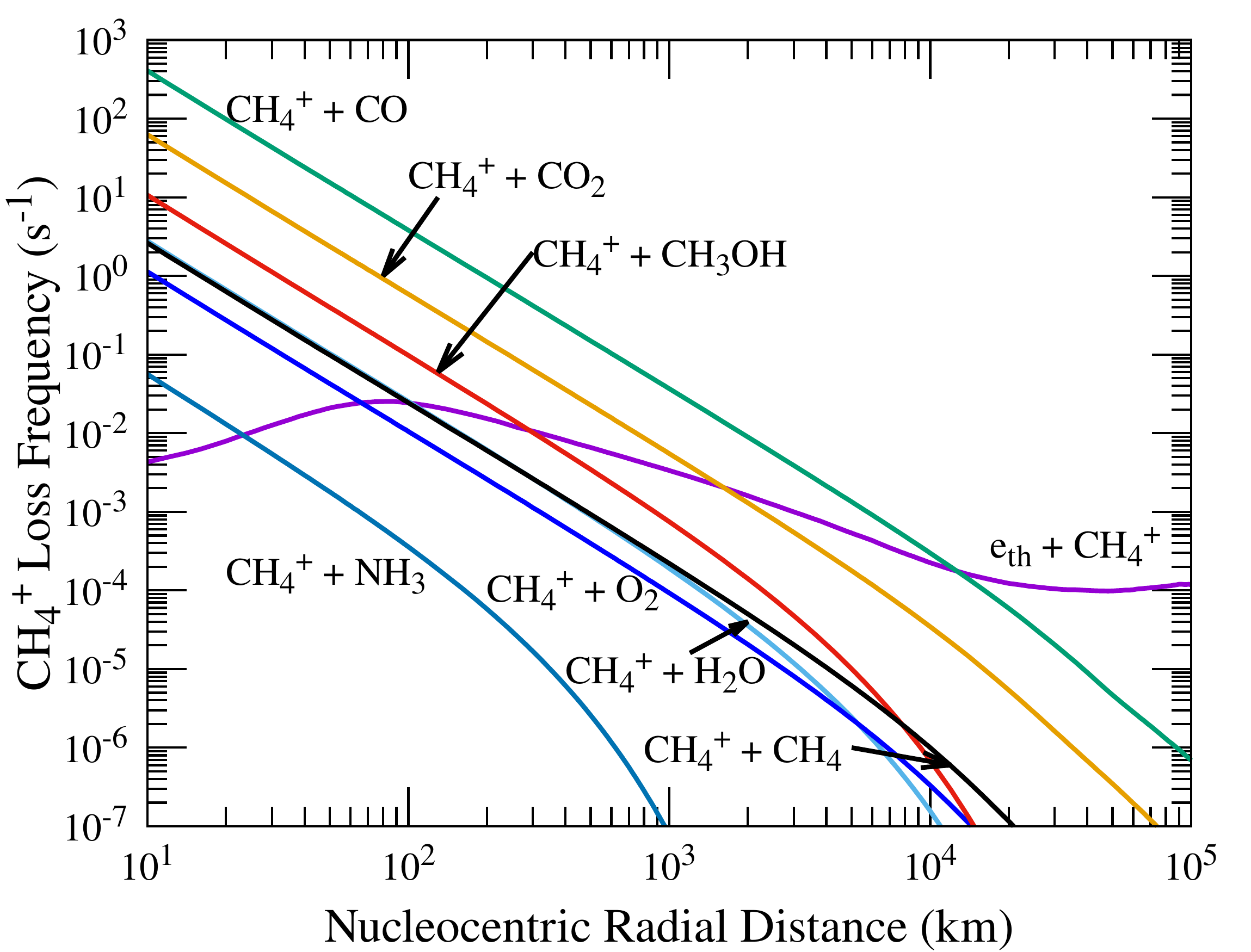}
	\vspace{-0.5cm}
	\caption{The production rates (top panel) and loss frequency profiles 
	(bottom 
		panel) of CH$_4^+$ in the coma of comet R2. Input conditions are same 
		as 	explained in Figure~\ref{fig:pr_co2p}. h$\nu$, e$_{ph}$, and 
		e$_{th}$ represent solar photon, photoelectron, and 
		thermal electron, respectively.  \label{fig:pr_ch4p}}
\end{figure}

The top panel of Figure~\ref{fig:pr_ch4p} shows that charge exchange between 
CO$_2^+$ and CH$_4$ is the major production source of CH$_4^+$ for radial 
distances below 1000 km and above this distance photoionization of CH$_4$ and
charge exchange between CO$^+$ and CH$_4$ are 
the most significant sources of this ion. Several collisional 
mechanisms are incorporated to determine the total chemical loss frequency of 
CH$_4^+$. The modelled loss frequency profiles of CH$_4^+$ in the bottom 
panel of Figure~\ref{fig:pr_ch4p} show that collisional reaction between 
CH$_4^+$ and CO is the dominant loss mechanism for the radial distances below  
10$^4$ km and above this distance thermal 
recombination is the major loss process for CH$_4^+$. Several other collisional 
processes play negligible role in the removal of CH$_4^+$ in the inner coma. 

\subsection{Production and loss  mechanisms of O$_2^+$}
\begin{figure}
	\includegraphics[trim=0 35  0 0, clip,
	width=\columnwidth]{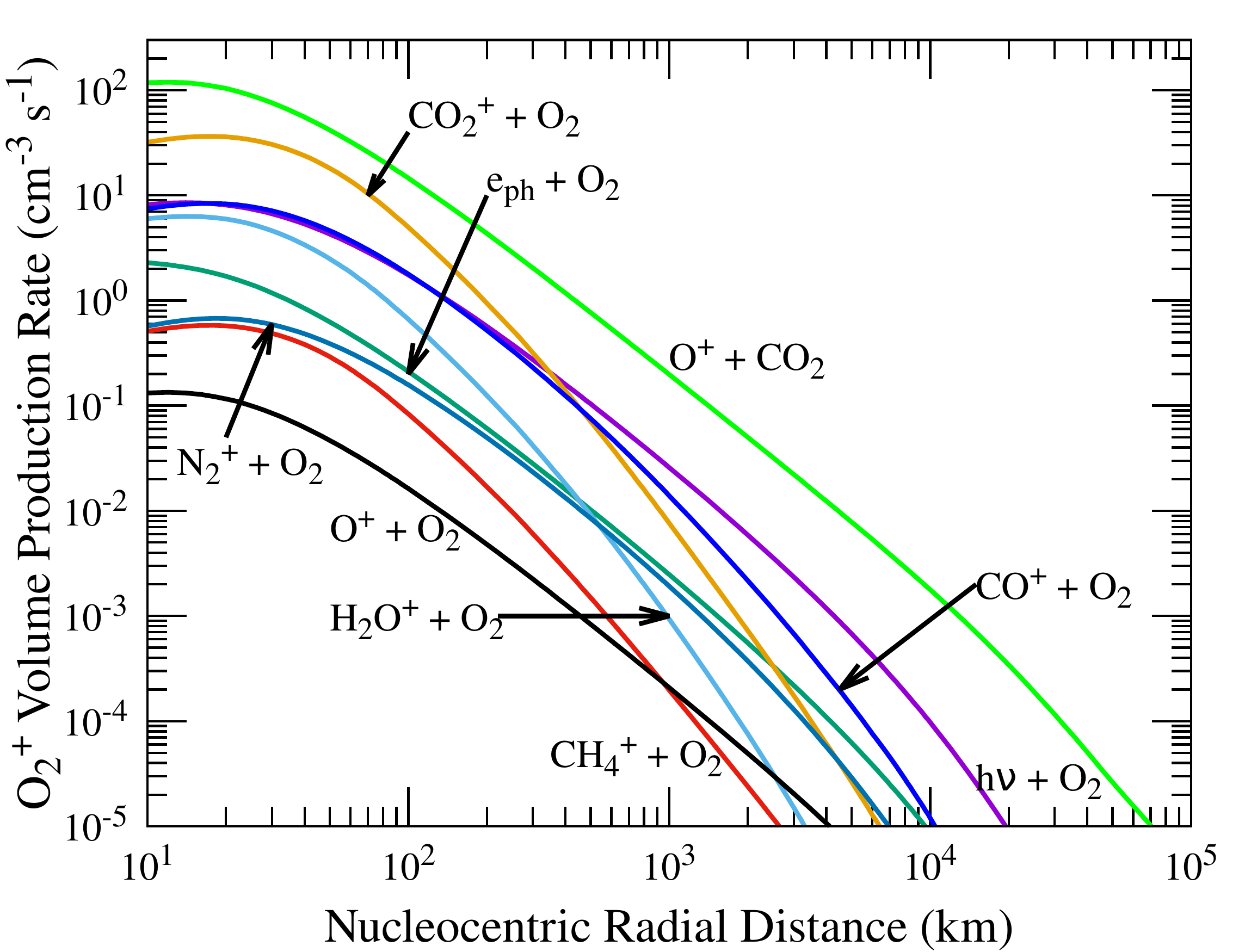}
	\includegraphics[width=\columnwidth]{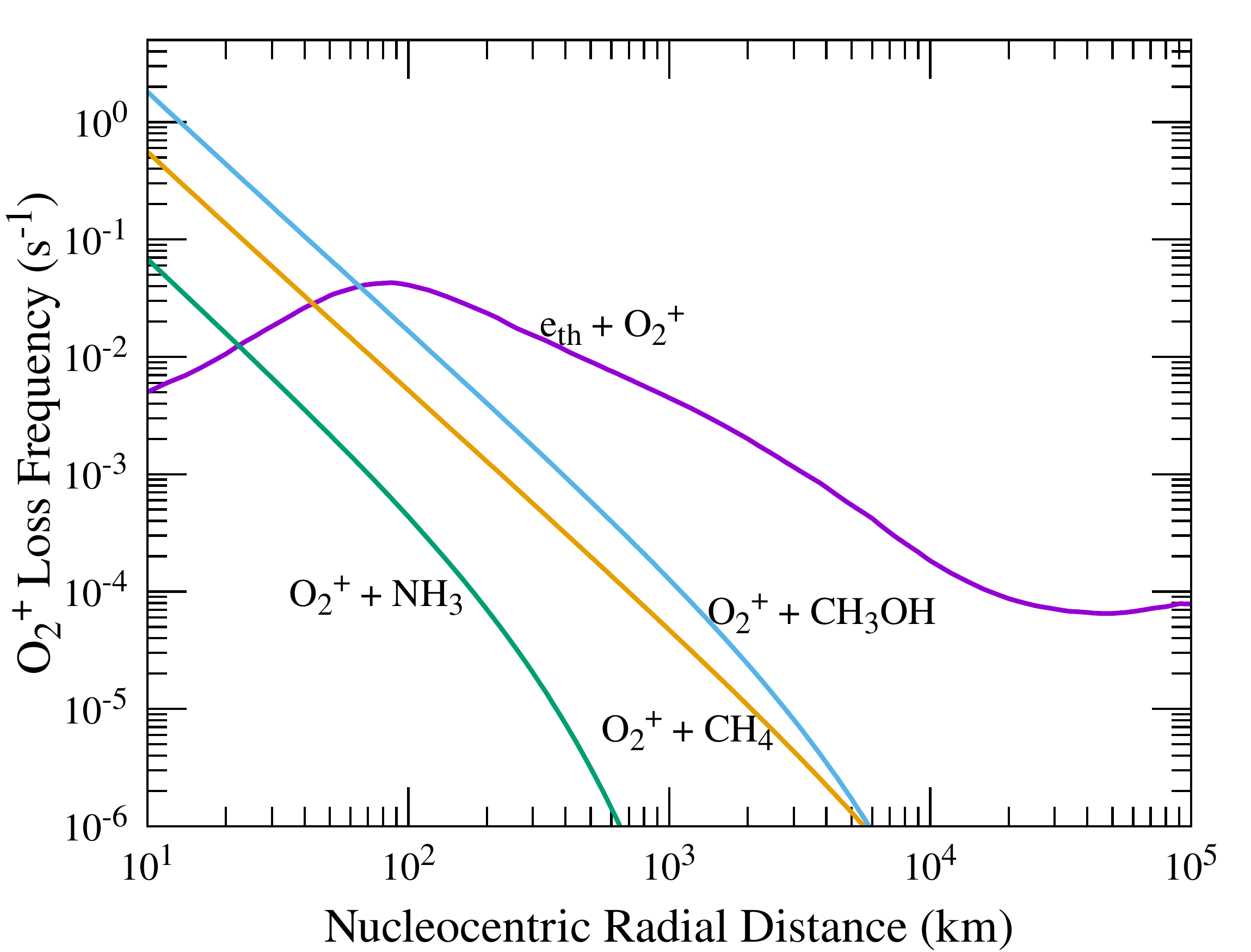}
	\vspace{-0.5cm}
	\caption{Modelled production rate (top panel) and loss frequency (bottom 
		panel) profiles of O$_2^+$ via different mechanisms in the coma of 
		comet C/2016 R2. Input conditions are same as explained in 
		Figure~\ref{fig:pr_co2p}. h$\nu$, e$_{ph}$, and e$_{th}$ represents 
		solar photon, photoelectron, and thermal electron, respectively.  
		\label{fig:pr_o2p}}
\end{figure}

The modelled formation and loss profiles of O$_2^+$ 
are presented in Figure~\ref{fig:pr_o2p}. The calculated production rates 
profiles in the top panel of this Figure show that collisional reaction between 
O$^+$ and CO$_2$ is the significant production source of O$_2^+$ in the inner 
coma rather 
than photoionization of O$_2$. Loss frequency profiles presented in 
the bottom panel of  Figure~\ref{fig:pr_o2p} show that collisions with CH$_3$OH 
significantly remove O$_2^+$ for radial distances below 100 
km and at larger radial distances thermal recombination takes over as the 
dominant 
loss mechanism. 

\subsection{Production and loss mechanisms of protonated ions H$_3$O$^+$, 
CH$_3$OH$_2^+$, and NH$_4^+$}
\begin{figure}
	\includegraphics[trim=0 35  0 0, clip,
	width=\columnwidth]{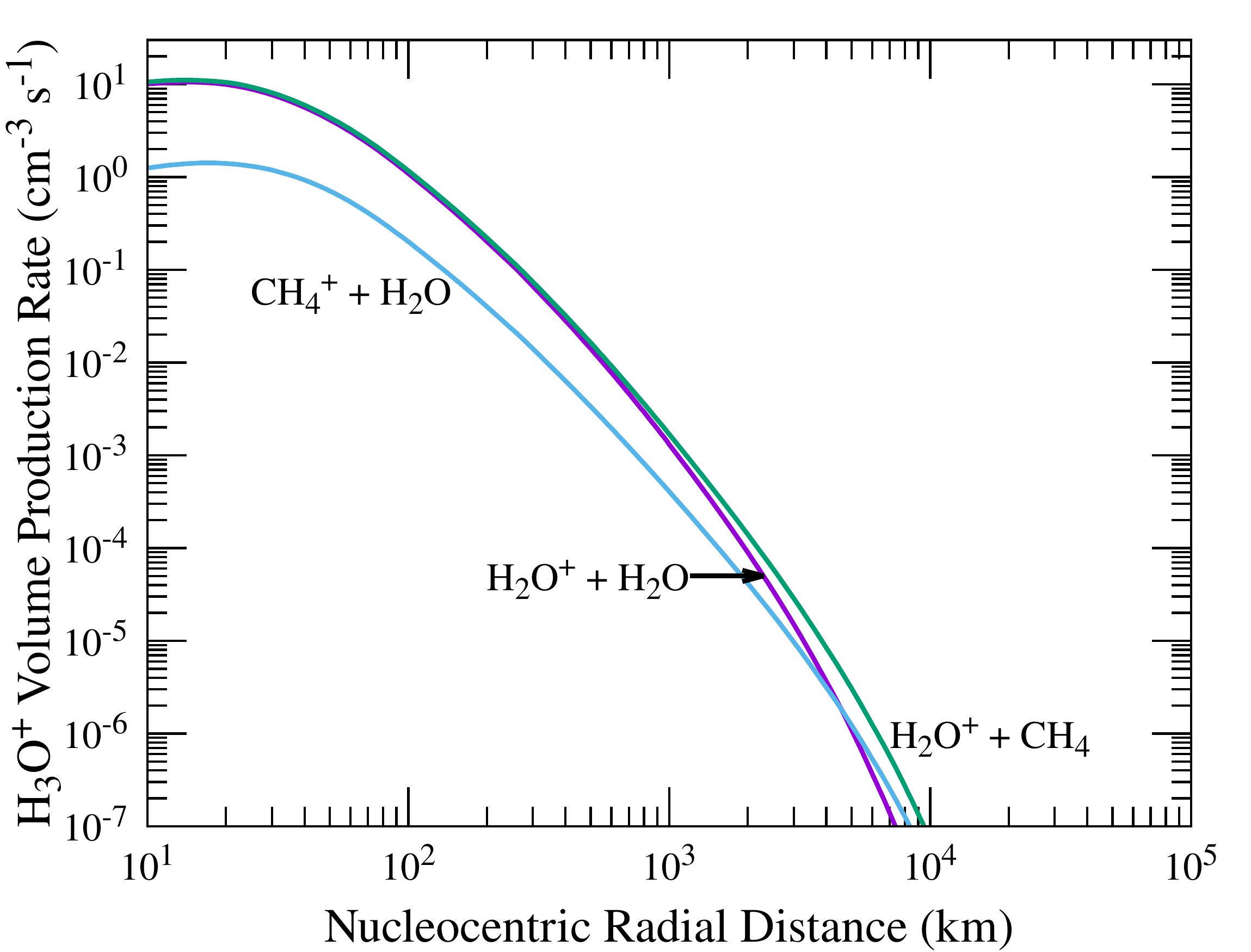}
	\includegraphics[width=\columnwidth]{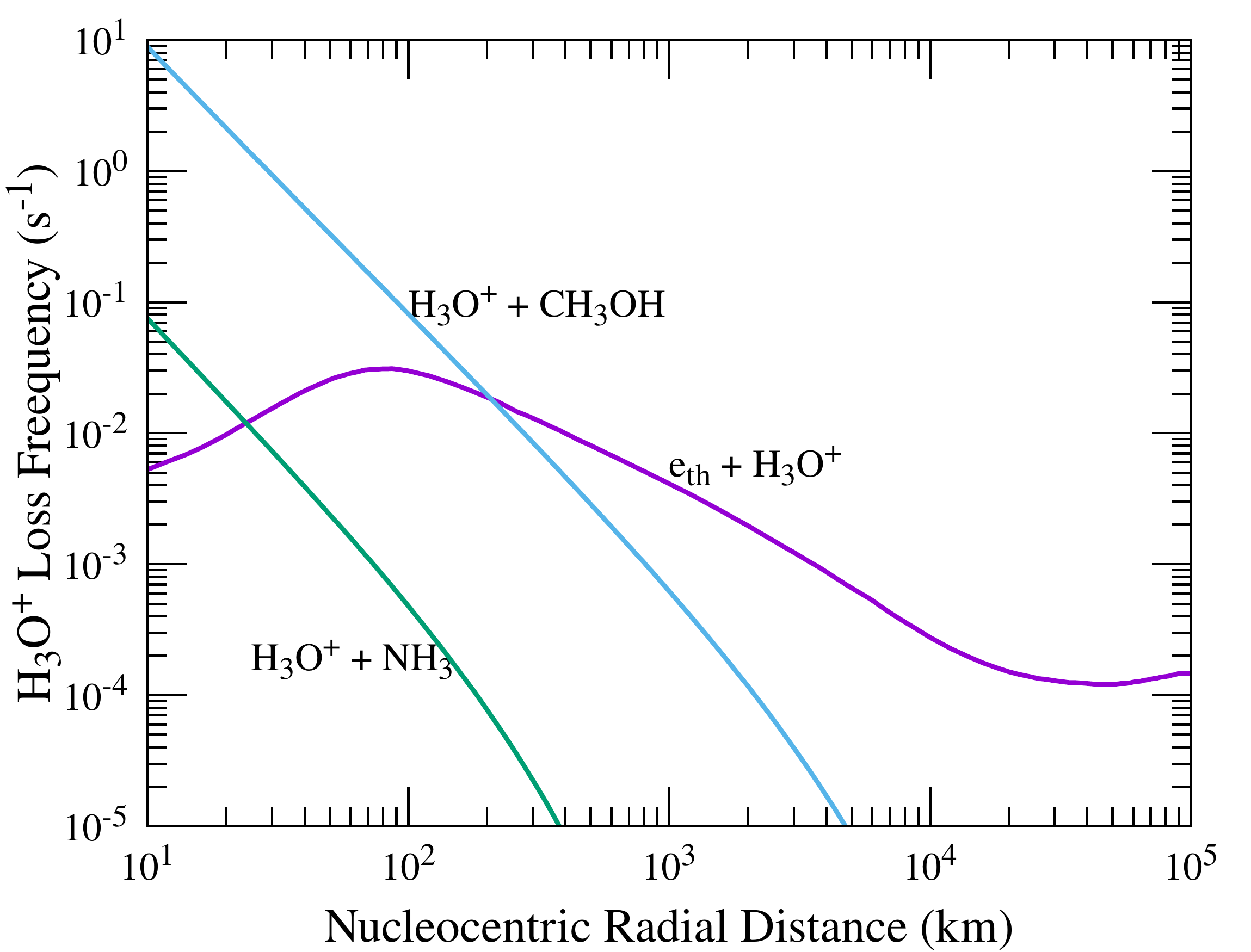}
	\vspace{-0.5cm}
	\caption{Modelled  production rate (top panel) and loss frequency 
		(bottom panel) profiles of H$_3$O$^+$ in the coma of comet C/2016 R2. 
		Input conditions are the same as explained in 		
		Figure~\ref{fig:pr_co2p}. h$\nu$, e$_{ph}$, and e$_{th}$ represent 
		solar photon, photoelectron, and thermal electron, respectively.  
		\label{fig:pr_h3op}}
\end{figure}

As shown in the top panel of Figure~\ref{fig:pr_h3op}, proton transfer 
reactions of  {H$_2$O$^+$ with H$_2$O and CH$_4$} ions produce H$_3$O$^+$ 
with 
equal production 
rates in 
the inner coma. At radial distances larger than 10$^4$ km proton transfer 
between 
H$_2$O
and CH$_4^+$ is also an important source of H$_3$O$^+$.   The modelled loss 
frequency 
profiles of H$_3$O$^+$ in the bottom panel of Figure~\ref{fig:pr_h3op} show 
that the proton transfer reaction between H$_3$O$^+$ and CH$_3$OH is the 
 {most significant} loss source of H$_3$O$^+$ for radial distances 
smaller than 
300 
km and above this radial  distance thermal recombination takes over as the 
main loss source of this ion.

\begin{figure} 
	\includegraphics[trim=0 35  0 0, clip,
	width=\columnwidth]{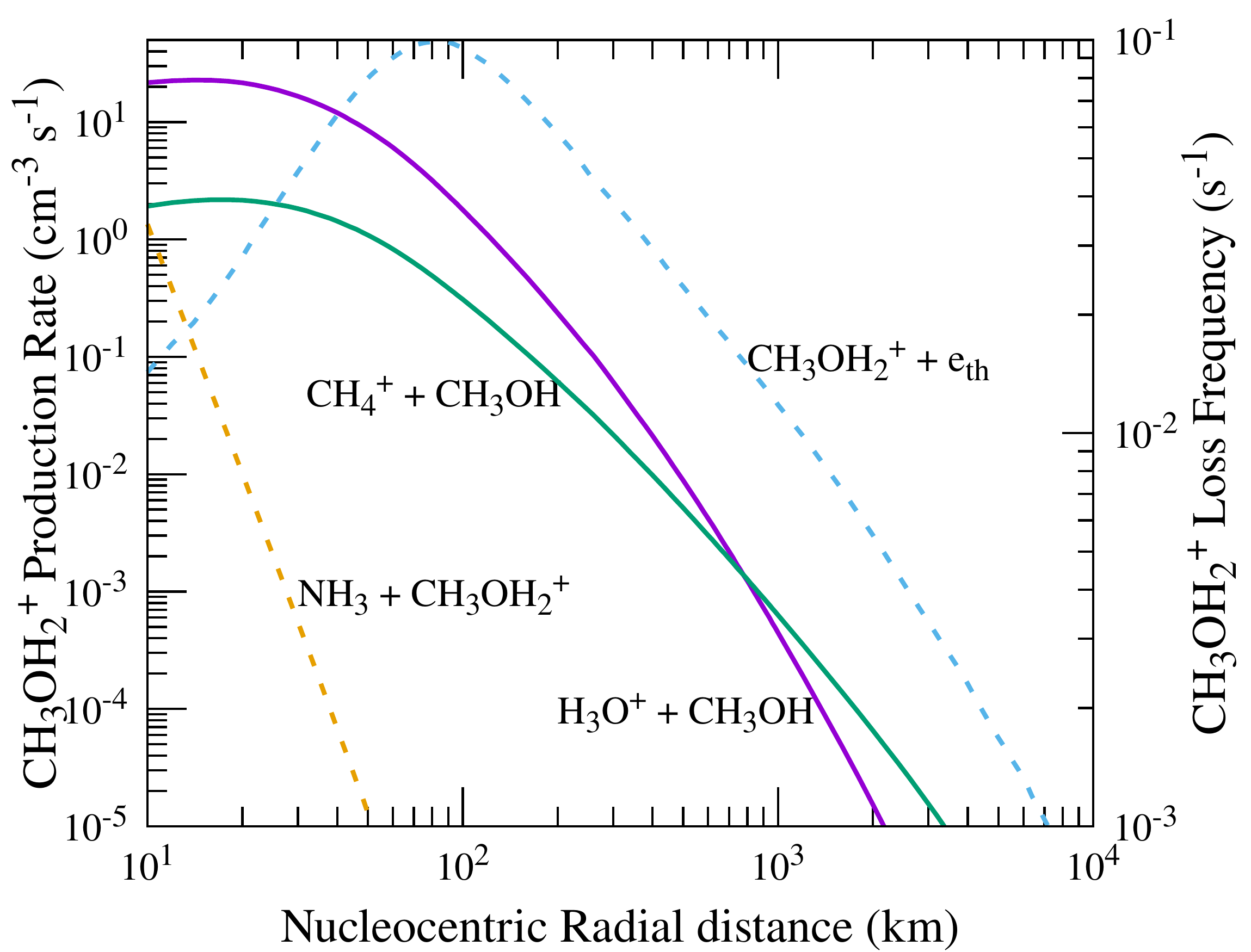}
	\includegraphics[width=\columnwidth]{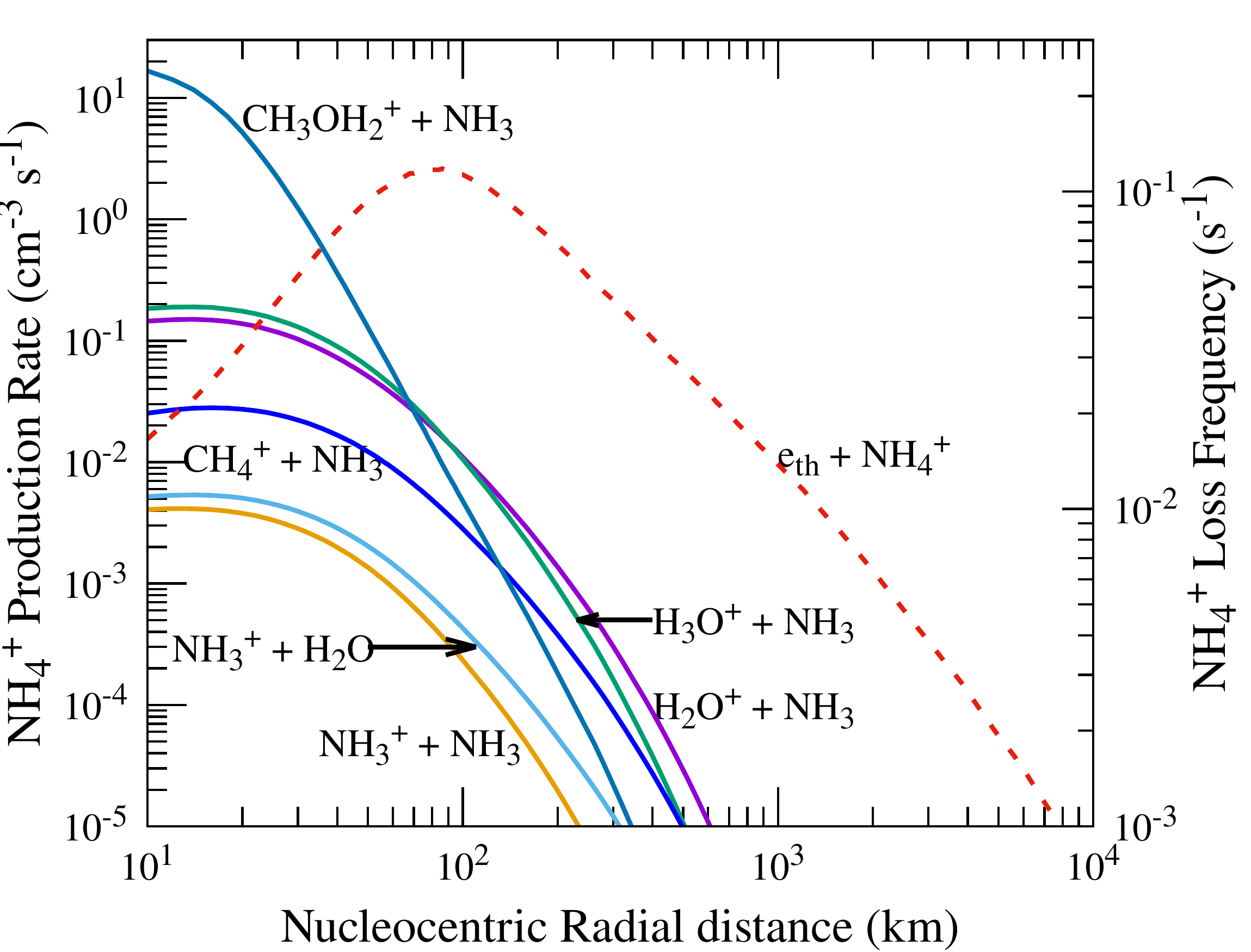}
	\vspace{-0.5cm}
	\caption{Calculated production rate and loss frequency profiles of 
	CH$_3$OH$_2^+$
		(top panel) and NH$_4^+$ (bottom panel). Loss frequency profiles are 
		plotted with dashed curves with scale on right y-axis. Input conditions 
		are the same as explained in 		
		Figure~\ref{fig:pr_co2p}.  e$_{th}$ represents the thermal electron.
		\label{fig:pr_ch3oh2p}}
\end{figure}

As shown in the top panel of Figure~\ref{fig:pr_ch3oh2p}, the formation of 
CH$_3$OH$_2^+$ is mainly due to proton transfer reaction of CH$_3$OH with 
H$_3$O$^+$ and CH$_4^+$ (see solid curves in this Figure). Close to the 
surface of nucleus, the loss
of this ion is significantly controlled by charge exchange between 
CH$_3$OH$_2^+$ and NH$_3$ which leads to NH$_4^+$ formation. Thermal 
recombination is the dominant loss source of this 
ion compared to collisional removal processes in the inner coma of comet C/2016 
R2. 

Similarly, as shown in the bottom panel of Figure~\ref{fig:pr_ch3oh2p}, 
several proton transfer reactions are involved in the production of NH$_4^+$.
As explained earlier, proton transfer between CH$_3$OH$_2^+$ and NH$_3$ leads 
to the significant formation of NH$_4^+$ for radial distances below 100 km. 
Above this radial distance the proton transfer reactions of H$_3$O$^+$, 
H$_2$O$^+$, and CH$_4^+$ with NH$_3$ are the  important formation sources of 
NH$_4^+$. Thermal 
recombination is the most significant loss source of this ion 
 {throughout} the 
cometary 
coma. 

\subsection{Production and loss mechanisms of atomic ions C$^+$ and O$^+$}
\begin{figure} 
	\includegraphics[trim=0 35  0 0, clip,
	width=\columnwidth]{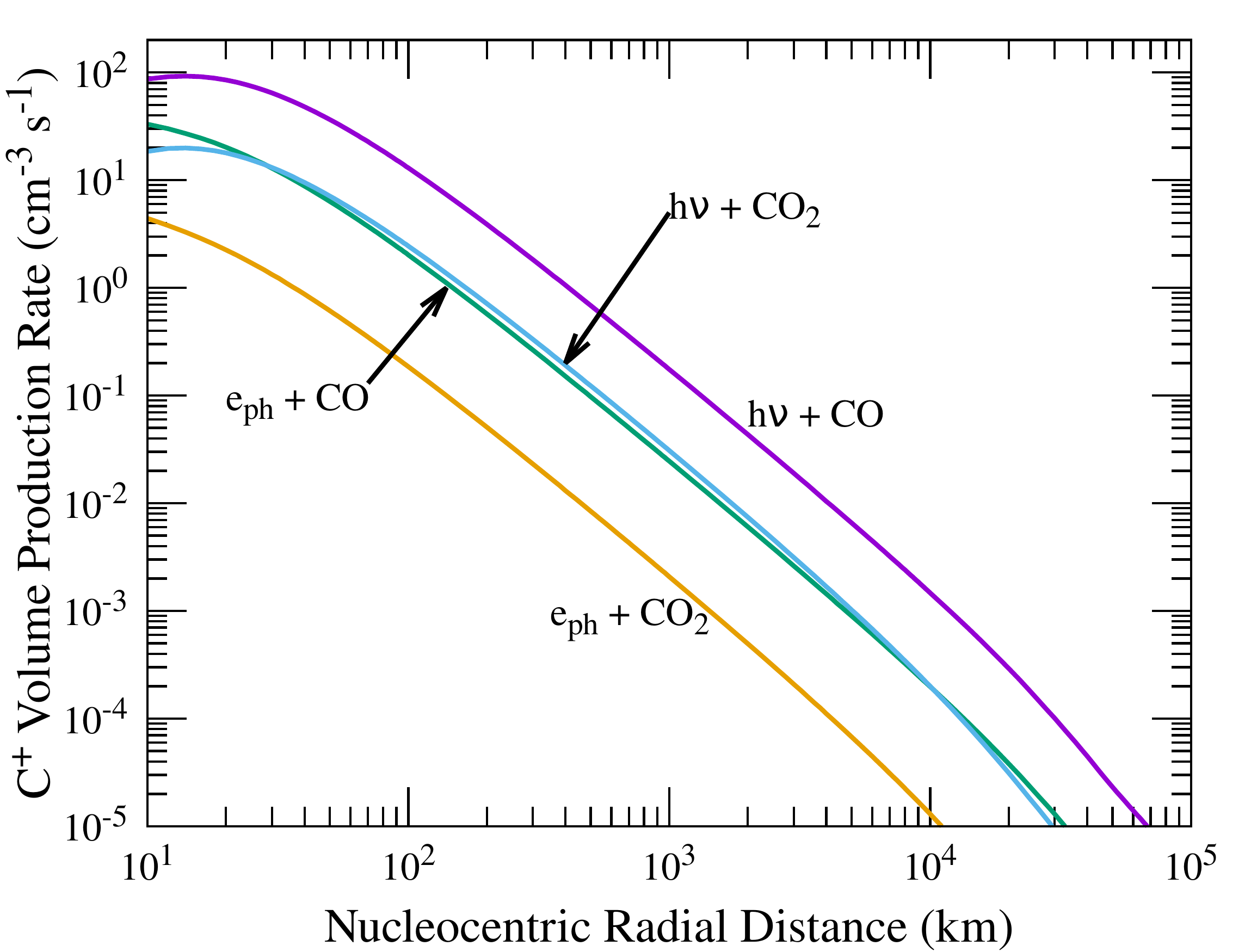}
	\includegraphics[width=\columnwidth]{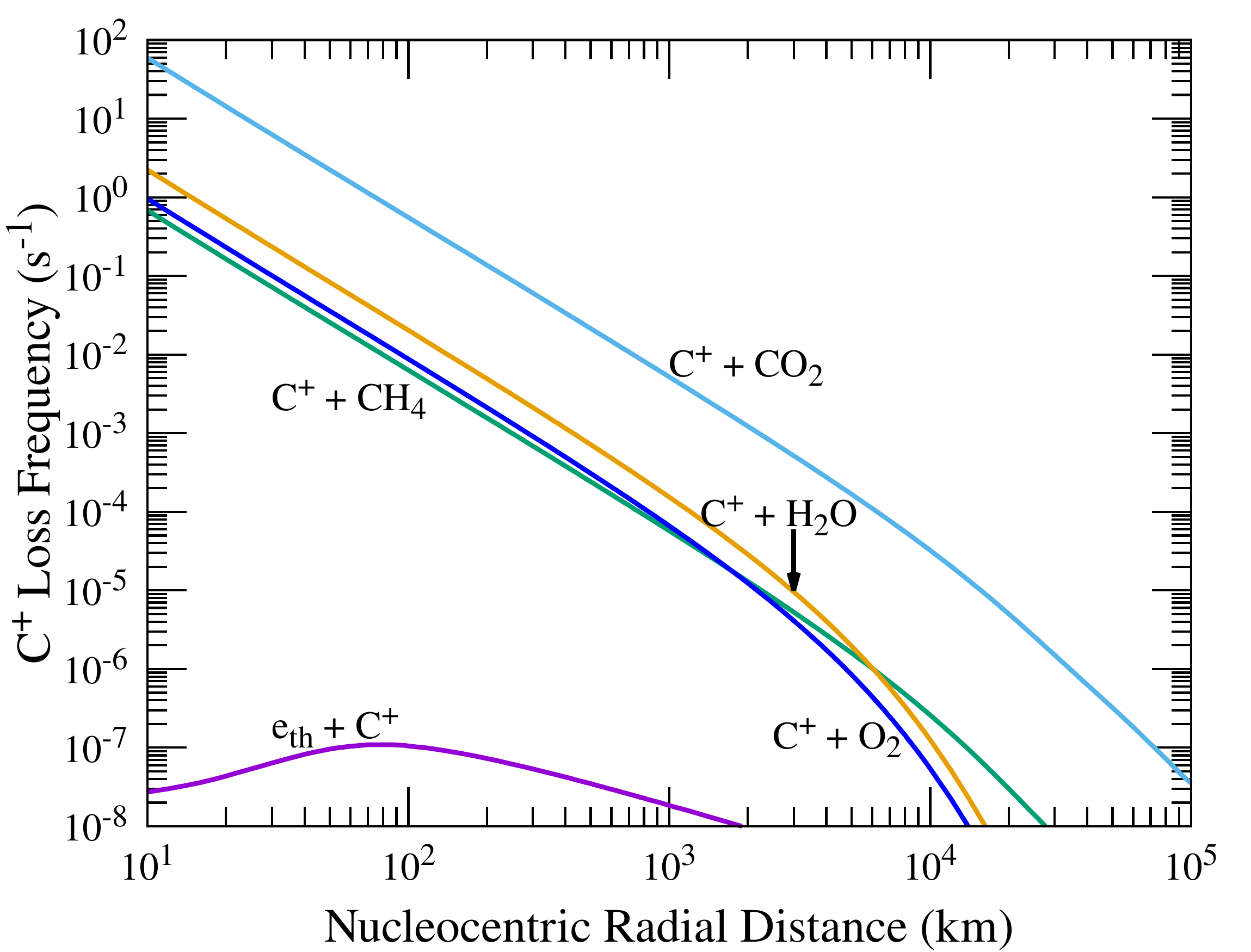}
	\vspace{-0.5cm}
	\caption{Modelled production rate (top panel) and loss frequency profiles 
	(bottom panel) of C$^+$ in comet C/2016 R2. Input conditions are the same 
	as explained in Figure~\ref{fig:pr_co2p}. h$\nu$, e$_{ph}$, and e$_{th}$ 
	represent solar photon, photoelectron, and thermal electron, respectively.  
	\label{fig:pr_cp}}
\end{figure}

The modelled formation and loss processes of C$^+$ in the inner coma of comet 
C/2016 R2 are presented in top and bottom panels of Figure~\ref{fig:pr_cp}, 
respectively. 
Calculations in this Figure show that the major formation of C$^+$ 
 {occurs  
due} to  the
photoionization of CO. Other production processes such as electron impact of CO 
and ionization of CO$_2$ by photons and photoelectrons together contribute to
about 50\% of the total C$^+$ production. The calculated loss frequency 
profiles in the bottom panel of Figure~\ref{fig:pr_cp} show that the collision 
with CO$_2$ is the significant loss source of this ion  {throughout} the 
inner 
coma.  

\begin{figure} 
	\includegraphics[trim=0 35  0 0, clip,
	width=\columnwidth]{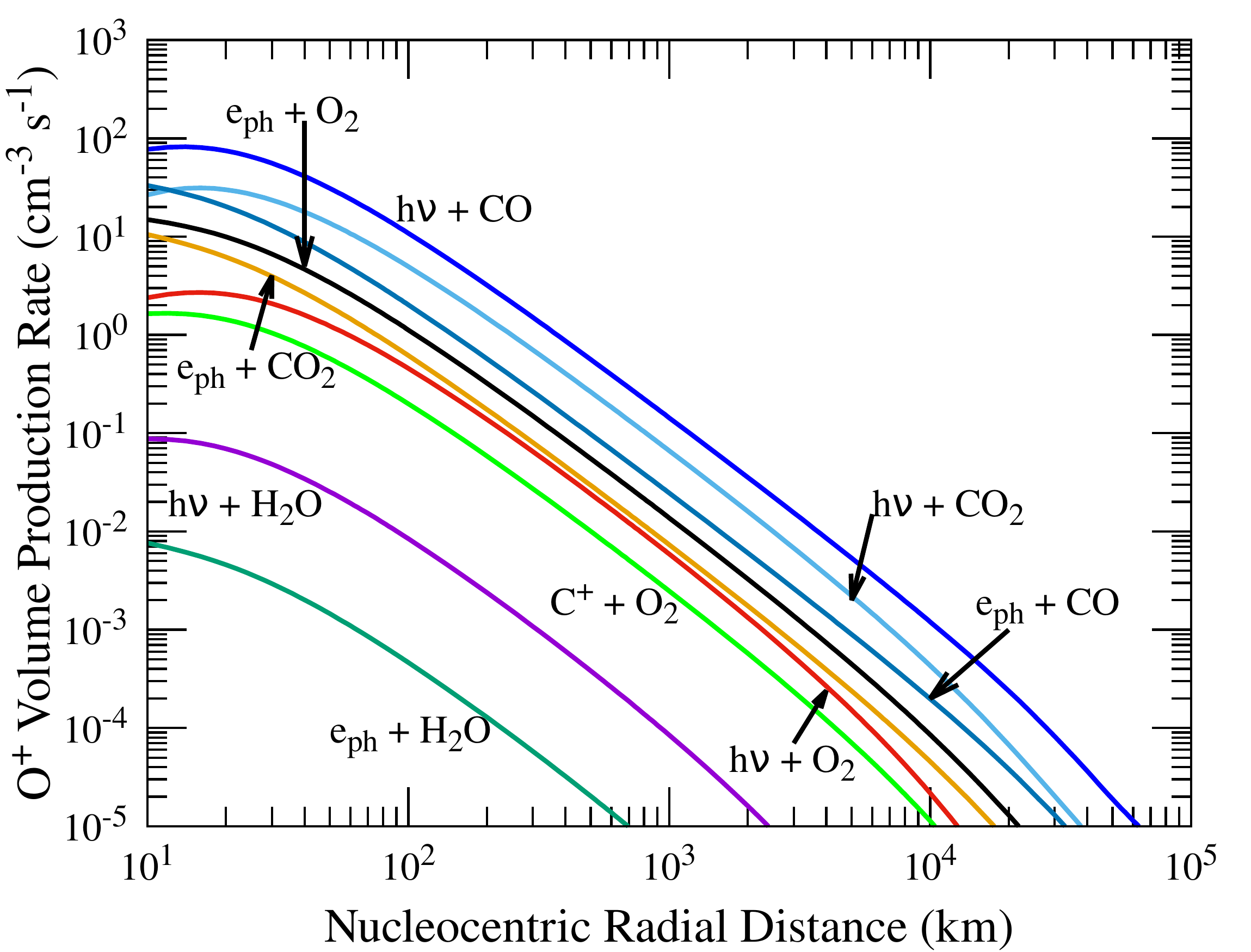}
	\includegraphics[width=\columnwidth]{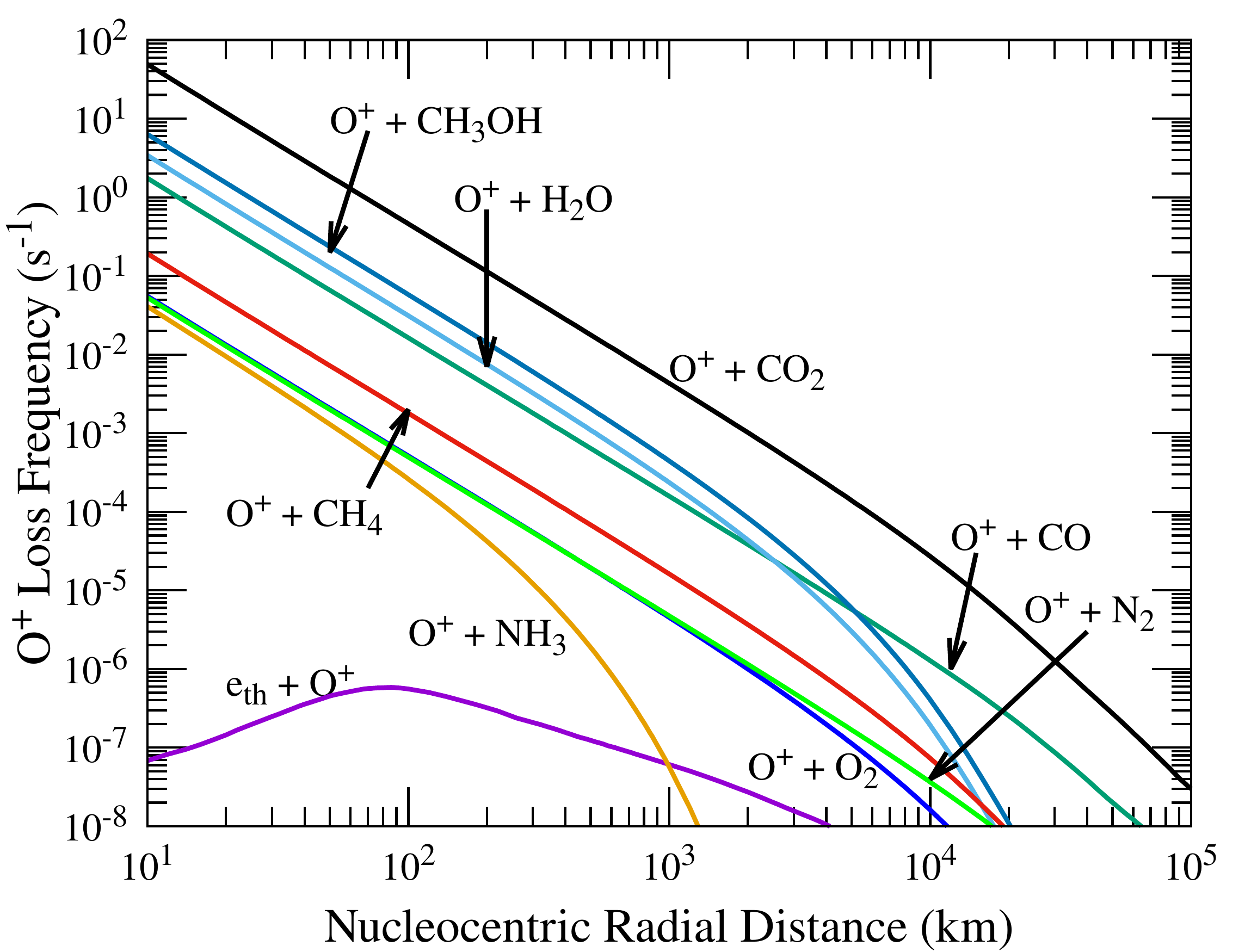}
	\vspace{-0.5cm}
	\caption{Calculated production rate (top panel) and loss frequency 
	profiles (bottom panel) of O$^+$ in comet C/2016 R2. Input conditions are 
	the same as explained in Figure~\ref{fig:pr_co2p}. h$\nu$, e$_{ph}$, and 
	e$_{th}$ represent solar photon, photoelectron, and thermal electron, 
	respectively. \label{fig:pr_op}}
\end{figure}

Similarly, the modelled formation rate and loss frequency profiles of O$^+$ are 
presented in top and bottom panels of  Figure~\ref{fig:pr_op}, respectively. 
The production of O$^+$ is majorly due to photoionization of CO followed by 
 {photodissociative ionization} of CO$_2$. The contribution from other 
sources to the O$^+$ 
production rate is negligible. As shown in the bottom panel of this figure, the 
loss of this ion is mainly due to collisions with CO$_2$ which leads to the 
formation of O$_2^+$. Thermal recombination   {plays} no significant role 
in the removal of this ion due to  slow reaction rate. 

\subsection{Time scales for ions}
\begin{figure} 
	\includegraphics[width=\columnwidth]{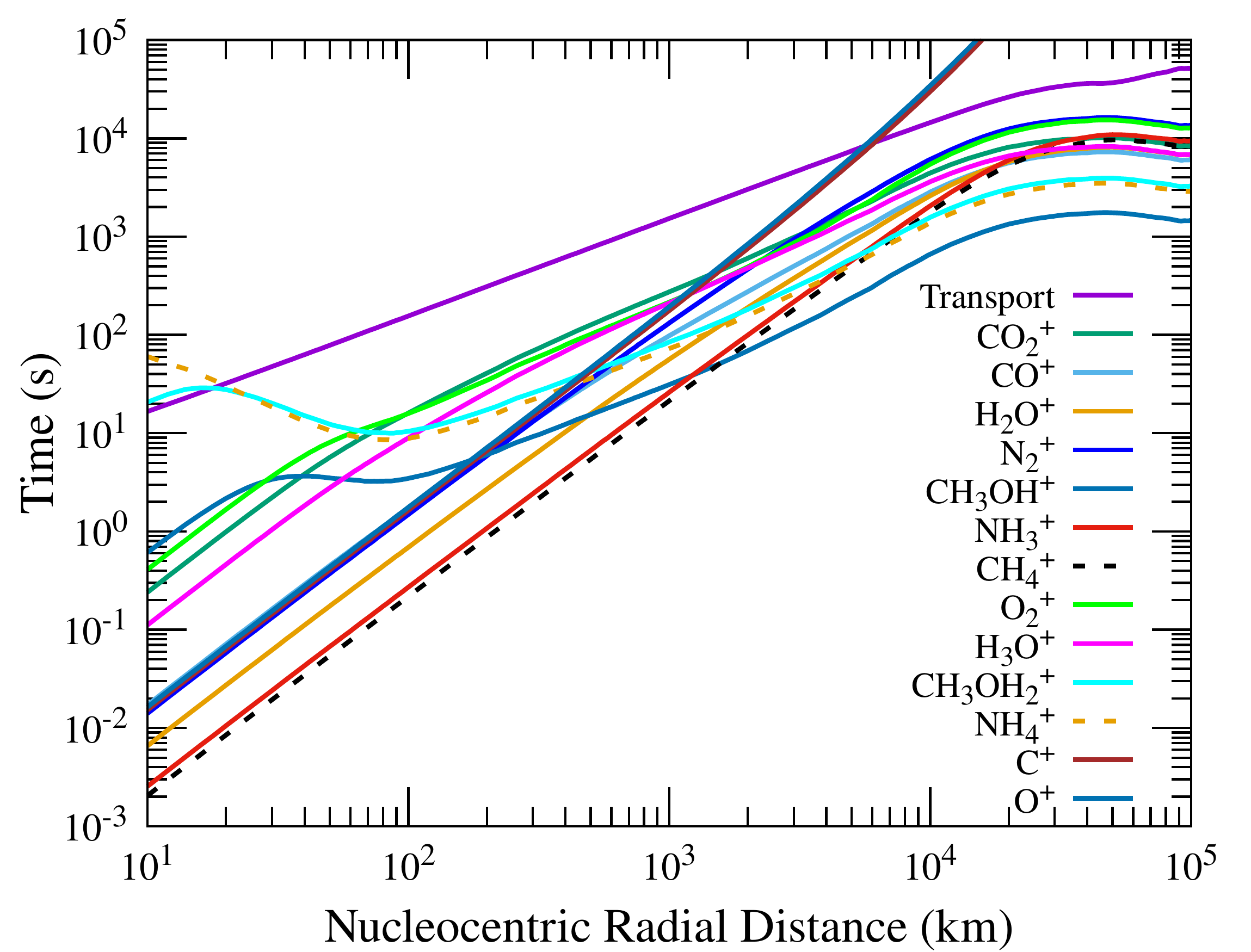}
	\vspace{-0.5cm}
	\caption{Calculated lifetime profiles of different ions due to 
	transport and chemical reactions in C/2016 R2. Input conditions are the 
	same as explained in Figure~\ref{fig:pr_co2p}.  \label{fig:lifetimes}}
\end{figure}

The calculated time scale profiles for the chemical loss of ions, which is due 
to collisions between species and thermal recombination, and transport due to 
advection are plotted in Figure~\ref{fig:lifetimes}. Since the chemical 
lifetimes of most of the ions are smaller than lifetime due to transport,  they 
are under photochemical equilibrium condition. For  radial distances below 50 
km, CH$_3$OH$_2^+$ and NH$_4^+$ have chemical lifetimes longer than the 
transport time scales  due to significantly lower collisional reaction rates 
with other species (see Figure~\ref{fig:pr_ch3oh2p}). The radial transport of 
C$^+$ and O$^+$ is important for radial distances above 5000 km.

%

\subsection{Ion density distribution}
\begin{figure} 
	\includegraphics[width=\columnwidth]{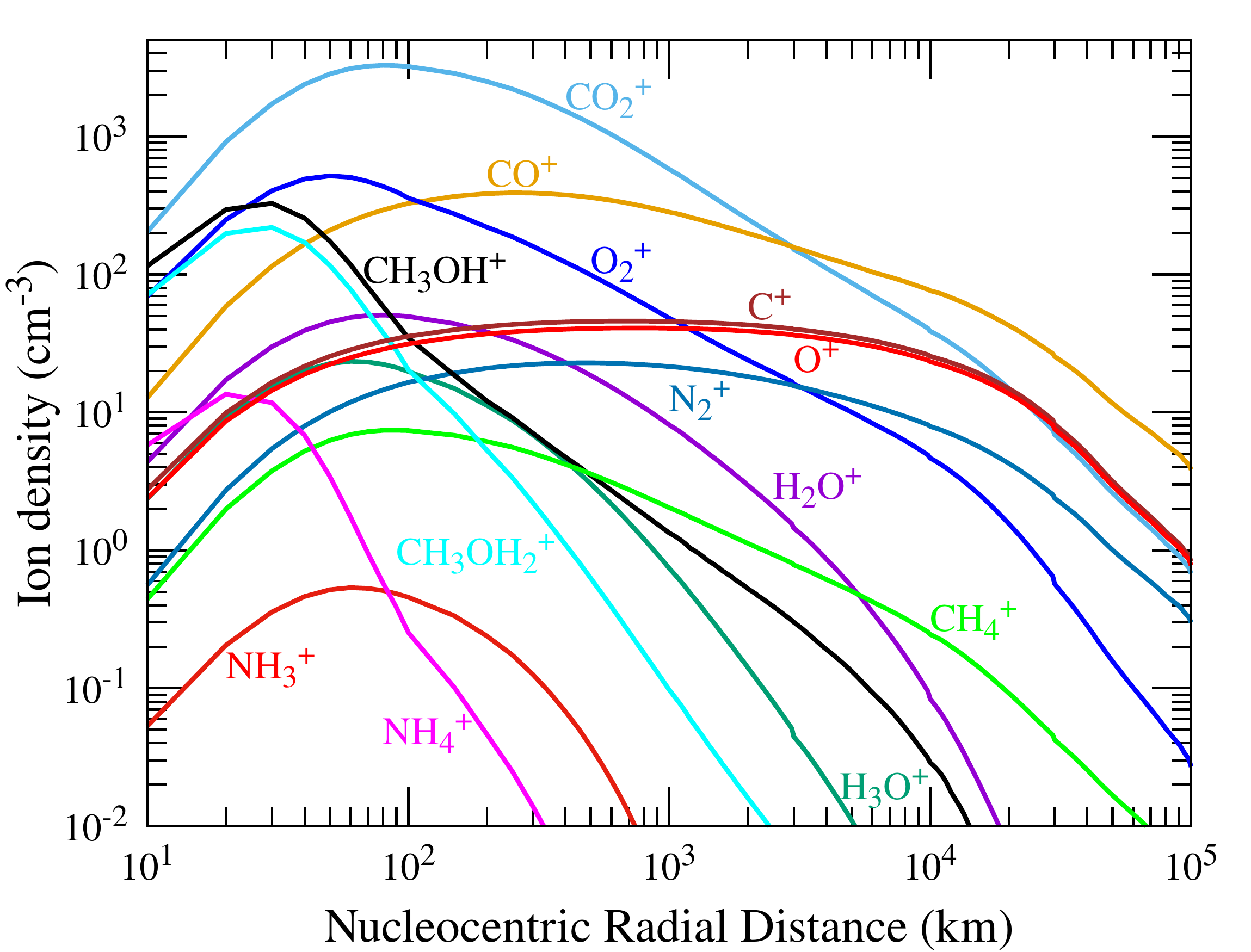}
	\vspace{-0.5cm}
	\caption{The modelled density profiles for different ions in the coma of 
	 comet  {C/2016 R2}. Input conditions are the same as 
	 explained in Figure~\ref{fig:pr_co2p}. \label{fig:ion_den}}
\end{figure}
By incorporating the previously discussed formation and destruction 
mechanisms the  ion density profiles in the inner coma of comet C/2016 
R2 are computed and presented in Figure~\ref{fig:ion_den}. These calculations 
show that CO$_2^+$ is the dominant ion for radial distances smaller than 
3000 km. But close to the nucleus, CH$_3$OH$^+$, CH$_3$OH$_2^+$, and O$_2^+$  
are also important ions. The peak density  {occurs} at  {a} radial 
distance 
around 100 
km with CO$_2^+$ as dominant ion. Above  {a radial distance of  3000 km} 
CO$^+$ is the 
dominant ion in the inner coma.

\subsection{Volume emission rates and emission intensity ratios}
\begin{figure*} 
	\includegraphics[trim=0 10 0 0, clip, 
	width=\columnwidth]{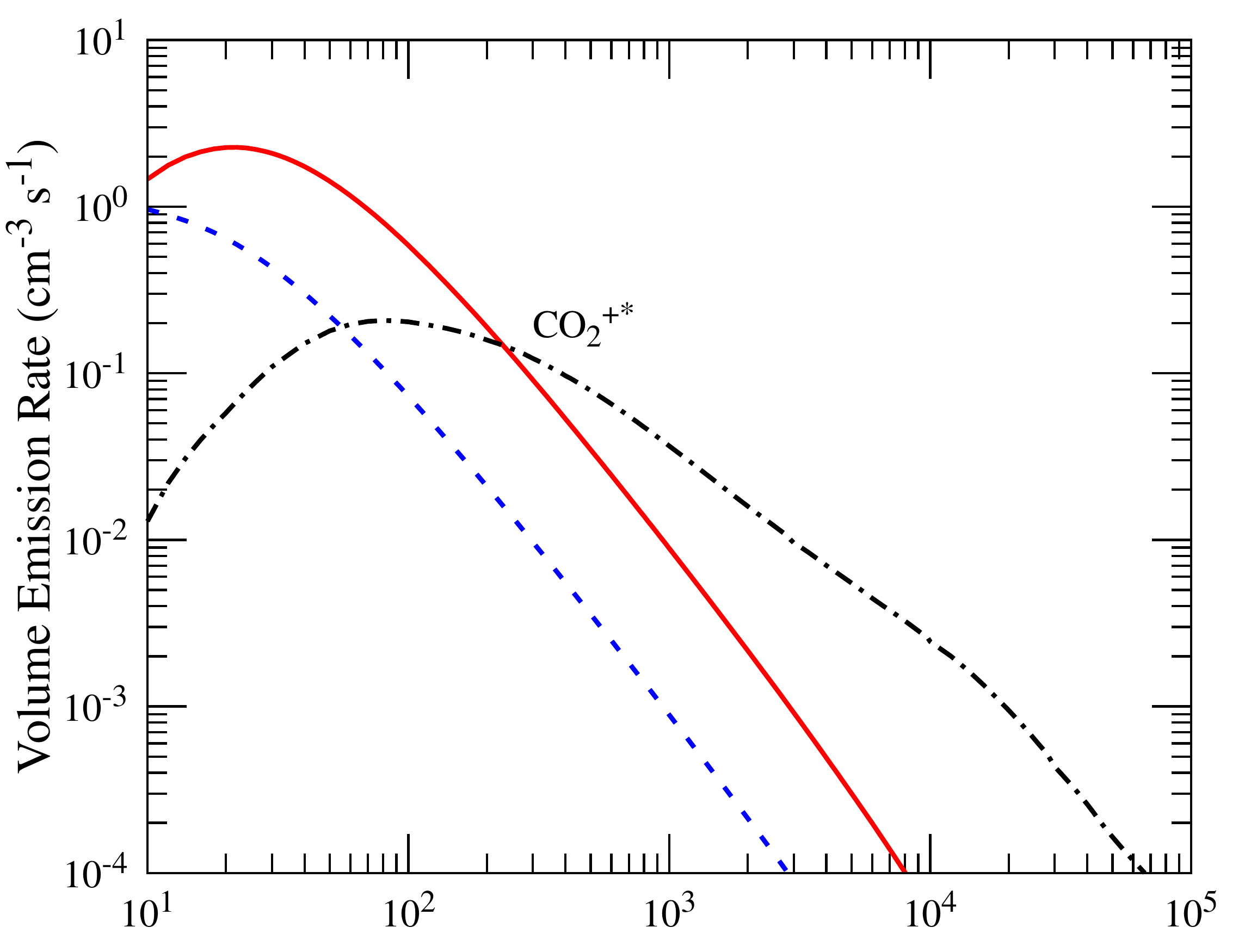}	
	\includegraphics[trim=0 10 0 0, 
	clip, width=\columnwidth]{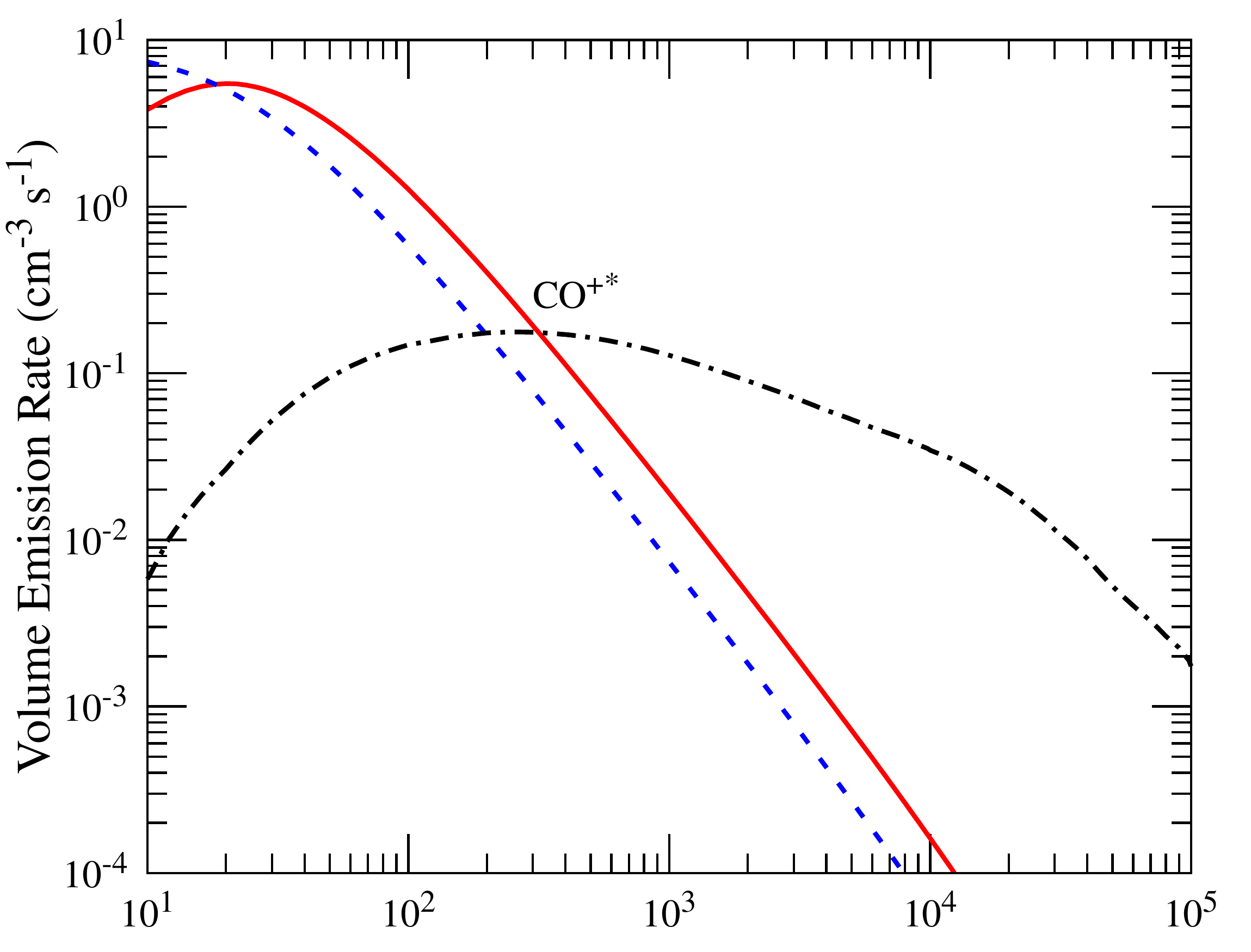}
		
	\includegraphics[width=\columnwidth]{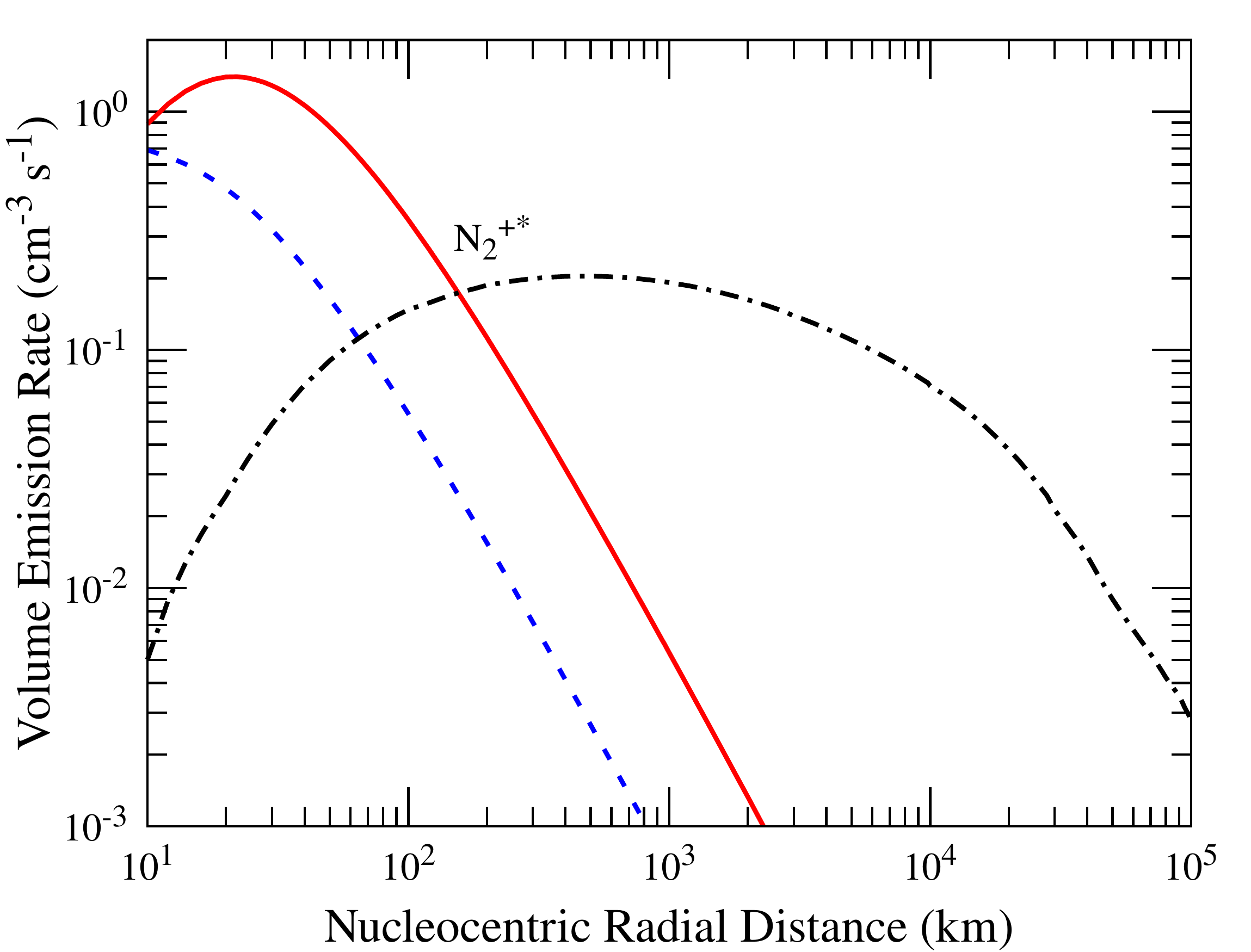}	
	\includegraphics[width=\columnwidth]{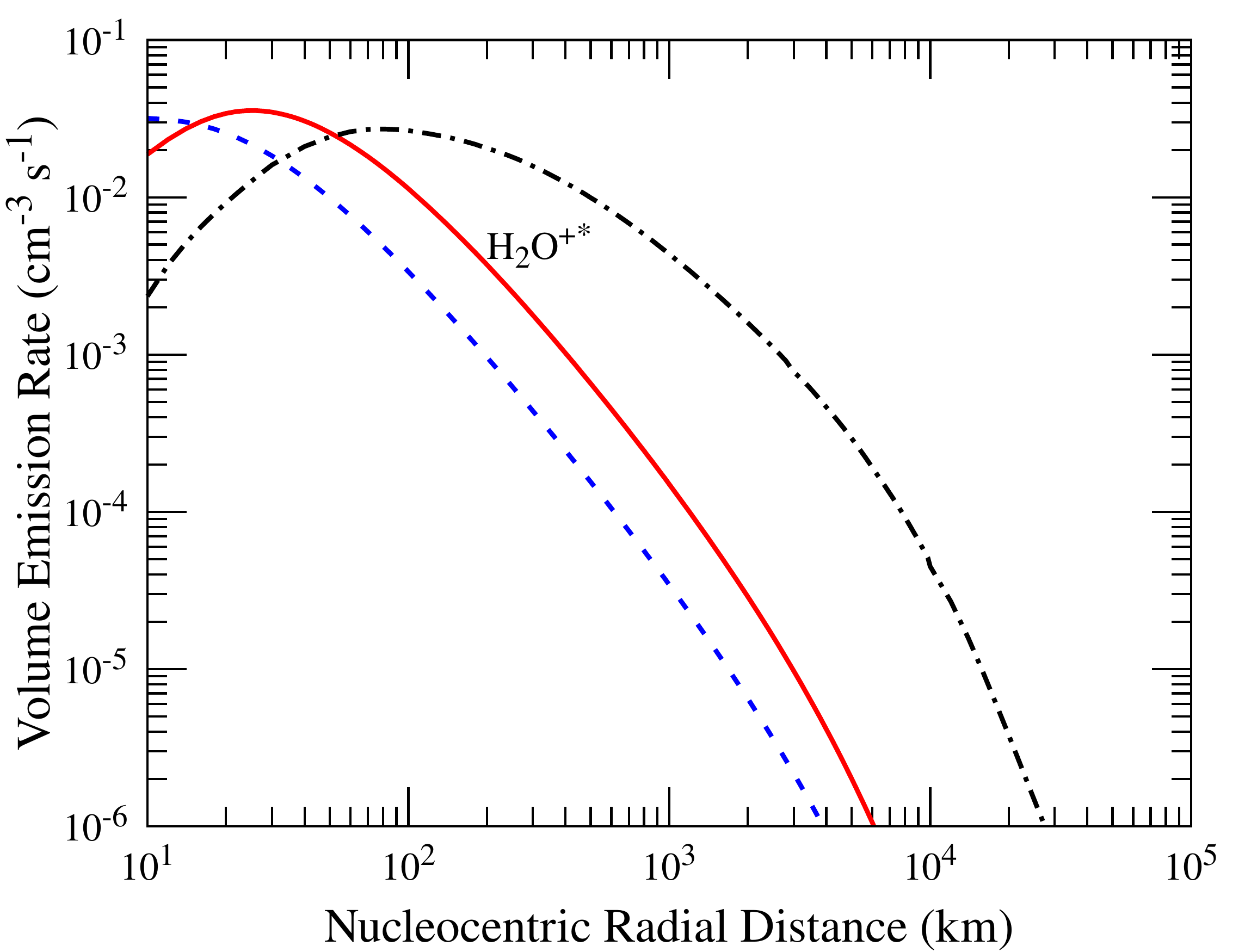}
	\vspace{-0.2cm}	
	\caption{ {Modelled volume emission rates for CO$_2^+$ (0-0)
		(top left), CO$^+$ (2-0) (top right), N$_2^+$(0-0) (bottom left), and 
		H$_2$O$^+$ (8-0) (bottom right) transitions. Solid red, dashed blue, 
		and dash-dotted black  curves represent the modelled volume emission 
		rate profiles due to photoionization, electron impact ionization, 
	    and solar resonance fluorescence of corresponding neutrals and ions,  
	    respectively.}  Input conditions are the same as 
		explained in Figure~\ref{fig:pr_co2p}.  \label{fig:vemis} }			
\end{figure*}

Figure~\ref{fig:vemis} shows the calculated volume 
emission rate profiles for different excited states of CO$_2^+$, CO$^+$, 
H$_2$O$^+$, and N$_2^+$.  {Photoionization and photoelectron impact 
ionization of corresponding neutrals are the important sources for the 
formation excited 
states of  these ions for radial distances smaller than 
300 km.}  Above these radial distances solar resonance 
fluorescence is the dominant excitation source. Close to the surface of the 
nucleus the electron impact ionization excitation is also a significant source 
for the excited states of ions.

%

\begin{figure} 
	\includegraphics[width=\columnwidth]{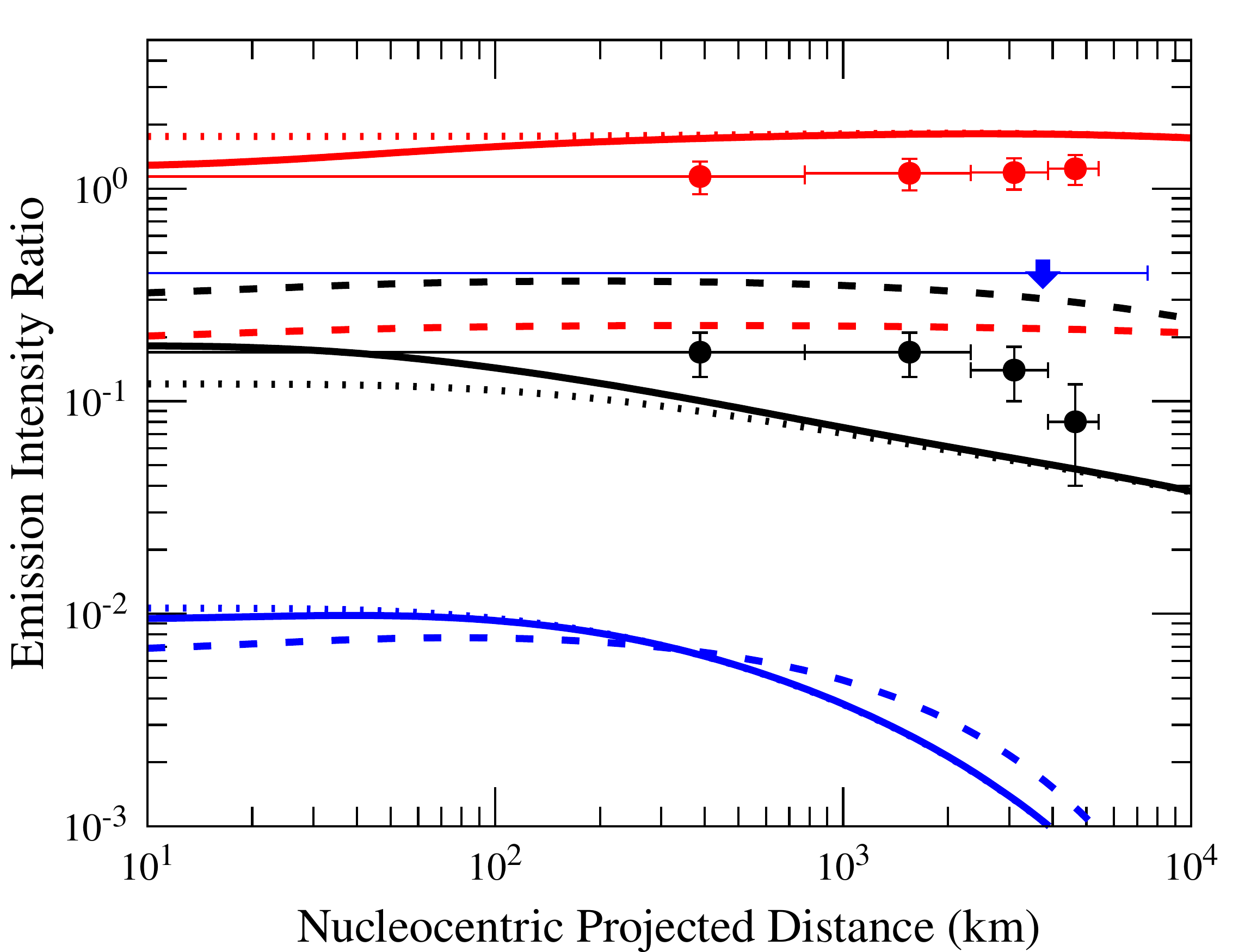}
	\vspace{-0.5cm}		
	\caption{Modelled emission intensity ratios of N$_2^+$/CO$^+$ (red), 
		CO$_2^+$/CO$^+$ (black), and H$_2$O$^+$/CO$^+$ (blue) as a function of 
		projected distance.
		Solid curves are the modelled ratio profiles by accounting for  
		ionization and	excitation neutral species by photons and 
		photoelectrons, and solar resonance fluorescence excitation mechanisms. 
		Dashed and 	dotted curves 
		represents the modelled emission intensity ratios profiles by
		accounting for only ionization of neutrals and only solar resonance 
		fluorescence mechanisms, 
		respectively. The observed flux ratios are plotted with corresponding 
		colours with vertical error bars. The blue horizontal line with a
		 {downward arrow} represents 
		the derived upper limit of H$_2$O$^+$/CO$^+$ emission intensity ratio.  
		Input 	
		conditions are the	same as explained in Figure~\ref{fig:pr_co2p}.  
		\label{fig:emis_ratios} 
	}			
\end{figure}

We present the measured and modelled  emission intensity ratios of
N$_2^+$/CO$^+$ and CO$_2^+$/CO$^+$ in comet C/2016 R2 as a function of the 
nucleocentric projected distance in Figure~\ref{fig:emis_ratios}. In this 
Figure, the  H$_2$O$^+$/CO$^+$ upper limit is also 
indicated.  The solid curves in this figure are the modelled 
emission intensity ratios by accounting for ionization of neutrals by solar 
photons and photoelectrons, and also resonance fluorescence excitation, which 
we call a standard case. The computed intensity ratios  (solid curves) are 
consistent with 
the observations within a factor 2 for CO$_2^+$/CO$^+$ and 50\% for 
N$_2^+$/CO$^+$. Our calculated H$_2$O$^+$/CO$^+$ 
emission intensity ratio profile is smaller by more than one  order of 
magnitude compared to the derived upper limit of the observation.

We  have also done several case studies to explore the impact of different 
excitation processes on the modelled ion emission intensity ratios. The 
implications of these calculations are discussed in the later section.  When we 
consider only photon and electron impact  ionization and excitation 
of neutrals,  the modelled  N$_2^+$/CO$^+$ emission intensity ratio  is 
significantly decreased by a factor of 5 or more compared to the standard case 
(see red dashed line in Figure~\ref{fig:emis_ratios}). In this case,  the 
modelled CO$_2^+$/CO$^+$ emission intensity ratio profile  is higher by a 
factor of about 2 for radial distances below 300 km, whereas, it is 
higher by an order of magnitude at large radial distances compared to standard 
case  (see black dashed curve in Figure~\ref{fig:emis_ratios}). By neglecting 
the resonance fluorescence excitation mechanism, the modelled 
H$_2$O$^+$/CO$^+$  emission intensity ratio profile is nearly consistent 
with the standard case (see blue dashed curve in Figure~\ref{fig:emis_ratios}). 
We  {also} calculated ion emission intensity ratio profiles by accounting 
for 
resonance fluorescence as the only excitation mechanism. As shown in Figure 
\ref{fig:emis_ratios}, the modelled H$_2$O$^+$/CO$^+$ (see blue dotted curve),
CO$_2^+$/CO$^+$ (see black dotted curve), N$_2^+$/CO$^+$ (see red dotted curve) 
emission intensity ratios are in agreement with our standard case values.

\vspace{-15pt}
\section{Discussion} \label{sec:discuss}
\subsection{Ion-neutral chemistry in CO-dominated coma}

In a water-dominated cometary coma, photoionization of H$_2$O produces
H$_2$O$^+$ and the collisions between the former and later species lead to the 
formation of H$_3$O$^+$.  Even with a small mixing ratio (about 
1\%), the molecules having higher proton affinity  can quickly react with 
H$_3$O$^+$ and produce NH$_4^+$ and CH$_3$OH$_2^+$ as the dominant ions in the 
inner coma \citep{Haider05, Vigren13, Heritier17}. But our modelled density 
profiles in Figure~\ref{fig:ion_den} show that the ion composition in the inner 
coma of  comet C/2016 R2  is completely different when compared to a regular 
water-dominated comet. This unusual ionospheric composition is mainly due to 
the peculiar neutral composition  of cometary coma. Unlike other comets, 
comet C/2016 R2 coma is dominantly composed of CO and remarkably depleted in 
water. In this peculiar  coma composition, the  modelled ion density 
profiles show that CO$_2^+$ is the dominant ion for radial distances below 3000 
km, with a peak ion density at around 100 km, whereas the CO$^+$ density is 
significant for radial distances above 5000 km (see Figure~\ref{fig:ion_den}).

The CO$_2^+$ and CO$^+$ ion distribution can be 
explained based on their 
corresponding  formation and destruction processes in the inner coma. The loss 
of CO$^+$  {mainly} depends on the collisions with CO$_2$, which is the 
major formation channel for CO$_2^+$ for radial distances smaller than 10$^3$ 
km (see Figures~\ref{fig:pr_cop}). But the loss of CO$_2^+$ significantly 
depends on the thermal recombination except for the radial distances smaller 
than 100 km. Our model calculation show that the destruction rate of CO$^+$ is 
more than an order of magnitude higher compared to that of CO$_2^+$  for radial 
distances smaller than 1000 km (see the lower panels of 
Figures~\ref{fig:pr_co2p} and \ref{fig:pr_cop}). These calculations suggest 
that the collisions between CO$^+$ and CO$_2$ strongly control both the CO$^+$ 
and CO$_2^+$ densities for radial distances smaller than 1000 km, whereas above 
this distance thermal recombination rates of these ions, which  are nearly 
 {the} same, determine the modelled ion densities. Our model calculations 
also show 
that the chemical  loss of CO$^+$, via CO$_2$ collisions,  produces CO$_2^+$ 
about an order of magnitude more efficiently  than photoionization of CO$_2$.  
Thus, the significant collisional loss of CO$^+$ substantially reduces its 
density in the 
coma and also leads to CO$_2^+$ as the dominant ion. Above 1000 km radial 
distance, CO$^+$ is the dominant ion  due to its higher formation rate compared 
 {to that} of CO$_2^+$. Due to its endothermic nature, the charge 
transfer reaction between CO$_2^+$ and CO does not occur hence, CO$_2^+$  can 
not  be removed by the collisions with CO in the cometary coma.

The modelled ion density profiles suggest that close to the nucleus surface, 
besides CO$_2^+$, CH$_3$OH$^+$, CH$_3$OH$_2^+$ 
and O$_2^+$ are also important ions with significant densities. The larger 
abundance of CO$_2$ in the coma and the collision reaction of O$^+$ with former 
species significantly produce O$_2^+$  below 100 km.  {Though we assumed 
a relatively large amount of O$_2$ compared to several other detected species 
in this 
comet (1\% of CO production rate), the calculations in top panel of
Figure~\ref{fig:pr_o2p} show that 
photoionization of O$_2$ plays no role in determining the O$_2^+$ 
density. For radial distances below 100 km, the formation rates of O$_2^+$ via 
charge exchange between CO$_2^+$ and O$_2$, and photoionization of O$_2$ are 
smaller by a factor 3 and 10 compared to formation rate via charge exchange 
between 
O$^+$ and CO$_2$, respectively (see Figure~\ref{fig:pr_o2p}). By varying the 
O$_2$ 
relative abundance between 1 and 
3\% with respect to the CO production rate no significant change in the 
modelled 
O$_2^+$ ion density is observed.   These 
calculations show that  the formation of O$_2^+$ is possible  due 
to the 
collisional chemistry even in the 
absence of O$_2$ in the  C/2016 R2
inner coma. Hence, our assumed O$_2$ relative 
abundance has no impact on 
the 
modelled O$_2^+$ ion density. }

For radial distances below 100 km, the 
contribution 
due to charge exchange between O$_2^+$ and CH$_3$OH  {leads to a 
significant amount of CH$_3$OH$^+$ formation when} compared to that due to 
the photoionization of CH$_3$OH (see Figure~\ref{fig:pr_ch3ohp}).  The strong 
proton affinity between CH$_3$OH$^+$ 
and H$_2$O  leads to CH$_3$OH$_2^+$ as an important ion close to the surface of 
the 
nucleus (see Figure~\ref{fig:pr_ch3oh2p}).  All these calculations suggest 
that for radial distances below 100 km, the CO$_2^+$, CH$_3$OH$^+$, 
CH$_3$OH$_2^+$, and O$_2^+$ ion densities are significantly determined by 
collisional chemistry among the neutral species and ions rather than photon and 
electron impact driven processes.

The ion density profiles of N$_2^+$, C$^+$, and O$^+$ are almost constant 
for  radial distances between 100 and 10$^4$ km. This can  be explained 
 {on the basis of} their corresponding production and loss mechanisms.
The ground state energy of N$_2^+$ is 15.57 eV, which is the largest 
among the considered ions. Hence, this ion can interact with 
most of the neutral species via  charge exchange mechanism. 
Our modelled loss frequency profiles show that N$_2^+$ ion interacts strongly 
with CO and CO$_2$ for the radial distances up to 3 $\times$ 10$^3$ km. Though 
CO$_2$ is less abundant than CO  {in the coma}, the modelled charge 
exchange rate of N$_2^+$ 
with CO$_2$  is comparable to 
that with CO (see the lower panel of Figure
\ref{fig:pr_n2p}). This is because the rate coefficient for 
the 
charge exchange between N$_2^+$ and CO$_2$ is higher by about an 
order 
of magnitude compared to that of N$_2^+$ and CO (see the reaction rates of R50 
and R55 
in 
Table~\ref{tab:chem-net}). Hence, the  strong collisional interaction of 
N$_2^+$ with major species in the coma makes the density profile of this ion 
almost constant  for radial distances between 100 and 10$^4$ km.
Similarly,  the constant C$^+$ and O$^+$ ion 
density profiles are also due to strong collisional interaction of these ions 
with CO$_2$ for  radial distances up to 10$^4$ km (see the lower panel 
of Figures~\ref{fig:pr_cp} and \ref{fig:pr_op}).

As explained earlier, in a water dominated comet, the protonated ions such 
as H$_3$O$^+$, NH$_4^+$, and CH$_3$OH$_2^+$ are significantly controlled by 
H$_2$O. Photoionization of H$_2$O is the primary reaction that produces 
H$_2$O$^+$ and further interaction of this ion with neutral species drives 
the collisional chemistry of protonated ions in the water-dominated coma. But 
in the case of comet C/2016 R2, which  is a water-poor comet, the
H$_2$O$^+$ ion density  for radial distances below 1000 km  {is}  
controlled by CO and CO$_2$ densities, rather than 
photoionization of H$_2$O. As shown by our calculations, for radial 
distances smaller than 1000km, the major formation H$_2$O$^+$ is due to the 
charge  {exchange} between  CO$_2^+$ and H$_2$O rather than  
photoionization of H$_2$O. Similarly, CO also  plays an important role 
in removing the H$_2$O$^+$  for radial distances smaller than 5000 km in the 
inner coma (see Figure~\ref{fig:pr_h2op}). This H$_2$O$^+$ ion chemistry, which 
is primarily driven by CO$_2$ and CO, further initiates the formation of 
H$_3$O$^+$ via proton transfer reactions of H$_2$O$^+$ with H$_2$O and CH$_4$.  
Due to high proton affinity, the collisions between CH$_3$OH and H$_3$O$^+$ 
results in CH$_3$OH$_2^+$ which further interact with NH$_3$ that leads to the 
formation of NH$_3^+$ for radial distances below 100 km (see 
Figures~\ref{fig:pr_h3op} and \ref{fig:pr_ch3oh2p}). Hence, unlike H$_2$O 
driven ion-neutral chemistry in normal comets, in comet C/2016 R2, CO and 
CO$_2$ play an important role in the formation of protonated ions.

The role of CO$_2$ is crucial in determining the 
ion-chemistry in this comet. Neutral density of CO$_2$ controls 
the total  loss 
frequencies and subsequently the ion  
densities of CO$^+$,  N$_2^+$, C$^+$, and O$^+$ for radial 
distances below 1000 km,  whereas 
the neutral density of CO significantly controls the loss rates of H$_2$O$^+$,
NH$_3^+$, and CH$_4^+$ ions.  Hence, any change in the abundances of CO$_2$ and 
CO  can significantly alter the total ion composition in the inner coma of 
this comet. 

\vspace{-10pt}
\subsection{Derivation of neutrals densities from the observed ion 	emission 
intensities}
Assuming solar resonance fluorescence is the only excitation mechanism, 
the observed flux ratios of ionic emissions are converted into ion density 
ratios and assumed to be equal to their 
respective neutral  {density ratios} in the C/2016 R2  coma  
\citep{Cochran18,Wierzchos18,Biver18,McKay19}. But none of these works 
account for the 
ionization and excitation of neutrals 
by photons and photoelectrons which can also produce the 
observed emissions. Here we discuss the derivation of neutral density ratio  
based on observed corresponding ion emission intensity ratio using our modelled 
emission processes in the inner coma.

\vspace{-10pt}

\subsubsection{Role of ionization and excitation processes in determining the 
ion emission intensity ratios}	

The ionization and excitation of neutrals by solar photons and photoelectrons 
 {can spontaneously} produce the observed emissions of 
ions. If 
these are the only emission 
mechanisms in the cometary coma, then the observed 
intensity ratios  {can be directly}  
used to constrain the corresponding neutral density ratio. If the
solar 
resonance fluorescence is the only excitation mechanism, then the
distribution of ion densities in the coma and the excitation factors 
(g-factors) determine the observed emission intensity ratio. Since the observed 
ion emissions are determined by both ionization and excitation of neutrals and 
resonance fluorescence of ions, it is 
essential 
to understand the photochemistry of cometary coma, which determines the 
ion density distribution, in order to derive the mixing ratios of neutral 
densities 
based on the 
observed ion emission intensity ratio.
Our calculations in Figure~\ref{fig:vemis} show that for radial distances 
smaller than 300 km, the modelled volume emission rates of  CO$^+$,  
CO$_2^+$, N$_2^+$, and H$_2$O$^+$ due to ionization and excitation of 
corresponding  neutrals by photons and photoelectrons are 
higher by several orders of magnitude compared to those due to solar resonance 
fluorescence. This calculation suggests that ionization and 
excitation of neutrals alone  can also determine the observed emission 
intensity ratio.  

When we account for ionization and excitation of neutrals and resonance 
fluorescence excitation mechanism,  our modelled emission intensity ratios of 
CO$_2^+$/CO$^+$ and N$_2^+$/CO$^+$
are consistent with the measured emission ratios within a factor of 2 (see  
solid red and black curves in Figure~\ref{fig:emis_ratios}).  {In the 
case of {H$_2$O$^+$/CO$^+$}, the modelled emission ratio } 
 is smaller by two orders of magnitude compared to 
the derived 
upper limit from the observation. Our modelled volume emissions rates in 
Figure~\ref{fig:vemis} show that the ionization and excitation of N$_2$ 
and CO can also produce the corresponding 
ionic emissions in the cometary coma. Hence, to assess the role of ionization 
and excitation of neutrals on the modelled ion 
emission ratios, we neglected the resonance fluorescence excitation. In this 
case, the modelled 
N$_2^+$/CO$^+$ emission intensity ratio is smaller by a factor of 5 compared to 
the observation (see red dashed line in Figure~\ref{fig:emis_ratios}). But when 
we account for resonance fluorescence as the only
excitation mechanisms, the modelled N$_2^+$/CO$^+$ emission intensity ratio is 
consistent with the observation (see red dotted curve in 
Figure~\ref{fig:emis_ratios}). This is mainly because the contribution from 
ionization and 
excitation of CO and N$_2$ to the total volume emission rate reduces rapidly 
for radial 
distances above 300 km whereas, the 
resonance fluorescence takes over the major excitation source (see 
Figure~\ref{fig:vemis}). Hence, the observed emission ratio  {is} 
significantly 
controlled by resonance fluorescence rather than  {the} remaining 
excitation 
processes.

By accounting for 
ionization and excitation of neutral and also resonance fluorescence 
mechanisms, we find 
the 
calculated CO$_2^+$/CO$^+$ emission intensity ratio decreases for projected 
distances above 300 km. This can be explained based on the modelled volume 
emission rates of these ions. As shown in Figure~\ref{fig:vemis}, at larger 
radial 
distances ($>$10$^3$ km), the  volume emission rate of CO$_2^+$ is smaller by 
more than an order of magnitude compared to that of CO$^+$, which is mainly 
due to the radial density distribution of these ions. Hence, the higher CO$^+$ 
ion density at larger radial 
distances reduces the modelled CO$_2^+$/CO$^+$ 
emission intensity ratio,  {as can be} noticed in the observation. 
When we only account for 
 ionization excitation processes of neutrals, the 
modelled ion emission intensity ratio of CO$_2^+$/CO$^+$  {is} closer to 
the 
observed profile for the radial distances below 2000 km (see black dashed curve 
in  Figure~\ref{fig:emis_ratios}). However,  the decrease in the observed 
 {CO$_2^+$/CO$^+$} emission intensity ratio for projected distances 
larger than 2000 km can not be explained by only ionization and excitation 
processes. All these calculations suggest that resonance fluorescence 
excitation plays a more significant role in determining the observed 
CO$_2^+$/CO$^+$ emission intensity ratio than ionization and excitation of 
neutrals at large radial distances.  

Due to lack of measured branching ratios, we assumed 50\% of 
electron impact ionization of  H$_2$O producing  {H$_2$O$^+$} in the
$\widetilde{A}^2$A$_1$ excited state.   
By increasing our assumed branching 
ratio to 100\%, no significant change in the modelled 
 {H$_2$O$^+$/CO$^+$} 
emission 
intensity ratio is found. This calculation suggests that solar photons 
significantly determine the ion emission intensity ratio rather than electron 
impact ionization and excitation of H$_2$O.
The calculation in Figure~\ref{fig:emis_ratios} show that the modelled  
H$_2$O$^+$/CO$^+$ emission intensity ratio does not vary significantly by 
accounting for 
only resonance fluorescence and/or ionization excitation of neutrals. This  
suggests that the ionization and 
excitation of neutrals play  {an} equal role  in determining the modelled 
H$_2$O$^+$/CO$^+$ emission intensity ratio compared to  {the role} of 
solar resonance 
fluorescence excitation mechanism.

\vspace{-10pt}
\subsubsection{Derivation of neutral density based on the N$_2^+$/CO$^+$ 
	emission intensity ratio}
\cite{Cochran18} converted the observed emission flux ratio of 
N$_2^+$/CO$^+$ to N$_2$/CO neutral density ratio based on the 
approach of  \cite{Lutz93} and 
\cite{Wyckoff03}. It should be noticed that \cite{Lutz93} and 
\cite{Wyckoff03} observed these ionic 
emissions on  cometary tail regions where the 
photoionization excitation processes are not significant compared to 
resonance fluorescence. Our modelled volume emission rate profiles also show 
that at larger radial distances emission intensity is  {mainly} 
determined by 
resonance fluorescence excitation mechanism.  
Hence, the observed N$_2^+$/CO$^+$ emission ion 
intensity ratio can be used to derive  their respective neutral density ratio. 
But for radial distances below 100 km, the photon and electron impact 
ionization and excitation of neutral species significantly 
controls the observed N$_2^+$/CO$^+$ ratio. Thus,  {only} the observed 
emission 
intensity ratios of these ions at larger radial distances are suitable to 
derive their respective neutral abundance ratio.

 {As discussed earlier, the densities of N$_2^+$ and CO$^+$ at large 
radial 
distances 
($>$10$^3$ km) 
are essentially governed by photoionization of corresponding neutrals and 
thermal recombination (see Figures~\ref{fig:pr_cop} and \ref{fig:pr_n2p}). The 
electron temperature, which determines the 
recombination rates of these ions and consequently their density, can influence 
the modelled emission intensity 
ratios. In the model we 
assumed 
electron temperature ($T_e$) in the coma  equivalent to that of neutral 
temperature, which is calculated by
\cite{Ip83}. This assumption causes a discrepancy between the modelled and 
observed N$_2^+$/CO$^+$ emission ratios of about a factor 2 (See 
Figure~\ref{fig:emis_ratios}).} Due to lower cooling rates, it is expected that 
the 
electron 
temperatures at larger radial distances can be much higher than our assumed 
value \citep{Gan90}. In order to evaluate the 
role of electron temperature on the modelled ion emission intensity ratio, we 
used the electron temperature profile for 1P/Halley from \cite{Eberhardt95}, 
which is determined  from Giotto measurements. 
We noticed that at larger 
radial distances, the \cite{Ip83} calculated neutral temperature is several 
orders of magnitude smaller  {than} the  electron temperature  derived by 
\cite{Eberhardt95}. 
By using the temperature profile of comet 1P/Halley in the model, we find that 
the 
modelled N$_2^+$/CO$^+$ emission intensity ratio profile  {decreases} by 
a factor 2 
and  {is then} 
consistent with the observation. This calculation suggests that electron 
temperature can also play an important role in determining the observed 
emission intensity ratio. Modelling the electron temperature profile in this 
CO-dominated 
cometary coma is beyond the scope of the present work. 

Using the modelled photoionization and recombination rates, we derive the 
following analytical expression to convert the observed flux ratio into 
neutral density ratio :

\begin{equation}
	\frac{[N_2]}{[CO]} =  \frac{I_{N_2^+}} {I_{CO^+}} \times {1.69 
	\times 10^{-2}} \times T_e^{0.16}  	
\end{equation}
The derivation of the above expression is provided in Appendix C.
By substituting our observed flux ratio (I$_{N_2^+}$/I$_{CO^+}$) in the above 
equation  { and for an electron temperature of 1000 K,} we obtained the 
neutral density ratio of [N$_2$]/[CO] as 0.06, which is consistent with the 
derived volume mixing ratio of N$_2$ from the other observations.  {It 
should be noted that larger electron temperature can lead to a higher 
abundance 
ratio of neutral species.}

\vspace{-10pt}
\subsubsection{Derivation of neutral density based on the CO$_2^+$/CO$^+$ 
	emission intensity ratio}
It should be noticed that by accounting for 18\% of CO$_2$ relative 
 to the CO production rate \citep[based 
 on the observations of][]{McKay19}, the 
	modelled  CO$_2^+$/CO$^+$ ion density ratio 
	significantly  {varies} for radial distances below 10$^4$ km (see 
	Figure \ref{fig:ion_den}). 
	As explained earlier, both the CO$_2^+$ and CO$^+$ ion densities are  
	significantly controlled by ion-neutral chemistry rather than 
	photoionization 
	and thermal recombination processes for the radial distances below 
	1000 km (see 
	Sections~\ref{sec:pl_co2p} and \ref{sec:pl_cop}). Hence, the conversion of 
	observed CO$_2^+$/CO$^+$ emission flux 
	ratio  to corresponding 
	neutral density ratio leads to an overestimation 
	of the CO$_2$ abundance in the coma.

{Our modelled volume emission rates of CO$^+$ and CO$_2^+$ for resonance 
fluorescence excitation are nearly the same for a radial distance smaller than 
500 km
 (see dash-dotted black curves in the top panels of Figure~\ref{fig:vemis}). 
This can be explained based on their g-factors and  ion density distribution in 
the 
 coma. It should be noticed that the g-factor of 
CO$_2^+$ is smaller  by an order of magnitude compared to that of  CO$^+$ (see 
the g-factor values for reactions E6 and E11 in Table~\ref{tab:photfrq}), 
whereas its density is higher by more than an order of magnitude 
compared to that of CO$^+$ for the radial distances below 500 km (see 
Figure~\ref{fig:ion_den}). In spite of the lower emission rate, the emission 
rate of CO$_2^+$  is comparable to that of CO$^+$  due its large density for 
these radial distances  in the coma. Above a radial 
distance of 1000 km the density of CO$^+$  is higher than that of CO$_2^+$ 
which results in larger volume emission rate. }

Assuming resonance fluorescence is the only primary excitation
source, \cite{Opitom19} converted the observed flux ratio 
(I$_{CO_2^+}$/I$_{CO^+}$) into the ion density 
ratio (CO$_2^+$/CO$^+$) using the following expression.
\begin{equation}
\frac{[CO_2^+]}{[CO^+] } = \frac{ g_{CO^+} I_{CO_2^+}}{g_{CO_2^+} I_{CO^+}}
\end{equation}
\cite{Opitom19} observed the average emission intensity ratio of 
CO$_2^+$/CO$^+$ of about 0.15, over the 
radial distance of 10$^4$ km, whereas the g-factors ratio 
(g$_{CO^+}$/g$_{CO_2^+}$) is about 7. 
Hence, the product of 
g-factors ratio and the observed flux ratio results in the  derived ion density 
as about 1. When  {we account} for 18\% CO$_2$ with respect to 
CO production rate and resonance fluorescence 
as  the only excitation mechanism for the CO$_2^+$ and CO$^+$ emissions, the 
modelled average emission intensity ratio over the projected distance of 5000 
km is about 0.1,  {which is} consistent with the observed ion emission 
intensity 
ratio (see Figure~\ref{fig:emis_ratios}).  {This agreement 
clearly suggests that the direct conversion of observed CO$_2^+$/CO$^+$  
emission ratio into their corresponding neutral density ratio leads to a wrong 
estimation of CO$_2$ abundance in the coma. The charge 
exchange between CO$^+$ and CO$_2$ significantly affects both the CO$^+$ and 
CO$_2^+$ ion 
densities in the coma and consequently the observed CO$_2^+$/CO$^+$  emission 
ratio. Hence, the ion-neutral chemistry  in this CO-dominated coma  plays an 
important role in modifying the ion composition in the coma and  explain 
the discrepancy between observed CO$_2$ abundance by \cite{McKay19} 
and  \cite{Opitom19} derived CO$_2^+$/CO$^+$ ion density ratios.}

\vspace{-10pt}
\subsubsection{Derivation of neutral density based on the H$_2$O$^+$/CO$^+$ 
	emission intensity ratio}
\cite{Opitom19}  derived an upper limit for the H$_2$O$^+$/CO$^+$ density   
of about 0.4 by accounting for solar resonance fluorescence as the 
only excitation source of the corresponding ionic emissions. Since the 
g-factors ratio 
(g$_{H_2O^+}$/g$_{CO^+}$) of these ionic emissions is close to one, the 
observed upper limit of the emission intensity ratio should also be about 0.4. 
Our modelled emission intensity ratio of these ions is smaller by more than two 
orders of 
magnitude compared to the derived upper limit of the observation (see 
Figure~\ref{fig:emis_ratios}). Moreover, 
our 
calculations in Figure~\ref{fig:pr_h2op} show that the formation and 
destruction of H$_2$O$^+$ are  majorly determined by  {the} CO$_2^+$ and 
CO 
distribution in the coma,  {respectively,} for radial distances below 
1000 km 
rather than 
photoionization of H$_2$O. Even 
at large radial distances the contribution from charge exchange reactions is 
significant in determining the H$_2$O$^+$ density. As discussed before,  no 
significant change is observed in the modelled H$_2$O$^+$/CO$^+$ emission 
intensity profiles by accounting for  only 
ionization and 
excitation mechanism or only solar resonance fluorescence excitation. This 
suggests that the these excitation processes  equally  {contribute}  in 
determining the emission intensity ratio.  Since this ion density is
majorly controlled by charge transfer reactions rather than photon and electron 
impact initiated reactions, we suggest that the observed emission intensity 
ratio is not 
suitable to constrain the density of H$_2$O in the cometary coma.  

\vspace{-15pt}

\section{Summary and Conclusions}
\label{sec:sum_con}
In the context of recent observations of  {the} CO-dominated and water 
poor comet 
C/2016 R2, we developed a physico-chemical model to study the ion density 
distribution and  {the emission processes of  excitation states of 
various ions}.
We have studied various formation and destruction mechanisms of different ions 
by incorporating the ionization of neutrals by photons and photoelectrons, 
charge exchange reactions, proton transfer reactions, and electron-ion thermal 
recombination reactions. Besides the fluorescence excitation mechanism, 
ionization and excitation of corresponding neutral species, which produce 
excited states of CO$^+$, H$_2$O$^+$, N$_2^+$, and CO$_2^+$, are also 
incorporated to study the emission mechanisms of these ions.  The major 
results of the present work can be summarized as follows. 

\renewcommand{\labelenumii}{\arabic{enumii}}
\begin{enumerate} [label={\arabic*.}]
		
	\item CO$_2^+$ is the major ion in this CO-dominated coma for the radial 
	distances smaller than 10$^3$ km. Above this distance CO$^+$ is 
	the most  {abundant one}. 
	
     \item The production of CO$_2^+$ is mainly controlled by charge transfer 
      between CO$^+$ and CO$_2$ for radial distances below 1000 km  
     and above this distance, photoionization of CO$_2$ is the major 
     source of CO$_2^+$.
     
     \item Photoionization of CO is the primary source of CO$^+$. The 
     collisions with CO$_2$ removes this ion for radial 
     distances smaller than 1000 km. Above this distance, thermal recombination 
     is the significant loss source of CO$^+$.
     
     \item The charge transfer reactions of CO$^+$, CO$_2^+$ and O$^+$ with 
     H$_2$O majorly determine  {the} H$_2$O$^+$ ion density in the inner 
     coma rather 
     than photoionization of H$_2$O. Collisions with CO remove H$_2$O$^+$ for 
     radial distances up to 5000 km, and above this distance thermal 
     recombination is the major loss processes.

     \item Photoionization of N$_2$ is the primary source of N$_2^+$, whereas 
     the loss of this ion is mainly due to collisions with CO$_2$ and CO for 
     radial distances up to 3000 km, above which thermal recombination
     takes over as primary loss source.
     
     \item The densities of protonated ions H$_3$O$^+$, NH$_4^+$, and 
     CH$_3$OH$_2^+$ are 
     majorly  linked with  {the} H$_2$O distribution in the inner coma 
     and the  
     formation chemistry of these ions is determined by CO and CO$_2$.

	\item Resonance fluorescence is the most significant excitation 
	mechanism for N$_2^+$ and CO$^+$ emissions, hence  the observed 
	N$_2^+$/CO$^+$ ion emission intensity ratio can be 
	used to derive their respective neutral density ratio in the cometary coma. 
	
	\item Since CO$_2^+$ and H$_2$O$^+$ ion densities in the 
	inner coma are significantly controlled by ion-neutral chemistry, rather 
	than photoionization of neutrals, the  observed  
	CO$_2^+$/CO$^+$ and H$_2$O$^+$/CO$^+$ emission intensity ratios can not be 
	used to derive their respective neutral densities. 
	
	\item  {If the ion densities in the coma are 
	  determined by ionization of corresponding neutrals and 
		thermal
		recombination, then the observed ion-emission intensity ratios can be 
		converted 
	into neutral density ratios. But if the ion density in the 
	coma is 
	essentially 
	determined by ion-neutral chemistry, then the conversion leads to a wrong
	estimation of neutral abundances in the coma}
	
    \item Our modelled emission intensity ratio profiles N$_2^+$/CO$^+$, 
    H$_2$O$^+$/CO$^+$, and CO$_2^+$/CO$^+$ are consistent with the recent 
    ground based observations.  
\end{enumerate}

\vspace{-15pt}
\section*{Acknowledgements}
SR is supported by Department of Science and Technology (DST) with 
Innovation in Science Pursuit for Inspired Research (INSPIRE) 
faculty award [Grant: DST/INSPIRE/04/2016/002687], and he would like 
to thank Physical Research Laboratory for facilitating conducive 
research environment. DH and EJ are FNRS Senior Research Associates. 
The authors would like to thank the anonymous reviewer for the valuable 
comments and 
suggestions that improved the manuscript.

\vspace{-15pt}

\section*{Data Availability}
Based on observations made with ESO Telescopes at the La Silla Paranal 
Observatory under program 2100.C-5035(A). The modelled data can be available on 
the reasonable request to the corresponding author.

\vspace{-15pt}




\appendix
\include{reaction_tables}


\twocolumn
\bsp	
\label{lastpage}
\end{document}

%% file: reaction_tables.tex
\onecolumn

\section{Chemical network of various ions}
\label{sec:appa}
\begin{table}  
	\centering
		\caption{The production and loss reactions of various ions considered 
		in the chemical network}
	\vspace{-0.25cm}
	 \label{tab:chem-net}
	 \resizebox{\columnwidth}{!}{%
	\begin{tabular}{lllllll}
		\hline\hline
	Number & \multicolumn{3}{c}{Reaction} & Rate (s$^{-1}$ or cm$^3$ s$^{-1}$) 
	& Reference \\
        \hline
%
	R1  & h$\nu$   + CO          & $\rightarrow$ & CO$^+$       + e  & 7.20
	    $\times$ 10$^{-7}$ & This work\\
	R2  & h$\nu$   + CO$_2$      & $\rightarrow$ & CO$^+$     + e    & 9.48 
	    $\times$ 10$^{-8}$  & This work\\
	R3  & h$\nu$   + CO$_2$      & $\rightarrow$ & CO$_2^+$     + e  & 1.25 
	    $\times$ 10$^{-6}$  & This work\\
	R4  & h$\nu$   + H$_2$O      & $\rightarrow$ & H$_2$O$^+$   + e  & 6.46 
	    $\times$ 10$^{-7}$ & This work\\
	R5  & h$\nu$   + N$_2$       & $\rightarrow$ & N$_2^+$      + e  & 6.95
	    $\times$ 10$^{-7}$ & This work\\
	R6  & h$\nu$   + O$_2$       & $\rightarrow$ & O$_2^+$      + e  & 9.10
	    $\times$ 10$^{-7}$ & This work\\
	R7  & h$\nu$   + CH$_3$OH    & $\rightarrow$ & CH$_3$OH$^+$ + e  & 1.10
	    $\times$ 10$^{-6}$ & This work\\
	R8  & h$\nu$   + CH$_4$      & $\rightarrow$ & CH$_4^+$     + e  & 7.20
	    $\times$ 10$^{-7}$ & This work\\
	
        R9  & h$\nu$   + CO          & $\rightarrow$ & O$^+$     + C + e  & 4.42
	    $\times$ 10$^{-8}$ & This work\\
	R10 & h$\nu$   + CO$_2$      & $\rightarrow$ & O$^+$     + CO + e    & 
	    1.18 $\times$ 10$^{-7}$  & This work\\
	R11  & h$\nu$   + H$_2$O      & $\rightarrow$ & O$^+$   + H$_2$ + e & 1.14
             $\times$ 10$^{-8}$ & This work\\
	R12 & h$\nu$   + O$_2$       & $\rightarrow$ & O$^+$   + O +  e  & 2.01
             $\times$ 10$^{-7}$ & This work\\   
        
        R13  & h$\nu$   + CO        & $\rightarrow$ & C$^+$     + O + e  & 5.45
              $\times$ 10$^{-8}$ & This work\\
        R14  & h$\nu$   + CO$_2$      & $\rightarrow$ & C$^+$     + O$_2$ + e
             &  5.42  $\times$ 10$^{-8}$  & This work\\             
        
	R15  & e$_{ph}$ + CO          & $\rightarrow$ & CO$^+$       + 2e & 
	     Calculated & This work\\
	R16  & e$_{ph}$ + CO$_2$     & $\rightarrow$ & CO$_2^+$     + 2e & 
	     Calculated & This work\\
	R17 & e$_{ph}$ + H$_2$O      & $\rightarrow$ & H$_2$O$^+$   + 2e & 
	     Calculated & This work\\
	R18 & e$_{ph}$ + N$_2$       & $\rightarrow$ & N$_2^+$      + 2e & 
	     Calculated & This work\\
	R19 & e$_{ph}$ + O$_2$       & $\rightarrow$ & O$_2^+$      + 2e & 
	     Calculated & This work\\ 
	R20 & e$_{ph}$ + CH$_3$OH    & $\rightarrow$ & CH$_3$OH$^+$ + 2e & 
	     Calculated & This work\\
	R21 & e$_{ph}$ + CH$_4$      & $\rightarrow$ & CH$_4^+$     + 2e & 
	     Calculated & This work\\ 
	
	R22  & e$_{ph}$  + CO          & $\rightarrow$ & O$^+$       + C + 2e  & 
	     Calculated & This work\\
	R23  & e$_{ph}$  + CO$_2$      & $\rightarrow$ & O$^+$     + CO + 2e    & 
	     Calculated  & This work\\
	R24  & e$_{ph}$   + H$_2$O      & $\rightarrow$ & O$^+$   + H$_2$ + 2e  &
	     Calculated & This work\\
	R25  & e$_{ph}$   + O$_2$       & $\rightarrow$ & O$^+$   + O +  2e  & 
	     Calculated & This work\\   
	
	R26  & e$_{ph}$  + CO          & $\rightarrow$ & C$^+$     + O +2e  &  
	     Calculated & This work\\
	R27  & e$_{ph}$   + CO$_2$      & $\rightarrow$ & C$^+$     + O$_2$ + 2e
	     &  Calculated  & This work\\
	
	R28 & H$_2$O$^+$ + H$_2$O    & $\rightarrow$ & H$_3$O$^+$  + OH  & 2.10 
    	$\times$ 10$^{-9}$ T$_n$  & \cite{Huntress73}\\
	R29 & H$_2$O$^+$ + CH$_4$    & $\rightarrow$ & H$_3$O$^+$  + CH$_3$    & 1.1
	    $\times$ 10$^{-9}$ T$_n$ & \cite{Huntress80}\\
	R30 & H$_2$O$^+$ + CO        & $\rightarrow$ & HCO$^+$     + OH    & 5.0 
	    $\times$ 10$^{-10}$        & \cite{Jones81}\\                  
	R31 & H$_2$O$^+$ + NH$_3$    & $\rightarrow$ & NH$_4^+$    + OH    & 9.45 
	    $\times$ 10$^{-10}$ T$_n$ & \cite{Anicich93}\\    
	R32 & H$_2$O$^+$ + NH$_3$    & $\rightarrow$ & NH$_3^+$    + H$_2$O    & 
	    2.21 $\times$ 10$^{-9}$ T$_n$ & \cite{Anicich93}\\    
	R33 & H$_2$O$^+$ + O$_2$    & $\rightarrow$ & O$_2^+$     + H$_2$O    & 4.6 
    	$\times$ 10$^{-10}$        & \cite{Rakshit80}\\
	
	R34 & H$_3$O$^+$ + NH$_3$   & $\rightarrow$ & NH$_4^+$       + H$_2$O   & 
        2.21 $\times$ 10$^{-9}$ T$_n$  & \cite{Smith80}\\
	R35 & H$_3$O$^+$ + CH$_3$OH & $\rightarrow$ & CH$_3$OH$_2^+$ + H$_2$O   & 
    	2.5 $\times$ 10$^{-9}$  T$_n$  & \cite{Anicich93}\\
	
	R36 & CO$^+$     + H$_2$O   & $\rightarrow$ & H$_2$O$^+$ + CO           & 
	    1.79 $\times$ 10$^{-9}$ T$_n$  & \cite{Huntress80}\\  
	R37 & CO$^+$     + H$_2$O   & $\rightarrow$ & HCO$^+$    + CO           & 
	    8.84 $\times$ 10$^{-10}$ T$_n$ & \cite{Huntress80}\\  
	R38 & CO$^+$     + NH$_3$   & $\rightarrow$ & NH$_3^+$   + CO           & 
	    2.02 $\times$ 10$^{-9}$ T$_n$  & \cite{Huntress80}\\  
	R39 & CO$^+$     + CO$_2$   & $\rightarrow$ & CO$_2^+$   + CO           & 
	    1.00 $\times$ 10$^{-9}$        & \cite{Adams78}\\
	R40 & CO$^+$     + O$_2$    & $\rightarrow$ & O$_2^+$    + CO           & 
	    2.00 $\times$ 10$^{-10}$       & \cite{Ferguson73}\\  
	R41 & CO$^+$     + CH$_4$   & $\rightarrow$ & CH$_4^+$ + CO               & 
	    7.93 $\times$ 10$^{-10}$       & \cite{Adams78}\\      
	R42 & CO$^+$     + CH$_4$  & $\rightarrow$ & CH$_3$CO$^+$ + H            & 
	    5.20 $\times$ 10$^{-11}$       & \cite{Adams78}\\           
	R43 & CO$^+$     + CH$_4$  & $\rightarrow$ & HCO$^+$ + CH$_3$            & 
	    4.55 $\times$ 10$^{-10}$       & \cite{Adams78}\\ 
	
	R44 & CO$_2^+$   + H$_2$O  & $\rightarrow$ & H$_2$O$^+$ + CO$_2$       & 
	    2.04 $\times$ 10$^{-9}$ T$_n$  & \cite{Karpas78} \\ 
	R45 & CO$_2^+$   + H$_2$O  & $\rightarrow$ & HCO$_2^+$ + OH       & 7.56 
	    $\times$ 10$^{-10}$ T$_n$  & \cite{Karpas78} \\      
	R46 & CO$_2^+$   + NH$_3$  &  $\rightarrow$ & NH$_3^+$   + CO$_2$       & 
	    1.90 $\times$ 10$^{-9}$ T$_n$  & \cite{Copp82}\\ 
	R47 & CO$_2^+$   + O$_2$   & $\rightarrow$ & O$_2^+$    + CO$_2$       & 
	    5.3 $\times$ 10$^{-11}$    & \cite{Copp82}\\
	R48 & CO$_2^+$  + CH$_4$  & $\rightarrow$ & CH$_4^+$ +       
	    CO$_2$ & 5.5 $\times$ 10$^{-10}$       & \cite{Copp82}\\
	\hline
	
\end{tabular}}
	
\end{table}

\begin{table}  
	\centering
	\contcaption{table}
	\vspace{-0.25cm}
        \label{tab:continued}
	\resizebox{\columnwidth}{!}{%
\begin{tabular}{lllllll}
	\hline \hline
	Number & \multicolumn{3}{c}{Reaction} & Rate (s$^{-1}$ or cm$^3$ 
	s$^{-1}$) 
	& Reference \\
	\hline
	R49 & N$_2^+$    + O$_2$   & $\rightarrow$ & O$_2^+$    + 
	    N$_2$  & 3.5 $\times$ 10$^{-10}$        & \cite{Anicich93}\\
	R50 & N$_2^+$    + CO$_2$  & $\rightarrow$ & CO$_2^+$   + 
	    N$_2$  & 7.7 $\times$ 10$^{-10}$         & \cite{Adams80}\\   
	R51 & N$_2^+$    + CH$_4$  & $\rightarrow$ & CH$_3^+$ + N$_2$ + 
	    H & 9.3 $\times$ 10$^{-10}$         & \cite{Adams80}\\ 
	R52 & N$_2^+$    + CH$_4$  & $\rightarrow$ & CH$_2^+$ + N$_2$ + 
	    H$_2$  & 7.0 $\times$ 10$^{-11}$         & \cite{Adams80}\\
	R53 & N$_2^+$    + H$_2$O  & $\rightarrow$ & H$_2$O$^+$ + 
	    N$_2$ & 2.8 $\times$ 10$^{-9}$ & 			    
	    \cite{Ferguson73}\\ 
	R54 & N$_2^+$    + H$_2$O   & $\rightarrow$ & N$_2$H$^+$ + 
	    OH   & 	2.12 $\times$ 10$^{-9}$  & 
	    \cite{Ferguson73}\\                   
	R55 & N$_2^+$    + CO     & $\rightarrow$ & CO$^+$     + 
	    N$_2$  & 7.40 	$\times$ 10$^{-11}$       & \cite{Adams80}\\ 
	R56 & N$_2^+$    + N$_2$  & $\rightarrow$ & N$_3^+$    + 
	    N  & 5.5 $\times$ 10$^{-11}$        & \cite{Bowers74}\\
	R57 & N$_2^+$    + NH$_3$  & $\rightarrow$ & NH$_3^+$    + 
	    N$_2$ & 1.9 $\times$ 10$^{-9}$T$_n$        & 
	    \cite{Adams80}\\
	R58 & O$_2^+$    + NH$_3$ & $\rightarrow$ & NH$_3^+$     + 
	    O$_2$      & 2.0 $\times$ 10$^{-9}$ T$_n$   & \cite{Adams80}\\  
	R59 & O$_2^+$    + CH$_4$  & $\rightarrow$ & HCOOH$_2^+$  + 
	    H  & 3.2 $\times$ 10$^{-10}$        & \cite{Rowe84}\\           
	R60 & O$_2^+$    + CH$_3$OH & $\rightarrow$ & CH$_3$OH$^+$ + 
	    O$_2$  & 5.0 $\times$ 10$^{-10}$ T$_n$  & \cite{Adams78}  \\
	R61 & CH$_4^+$ + O$_2$ & $\rightarrow$ & O$_2^+$  +	CH$_4$  
	    & 3.9 $\times$ 10$^{-10}$        & \cite{Anicich93} \\
	R62 & CH$_4^+$   + CO$_2$  & $\rightarrow$ & HCO$_2^+$    + 
	      CH$_3$  & 1.2 $\times$ 10$^{-9}$         & \cite{Smith77} \\
	R63 & CH$_4^+$   + CO      & $\rightarrow$ & HCO$^+$      + 
	    CH$_3$     & 1.4 $\times$ 10$^{-9}$         & \cite{Smith77} \\
	R64 & CH$_4^+$   + H$_2$O  & $\rightarrow$ & H$_3$O$^+$   + 
	    CH$_3$   & 	2.6 $\times$ 10$^{-9}$ T$_n$   & \cite{Smith77} \\
	R65 & CH$_4^+$   + NH$_3$  & $\rightarrow$ & NH$_4^+$     + 
	    CH$_3$  & 1.15 	$\times$ 10$^{-9}$ T$_n$  & \cite{Smith77} \\
	R66 & CH$_4^+$   + CH$_3$OH & $\rightarrow$ & CH$_3$OH$_2^+$ + 
	    CH$_3$  & 1.20 	$\times$ 10$^{-9}$ T$_n$  & \cite{Adams78} \\
	R67 & CH$_4^+$   + CH$_3$OH & $\rightarrow$ & CH$_3$OH$^+$ + 
	    CH$_4$  & 1.80 	$\times$ 10$^{-9}$ T$_n$  & \cite{Adams78} \\		    
	R68 & CH$_4^+$   + CH$_4$   & $\rightarrow$ & CH$_5^+$     + 
	    CH$_3$  & 1.5 $\times$ 10$^{-9}$         & \cite{Smith77} \\
	R69 & CH$_3$OH$^+$   + H$_2$O  &$\rightarrow$ & H$_2$O$^+$     + 
	    CH$_3$OH & 1.5 	$\times$ 10$^{-9}$  & \cite{Haider05} \\
	R70 & NH$_3^+$   + H$_2$O  &  $\rightarrow$ & NH$_4^+$ + OH
        & 1.1 $\times$ 10$^{-10}$        & \cite{Anicich77} \\
	R71 & NH$_3^+$   + NH$_3$  &  $\rightarrow$ & NH$_4^+$ + N
	    & 2.2 $\times$ 10$^{-9}$ T$_n$  & \cite{Adams80} \\
	R72 & NH$_3^+$   + CO      & $\rightarrow$ & CO$^+$ + NH$_3$  
	    & 1.1 $\times$ 10$^{-9}$ T$_n$   & \cite{Adams80} \\
	R73 & CH$_3$OH$_2^+$   + NH$_3$  &$\rightarrow$& NH$_4^+$ + CH$_3$OH  
	    & 1.1 $\times$ 10$^{-9}$ T$_n$   & \cite{McElroy13} \\
    R74 & C$^+$  + CH$_4$ & $\rightarrow$ &  C$_2$H$_2^+$ + H$_2$ 
        & 3.89 $\times$  10$^{-10}$	& \cite{Schiff79} \\
    R75 & C$^+$  + CH$_4$ & $\rightarrow$ &  C$_2$H$_3^+$ + H	   
        & 1.00 $\times$ 10$^{-9}$	& \cite{Schiff79}\\
    R76 & C$^+$  + CO$_2$ & $\rightarrow$ &  CO$^+$	    + CO   
        & 1.10 $\times$ 10$^{-9}$	& \cite{Fahey81} \\
    R77 & C$^+$  + H$_2$O & $\rightarrow$ &  HOC$^+$	    + H	  
        & 2.09 $\times$ 10$^{-9}$	& \cite{Anicich76} \\
    R78 & C$^+$  + H$_2$O & $\rightarrow$ &  HCO$^+$	    + H	   
        & 9.00 $\times$ 10$^{-10}$	& \cite{Anicich76} \\
    R79 & C$^+$  + NH$_3$ & $\rightarrow$ &  H$_2$NC$^+$  + H	   
        & 1.61 $\times$ 10$^{-9}$	& \cite{Smith77}\\
    R80 & C$^+$  + NH$_3$ & $\rightarrow$ &  NH$_3^+$  + C	   
        & 6.72 $\times$ 10$^{-10}$	& \cite{Smith77}\\
    R81 & C$^+$  + O$_2$  & $\rightarrow$ &  CO$^+$	    + O   
        & 3.40 $\times$ 10$^{-10}$	& \cite{Smith77} \\
    R82 & C$^+$  + O$_2$  & $\rightarrow$ &  O$^+$	    + CO    
        & 4.50 $\times$ 10$^{-10}$	& \cite{Smith77}\\
    R83 & O$^+$   +	CO	& $\rightarrow$ & CO$^+$     +	O   
        &  4.90 $\times$ 10$^{-12}$	T$_n$ & \cite{Petuchowski89} \\
    R84 & O$^+$   +	H$_2$O	& $\rightarrow$ & H$_2$O$^+$ +	O   
        &  3.20 $\times$ 10$^{-9}$	T$_n$ & \cite{Adams80} \\
    R85 & O$^+$   +	NH$_3$	& $\rightarrow$ & NH$_3^+$   +	O   
        &  1.20 $\times$ 10$^{-9}$	T$_n$ & \cite{Adams80} \\
    R86 & O$^+$   +	O$_2$	& $\rightarrow$ & O$_2^+$    +	O   
        &  1.90 $\times$ 10$^{-11}$	   & \cite{Adams80} \\
    R87 & O$^+$   +	CH$_4$OH& $\rightarrow$ & H$_3$CO$^+$ + OH  
        &  1.33 $\times$ 10$^{-9}$ T$_n$  & \cite{Adams80} \\
    R88 & O$^+$   +	CH$_4$	& $\rightarrow$ & CH$_3^+$   +	OH  
        &  1.10 $\times$ 10$^{-10}$	   & \cite{Adams80} \\
    R89 & O$^+$   +	CO$_2$	& $\rightarrow$ & O$_2^+$    +	CO 
        &  9.40 $\times$ 10$^{-10}$	   & \cite{Adams80} \\
    R90 & O$^+$   +	N$_2$	& $\rightarrow$ & NO$^+$     +	N   
        &  2.40 $\times$ 10$^{-12}$	T$_n$ & \cite{Adams80} \\
    R91 & O$^+$   +	CH$_4$OH& $\rightarrow$ & CH$_3$OH$^+$ + O 
        &  4.75 $\times$ 10$^{-10}$ T$_n$  & \cite{Adams80} \\  
	\hline
\end{tabular} }
{Photoionization rates presented in this table 
are calculated at 1 au heliocentric distance; T$_n$ = (300/T)$^{0.5}$; \\ 
h$\nu$ and e$_{ph}$ are 
solar photon and 
photoelectron,	respectively.}
\end{table}


\begin{table}
	\centering
	\caption{Thermal recombination reactions of various ions 
			considered in the chemical network. }
			\vspace{-0.25cm}
	\label{tab:recomb_rect} 
	\begin{tabular}{llllllllll}
    \hline\hline
	Number & \multicolumn{3}{c}{Reaction} & Rate (cm$^3$ s$^{-1}$) & Reference\\
	\hline

	L1& H$_3$O$^+$ + e$_{th}$ & $\rightarrow$ & Products & 4.30 $\times$ 
	10$^{-7}$  
	$\times$ (300/T)$^{0.83}$ 
	& 
	\cite{Novotny10}\\	
	L2& H$_2$O$^+$ + e$_{th}$ & $\rightarrow$ & Products & 4.30 $\times$ 
	10$^{-7}$  
	$\times$ (300/T)$^{0.50}$ 
	& \cite{Rosen00}\\
	L3& CO$_2^+$ + e$_{th}$ & $\rightarrow$ & Products & 3.50 $\times$ 
	10$^{-7}$  
	$\times$ (300/T)$^{0.50}$ 
	& \cite{Geoghegan91}\\
	L4& CO$^+$ + e$_{th}$ & $\rightarrow$ & Products   & 2.75 $\times$ 
	10$^{-7}$  
	$\times$ (300/T)$^{0.50}$ 
	& \cite{Rosen98}\\
	L5& N$_2^+$ + e$_{th}$ & $\rightarrow$ & Products   & 2.20 $\times$ 
	10$^{-7}$  
	$\times$ (300/T)$^{0.39}$ 
	& \cite{Sheehan04}\\
	L6& O$_2^+$ + e$_{th}$ & $\rightarrow$ & Products   & 1.95 $\times$ 
	10$^{-7}$  
	$\times$ (300/T)$^{0.70}$ 
	& \cite{Sheehan04}\\
	L7& NH$_4^+$ + e$_{th}$ & $\rightarrow$ & Products  & 9.34 $\times$ 
	10$^{-7}$  
	$\times$ (300/T)$^{0.60}$ 
	& \cite{Ojekull04}\\
	L8& NH$_3^+$ + e$_{th}$ & $\rightarrow$ & Products  & 3.10 $\times$ 
	10$^{-7}$  
	$\times$ (300/T)$^{0.60}$ 
	& \cite{Brian90}\\
	L9& CH$_3$OH$^+$ + e$_{th}$ & $\rightarrow$ & Products  & 1.80 $\times$ 
	10$^{-6}$  
	$\times$ 
	(300/T)$^{0.66}$ 
	& \cite{Hamberg07}\\
	L10& CH$_3$OH$_2^+$ + e$_{th}$ & $\rightarrow$ & Products  & 8.36 $\times$ 
	10$^{-7}$  
	$\times$ (300/T)$^{0.66}$ & \cite{Geppart06}\\
	L11& CH$_4^+$ + e$_{th}$ & $\rightarrow$ & Products   & 3.50 $\times$ 
	10$^{-7}$  
	$\times$ 
	(300/T)$^{0.50}$ & \cite{Mitchell90}\\
    \hline

	\end{tabular}

    {e$_{th}$ and T are thermal electron and electron 
	temperature, respectively.}
\end{table}


\section{Chemical reactions for various ions in the excited states}
\label{sec:appb}

\begin{table}
		\centering
		\caption{Reactions for the formation of different excited states of 
		ions via photoionization, electron impact ionization of neutrals and 	
		resonance fluorescence excitation of ion.}
    	\label{tab:photfrq}
		\vspace{-0.25cm}
\begin{tabular}{llllllllllll}
    \hline \hline
	Number & \multicolumn{3}{c}{Reaction}  & Frequency & 
	Reference\\
	 &	 \multicolumn{3}{l}{}  &  (s$^{-1}$ or photons s$^{-1}$ 
molecule$^{-1}$) & 
	\\
	\hline
	E1 & h$\nu$ + H$_2$O & $\rightarrow$ &
	H$_2$O$^+$($\widetilde{A}^2$A$_1$) + e$_{ph}$
	& 1.04 $\times$ 10$^{-7}$ & This work \\
	E2 & e$_{ph}$ + H$_2$O  & $\rightarrow$ & 
	H$_2$O$^+$($\widetilde{A}^2$A$_1$) + 2e      & Calculated & This work \\
	E3 & h$\nu$ + H$_2$O$^+$ & $\rightarrow$ & 
	H$_2$O$^+$($\widetilde{A}^2$A$_1$) 
	& 
	4.2 $\times$ 10$^{-3}$ & \cite{Lutz93} \\ 
	E4 &  h$\nu$ + CO$_2$ & $\rightarrow$  &CO$_2^+$($\widetilde{A}^2\Pi_u$) + 
	e$_{ph}$
	& 2.50 $\times$ 10$^{-7}$ & This work \\
	E5 &  e$_{ph}$ + CO$_2$ & $\rightarrow$ & CO$_2^+$($\widetilde{A}^2\Pi_u$) 
	+ 2e 
	& Calculated & This work \\ 
	E6 &  h$\nu$ + CO$_2^+$ & $\rightarrow$ & CO$_2^+$($\widetilde{A}^2\Pi_u$) 
	&   
	4.96 	$\times$ 10$^{-4}$  & \cite{Kim99}  \\
	E7 &  h$\nu$ + CO$_2$ & $\rightarrow$ & CO$^+$(A$^2\Pi$) + e$_{ph}$
	& 9.05 $\times$ 10$^{-8}$ & This work \\     
	E8 &  h$\nu$ + CO & $\rightarrow$ & CO$^+$(A$^2\Pi$) + e$_{ph}$
	& 2.44 $\times$ 10$^{-7}$ & This work \\
	E9 &  e$_{ph}$ + CO$_2$ & $\rightarrow$ & CO$^+$(A$^2\Pi$) + 2e 
	& Calculated & This work \\     
	E10 &  e$_{ph}$ + CO & $\rightarrow$ & CO$^+$(A$^2\Pi$) + 2e 
	& Calculated & This work \\
	E11 &  h$\nu$ + CO$^+$ & $\rightarrow$ & CO$^+$(A$^2\Pi$) &  3.55 $\times$ 
	10$^{-3}$  	& \cite{Magnani86}  \\ 
	E12 &  	h$\nu$ + N$_2$ & $\rightarrow$ & N$_2^+$(B$^2\Sigma^+_u$) + e$_{ph}$
	& 5.48 $\times$ 10$^{-8}$ & This work  \\   
	E13 &  e$_{ph}$ + N$_2$ & $\rightarrow$ & N$_2^+$(B$^2\Sigma^+_u$) + 2e 
	& Calculated& This work  \\
	E14 &  h$\nu$ + N$_2^+$ & $\rightarrow$ & N$_2^+$(B$^2\Sigma^+_u$) &  7  
	$\times$ 
	10$^{-2}$ & \cite{Lutz93} \\
	\hline
\end{tabular}

	{h$\nu$, e$_{ph}$, and e are solar photon and photoelectron, 
	and electron, respectively. The photoionization frequencies and		
	g-factors 
	presented in this table are  at 1 au heliocentric distance.}
\end{table}

\twocolumn
\section{Derivation of N$_2$/CO density ratio based on the observed ion 
	emission ratio}
%
\begin{equation}
\begin{split}
\frac{I_{N_2^+}}{I_{CO^+}} & = \frac{g_{N_2^+}}{g_{CO^+}} 
\frac{[N_2^+]}{[CO^+]}\\
& = \frac{g_{N_2^+}}{g_{CO^+}} \frac{[N_2]q_{n_2^+}}{n_e R_{n_2^+}} 
\frac{n_e R_{co^+}}{[CO]q_{co^+}}\\
& = \frac{g_{N_2^+}}{g_{CO^+}} \frac{[N_2]}{[CO]} \frac{q_{n_2^+}} {q_{co^+}} 
\frac{R_{co^+}}{R_{n_2^+}}\\ 
\end{split} 
\end{equation}

\begin{equation}
\begin{split}
\frac{[N_2]}{[CO]} = & \frac{I_{N_2^+}}{I_{CO^+}}  
\frac{g_{CO^+} q_{co^+} R_{n_2^+} }{g_{N_2^+} q_{n_2^+} R_{co^+}} \\
\end{split}
\label{eq:deriv} 
\end{equation}

here $q_{co^+}$ and $q_{n_2^+}$ are the photoionization frequencies of CO and 
N$_2$ producing CO$^+$ and N$_2^+$ ions, respectively (see the calculated 
values for reactions R1 and R5 in Appendix Table~\ref{tab:chem-net}). 
$R_{co^+}$ and $R_{n_2^+}$ are recombination rates of CO$^+$ and N$_2^+$ ions, 
respectively (see reaction rates L4 and L5 in Appendix 
Table~\ref{tab:recomb_rect}). 
$n_e$ is total electron density. Substituting the constant values in equation 
\ref{eq:deriv}, we get 
\begin{equation}
\begin{split}
\frac{[N_2]}{[CO]} = & \frac{I_{N_2^+}}{I_{CO^+}} \times 1.69 \times 10^{-2} 
\times 
T_e^{0.16} \\
\end{split} 
\end{equation}

here $T_e$ is electron temperature.
